\documentclass[fleqn,usenatbib,useAMS]{mnras}

\usepackage{graphicx}	
\usepackage{amsmath}	
\usepackage{amssymb}	
\usepackage{multicol}        
\usepackage{bm}		
\usepackage{pdflscape}	

\usepackage[normalem]{ulem} 

\newcommand{\kms}{\,km\,s$^{-1}$} 

\newcommand{\second}{2^{\rm nd}}
\newcommand{\fenr}{f_{\rm poll}}

\def\ncolldot{\dot{N}_{\rm coll}}
\def\ncoll{N_{\rm coll}}
\def\Eorb{\epsilon_{\rm orb}}

\def\dr{{\rm d}}
\def\feh{{\rm [Fe/H]}}
\def\feh{{\rm [Fe/H]}}
\def\fch{f_{\rm ch}}

\def\vesc{v_{\rm esc,\, SMS}}
\def\vesccl{V_{\rm esc}}
\def\vinf{v_\infty}

\def\finit{F_0}
\def\sigmae{\sigma_{\rm e}}

\def\smssub{{\rm  SMS}}
\def\rsms{r_{\smssub}}
\def\rsmsn{r_{\smssub0}}

\def\msms{m_{\smssub}}
\def\mproc{m_{\rm proc}}
\def\lsms{l_{\smssub}}
\def\msmsn{m_{\smssub0}}
\def\msmsdot{\dot{m}_{\smssub}}
\def\mwind{m_{\rm wind}}
\def\mdotsms{\dot{m}_{\smssub}}

\def\mdotsev{\dot{M}_{\rm sev}}
\def\mgcdotacc{\dot{M}_{\rm acc}}

\newcommand{\Teff}{\mbox{$T_{\rm{eff}}$}}

\def\mdotsmsrel{\dot{m}_{\smssub}^{\rm coll,\,rel}}
\def\mdotsmscoll{\dot{m}_{\smssub}^{\rm coll}}
\def\mdotacc{\dot{m}_{\rm acc}}
\def\mdotaccav{\langle\dot{m}_{\rm acc}\rangle}

\def\mdotwind{\dot{m}_{\rm wind}}

\def\Edotcoll{\dot{E}_{\rm bin}}

\def\rhoc{\rho_{\rm c}}

\def\rh{R_{\rm h}}
\def\rhn{R_{\rm h0}}
\def\rstall{R_{\rm stall}}

\def\rhoh{\rho_{\rm h}}

\def\rhohn{\rho_{\rm h0}}

\def\rhdot{\dot{R}_{\rm h}}

\def\tkh{\tau_{\rm KH}}
\def\thelium{\tau_{\rm He}}

\def\trh{\tau_{\rm rh}}
\def\trhn{\tau_{\rm rh0}}

\def\tacc{\tau_{\dot{M}}}
\def\taccn{\tau_{\dot{M}0}}

\def\tsms{\tau_{\dot{m}_{\smssub}}}
\def\tsmsn{\tau_{\dot{m}_{\smssub}0}}

\def\tdyn{\tau_{\rm dyn}}
\def\tdynn{\tau_{\rm dyn0}}

\def\tcontr{t_{\rm contr}}
\def\tinf{t_{\infty}}

\def\nrel{N_{\rm rel}}

\def\vrms{V_{\rm rms}}

\def\msun{{\rm M}_\odot}
\newcommand{\msunyr}{{\rm M}_{\odot}/{\rm yr}}

\def\rsun{{\rm R}_\odot}
\def\lsun{{\rm L}_\odot}
\def\ledd{l_{\rm Edd}}
\def\yr{{\rm yr}}

\def\myr{{\rm Myr}}
\def\gyr{{\rm Gyr}}

\def\kms{{\rm{km}\,{\rm s}^{-1}}}

\def\pc{{\rm{pc}}}

\def\xenr{X_{\rm processed}}
\def\xprist{X_{\rm pristine}}
\def\xmixed{X_{\rm mixed}}

\def\scipy{{\textsc{scipy}}}
\def\integrate{\textsc{integrate}}

\usepackage[T1]{fontenc}
\usepackage{ae,aecompl}

\usepackage{txfonts,color}
\usepackage{soul}
\usepackage{subfig}

\title[Concurrent formation of supermassive stars and GCs]
{Concurrent formation of supermassive stars and globular clusters: implications for early self-enrichment }
\author[Gieles et al. ]{Mark Gieles$^{1}$\thanks{Contact e-mail:
    \href{mailto:m.gieles@surrey.ac.uk}{m.gieles@surrey.ac.uk}},
  Corinne Charbonnel$^{2,3}$, Martin G.H. Krause$^4$, Vincent
  H\'{e}nault-Brunet$^{5,6}$, \newauthor Oscar Agertz$^{7}$, Henny
  J.G.L.M. Lamers$^8$, Nathan Bastian$^9$, Alessia
  Gualandris$^1$,\newauthor Alice Zocchi$^{10,11}$, James A. Petts$^1$
  \\
$^{1}$Department of Physics, University of Surrey, Guildford, GU2 7XH, UK\\
$^2$ Department of Astronomy, University of Geneva, Chemin des Maillettes 51, 1290, Versoix, Switzerland\\
$^3$ IRAP, UMR~5277, CNRS and Universit\'e de Toulouse,
14, avenue \'{E}douard Belin, 31400 Toulouse, France \\
$^4$ Centre for Astrophysics Research, School of Physics, Astronomy and Mathematics, University of Hertfordshire, College Lane, Hatfield AL10 9AB, UK\\
$^5$ National Research Council, Herzberg Astronomy \& Astrophysics, 5071 West Saanich Road, Victoria, BC, V9E 2E7, Canada\\
$^6$ Department of Astrophysics/IMAPP, Radboud University, PO Box 9010, 6500 GL, Nijmegen, The Netherlands\\
$^7$ Lund Observatory, Department of Astronomy and Theoretical Physics, Lund University, Box 43, SE-221 00 Lund, Sweden\\
$^8$ Astronomical Institute Anton Pannekoek, University of Amsterdam, Science Park 904, NL-1098XH, Amsterdam, The Netherlands\\
$^9$ Astrophysics Research Institute, Liverpool John Moores University, 146 Brownlow Hill, Liverpool L3 5RF, UK\\
$^{10}$ Dipartimento di Fisica e Astronomia, Universit{\`a} degli Studi di Bologna, viale Berti Pichat 6/2, I40127, Bologna, Italy\\
$^{11}$ European Space Research and Technology Centre, Keplerlaan 1, 2200 AG Noordwijk, Netherlands
}

\date{Accepted 2018 April 23. Received 2018 April 23; in original form 2017 November 24}

\pubyear{2018}

\begin{document}
\label{firstpage}
\pagerange{\pageref{firstpage}--\pageref{lastpage}}
\maketitle

\begin{abstract}
We present a model for the concurrent formation of globular clusters
(GCs) and supermassive stars (SMSs, $\gtrsim 10^3\,\msun$) to address
the origin of the HeCNONaMgAl abundance anomalies in GCs. GCs form in
converging gas flows and accumulate low-angular momentum gas, which
accretes onto protostars. This leads to an adiabatic contraction of
the cluster and an increase of the stellar collision rate. A SMS can
form via runaway collisions if the cluster reaches sufficiently high
density before two-body relaxation halts the contraction. This
condition is met if the number of stars $\gtrsim10^6$ and the gas
accretion rate $\gtrsim 10^5\,\msun/\myr$, reminiscent of GC formation
in high gas-density environments, such as -- but not restricted to --
the early Universe. The strong SMS wind mixes with the inflowing
pristine gas, such that the protostars accrete diluted hot-hydrogen
burning yields of the SMS. Because of continuous rejuvenation, the
amount of processed material liberated by the SMS can be an order of
magnitude higher than its maximum mass. This `conveyor-belt'
production of hot-hydrogen burning products provides a solution to the
mass budget problem that plagues other scenarios. Additionally, the
liberated material is mildly enriched in helium and relatively rich in
other hot-hydrogen burning products, in agreement with abundances of
GCs today. Finally, we find a super-linear scaling between the amount
of processed material and cluster mass, providing an explanation for
the observed increase of the fraction of processed material with GC
mass.  We discuss open questions of this new GC enrichment scenario
and propose observational tests.
 \end{abstract}




\section{Introduction}
\label{sec:intro}
 What started as a curiosity of the horizontal branch morphology of
 GCs -- the so-called `$\second$ parameter problem'
 \citep[][]{1967ApJ...150..469S} -- has become the largest unsolved
 problem of GC stellar populations. Nearly all old and massive GCs
 ($\gtrsim10\,\gyr,\,\gtrsim10^5\,\msun$) display anti-correlated C--N
 and O--Na abundances \citep[e.g.][]{2009A&A...505..117C,
   2010A&A...516A..55C}.  A fraction of the GCs (preferentially the
 most massive and most metal-poor ones, with some exceptions) also
 display anti-correlated Mg--Al abundances
 (\citealt{2009A&A...505..139C};
 \citealt{2015AJ....149..153M,2017A&A...601A.112P}).  Recently,
 \citet{2017MNRAS.465L..39H} found spectroscopic evidence for N
 enhancement in the 8\,\gyr\ old Small Magellanic Cloud (SMC) cluster
 Lindsay~1.

In addition to these spectroscopic peculiarities, most GCs present
photometric signatures of the presence of multiple stellar populations
(MSPs), e.g. broadened or multiple sequences in different areas of the
colour-magnitude diagram (e.g. \citealt{2002ASPC..265...87A};
\citealt{2004ApJ...605L.125B}; \citealt{2015AJ....149...91P,
  2007ApJ...661L..53P,2012ApJ...745...27M,
  2013ApJ...767..120M,2018MNRAS.473.2688M}).  Using {\it Hubble Space
  Telescope (HST)} imaging in filters that are sensitive to C, N and O
variations, \citet{2017MNRAS.465.4159N} found N enhancement in stars
in three clusters with ages of $6-8\,\gyr$ in the SMC. This was also
found in the 2 Gyr old cluster NGC\,1978 in the Large Magellanic Cloud
\citep[][]{2018MNRAS.473.2688M}.  So far, no evidence for MSPs has
been found in clusters younger than $\sim2\,\gyr$
\citep[e.g.][]{2008AJ....136..375M,2015MNRAS.448.2224C,2016MNRAS.460.1869C,2018MNRAS.473.2688M}.
There is currently no explanation for the origin of these ubiquitous
MSPs in old star clusters, but the consensus is that a fraction of the
GC stars contain products of hot-hydrogen burning
\citep*{2017A&A...608A..28} which, via the CNO-cycle
($\gtrsim20\,$MK), the NeNa-chain ($\gtrsim45\,$MK) and the MgAl-chain
($\gtrsim70\,$MK), gives rise to anti-correlations between C--N, O--Na
and Mg--Al, respectively (\citealt*{1990SvAL...16..275D,
  1993PASP..105..301L}; \citealt{2001ApJ...550L..65V};
\citealt*{2007A&A...470..179P}; \citealt{2016EAS....80..177C,
  2017A&A...608A..28}). The main sequence broadening in optical
filters is thought to be due to a spread in helium abundance ($\Delta
Y$, with $Y$ being the helium mass fraction; e.g. \citealt[][and
  references therein]{2004ApJ...612L..25N,2005ApJ...631..868D,
  2016EAS....80..177C}).

Apart from a few exceptions -- such as Omega Centauri ($\omega$ Cen)
and M54, which are among the most massive clusters and may be (former)
nuclear clusters -- most GCs show no spread in iron abundance, meaning
that enrichment from supernova explosions needs to be
avoided. Finally, the maximum Na enhancement is similar in all GCs,
but the $\Delta Y$ varies from cluster to cluster
\citep{2015MNRAS.449.3333B}. In this paper, we focus on these
``Fe-normal" GCs (the large majority) which show light element
variations.

Three possible polluters that reach the required temperatures to
explain the properties of these GCs have been put forward: Asymptotic
Giant Branch (AGB) stars \citep{2001ApJ...550L..65V}, massive stars
\citep[$\gtrsim20\,\msun$,
][]{2006A&A...448L..37M,2006A&A...458..135P, 2009A&A...507L...1D}, and
supermassive stars \citep[SMSs,
  $\gtrsim10^3\,\msun$,][]{2014MNRAS.437L..21D}.  The models that try
to explain the GC abundance anomalies and invoke AGB stars (massive
enough to undergo hot-bottom burning;
\citealt{2001ApJ...550L..65V,2013MNRAS.431.3642V,2008MNRAS.391..825D})
or fast-rotating massive stars (FRMS, i.e. main sequence and luminous
blue variable (LBV) massive stars rotating at or near critical speed;
\citealt{2007A&A...464.1029D,2013A&A...552A.121K}) assume that a
second generation of stars forms from the yields of a first
generation. However, none of these sources is able to satisfy all the
nucleosynthesis constraints (\citealt*{2015MNRAS.449.3333B};
\citealt{2017A&A...608A..28}). AGB nucleosynthesis builds an O--Na
correlation instead of the observed anti-correlation
\citep{1997A&AS..123..241F,2003ApJ...590L..99D,2007PASA...24..103K,2010A&A...512A..10S,2013MNRAS.431.3642V,2014MNRAS.437..195D},
and it releases He-burning products, thus predicting total C+N+O
variations that are not observed in GCs
(\citealt{2006ApJ...652.1240K,2009A&A...505..727D}, but see
\citealt*{2015MNRAS.446.3319Y}).  On the other hand, the FRMS model
predictions hardly reach Mg-burning temperature required to fit the
observed Mg--Al anti-correlation without predicting strong He
enrichment \citep{2007A&A...464.1029D}.  Predicting the correct
\mbox{CNONaMgAl} abundances without over-predicting $\Delta Y$ is a
general problem for all the polluters \citep{2015MNRAS.449.3333B},
because recent results of {\it HST} photometry
\citep[e.g.][]{2015MNRAS.451..312N,2015ApJ...808...51M} show that
$\Delta Y$ is generally low.  However, current SMS models with masses
between $\sim2\times10^3\,\msun$ and $2\times10^4\,\msun$ reach the
required central temperature of $\sim72-78$\,MK already at the very
beginning of the evolution on the main sequence, when He enrichment is
minute \citep{2017A&A...608A..28}. Consequently, at that early
evolution phase the H-burning products of SMSs show remarkable
agreement with the various observed abundance anti-correlations
\citep[see figure 1 in ][]{2015MNRAS.448.3314D} and Mg isotopic ratios
\citep{2014MNRAS.437L..21D}.  As of today, SMS thus appear to be the
most appealing candidate from the nucleosynthesis point of view,
assuming that these fully convective objects release their entire
material at the very beginning of the main sequence to avoid
overproduction of He.

Getting the abundance patterns right is an important step, but a
successful model for the origin of the abundance anomalies in GCs
should also be able to explain how the required amount of material can
be  produced, and acquired by the low-mass stars that survive
until today.  Models that invoke a second burst of star formation from
the yields of a first generation -- hereafter referred to as multiple
generation models (MGMs) -- struggle to produce the required amount of
processed material and need either a first generation with a
top-heavy stellar initial mass function
\citep[IMF,][]{2006A&A...458..135P,2006ApJ...652.1240K}, or need the
cluster to lose $\gtrsim\!90\%$ of the first generation of stars
\citep{2006A&A...458..135P,2008MNRAS.391..825D,2011MNRAS.413.2297S} to
end up with a significant fraction of polluted stars ($\gtrsim50-90\%$
as observed in the most massive GCs). The latter scenario makes strong
predictions for the fraction of stars in GCs relative to the field,
which is in tension with empirical estimates of this (high) ratio in
dwarf galaxies \citep{2012A&A...544L..14L, 2014A&A...565A..98L}.
Additionally, common stellar feedback processes are not able to
achieve this via gas expulsion \citep{2012A&A...546L...5K} and need
careful fine-tuning to avoid the cluster to disperse completely
\citep{2010A&A...516A..73D,2015MNRAS.452..924K}. Finally, relying on
internal processes to expel a large amount of pristine stars generally
leads to a decrease in the remaining fraction of polluted stars with
GC mass \citep{2015MNRAS.453..357B}, which is not observed.  In fact,
both the inferred $\Delta Y$ and the fraction of polluted stars
($\fenr$) correlate with GC mass \citep[][]{2014ApJ...785...21M,
  2017MNRAS.464.3636M}. These trends imply that more polluted material
is required\ per unit of cluster mass in massive GCs, requiring fine
tuning in MGMs \citep[for recent reviews,
  see][]{2015MNRAS.454.4197R,2016EAS....80..177C, 2017IAUS..316..302B,
  BL18}. The inability of any existing MGM to create sufficient
polluted material and produce the correct trends with GC mass, is
commonly referred to as the `mass budget problem'.

Given the challenges with the MGMs, a model requiring only a single
generation of stars is therefore more attractive.
\citet[][]{2013MNRAS.436.2398B} present such a scenario, in which
low-mass pre-main sequence stars with large discs sweep up polluted
material released by interacting, massive binary stars. However, full
mixing of the polluted material with the proto-stellar seeds is
required to explain all the abundance patterns, which requires
that the accretion occurs at the very beginning of the pre-main
sequence when the contracting stars are still entirely convective,
i.e. on timescales shorter than $\sim1-3\,\myr$
\citep{2014A&A...566A.109S,2014MNRAS.443.3302D}. In addition,
accretion of hot and tenuous, low-angular momentum gas on a disc
causes the disc to rapidly shrink and accrete on the star, thereby
limiting the cross section for further accretion
\citep{2016A&A...594A..30W}. This early disc accretion is, therefore,
not efficient enough to explain the amount of pollution that is
observed.  Alternatively, \citet{2014A&A...569L...6C} propose that
only massive stars formed out of pristine GC material, and that only
low-mass stars formed from contaminated gas in the immediate vicinity
of the FRMS polluters during their very short lifetime
($\sim3-8\,\myr$).  Assuming 100$\%$ recycling of the FRMS ejecta and
accounting for dilution as required to explain the presence of Li in
polluted stars, the mass initially locked in the massive star
polluters should have been only two to four times the present-day
stellar mass, which strongly alleviates the mass budget
problem. Nevertheless, the required IMF is contrived and this idea
does not give rise to the  observed trends with GC mass.

Because of the promising results of the SMS yields and the problems
with the MGMs \citep{2015MNRAS.454.4197R,2016EAS....80..177C,
  2017IAUS..316..302B, BL18}, we here search for a solution in which a
SMS forms simultaneously with the GC and immediately pollutes the
cluster gas and eventually low-mass protostars during the cluster
formation process, i.e. without relying on multiple starbursts. We
focus in particular on overcoming the mass budget problem, and
understanding the observed correlations of $\Delta Y$ and $\fenr$ with
GC mass and the relative abundances of \mbox{CNONaMgAl} and
He. Because the MSP phenomenon is found in GCs with different [Fe/H],
the formation of the SMS can not rely on inefficient gas cooling in
metal-free initial conditions, as invoked in models of the formation
of massive Population~III stars
\citep*[e.g.][]{2002Sci...295...93A,2009MNRAS.396..343R}.  Instead,
the model presented in this work relies on the stellar dynamical
behaviour of proto-clusters in the gas accretion phase, i.e. physics
that is largely independent of metallicity (but see
Sections~\ref{ssec:mod:winds} and \ref{ssec:obspred} for a discussion
on metallicity dependence). Several of the ingredients of the
dynamical model are based on the theoretical work by
\citet*{1998MNRAS.298...93B} and \citet{2008MNRAS.388.1171C}, and the
numerical work by \citet*{2010MNRAS.407..381D} and
\citet{2011MNRAS.410.2799M}. These studies discuss the relative
importance of gas accretion and stellar collisions in massive star
formation. In this work we push this into the regime of GCs and show
that this is where SMSs can form via stellar collisions. We also
provide scaling relations for the dependence of the mass of the SMS on
cluster mass.

In Section~\ref{sec:analytics} we present a framework for the
formation of a SMS during GC formation. In Section~\ref{sec:model} we
use this new SMS formation model to put forward a new GC
self-enrichment scenario to explain the observed abundance anomalies
of light elements (HeCNONaMgAl) in Fe-normal GCs. In
Section~\ref{sec:discussion} we discuss the model uncertainties and
present predictions and observational tests that can verify this
scenario.  Our conclusions are given in Section~\ref{sec:conclusions}.


\section{SMS formation  during GC formation}
\label{sec:analytics}
In this section we introduce a simple model for the formation of a
SMS via stellar collisions during GC formation.  We consider the
effect of gas accretion and subsequent contraction of the cluster,
collisions between stars and the effect of two-body relaxation. We
derive the condition for SMS formation and scaling relations for the
rate of growth of the SMS. In section~\ref{sec:model} we add the mass
loss from the SMS as the result of a stellar wind.

\subsection{Typical initial conditions}
\label{ssec:typical}
Consider the very first phases of star formation in dense and
turbulent molecular clumps, in which a proto-stellar core mass
function with a slope close to the Salpeter value
\citep{1955ApJ...121..161S} develops via gravoturbulent fragmentation
\citep{2002ApJ...576..870P, 2008ApJ...684..395H}  and/or competitive
accretion \citep{1982NYASA.395..226Z,1998MNRAS.298...93B,
  2015MNRAS.452..566B}. Protostars form quickly (less than a core
free-fall time) from a seed mass \citep{2015MNRAS.450.4137G}, which
then continues accreting from the global mass reservoir and
{/or} their
local collapsing cores.  The protostars form with a mass spectrum, and
have typical initial masses of\footnote{From hereon we denote all
  quantities with subscripts $0$ when referring to their value at the
  start of gas accretion and we use capital symbols for the cluster
  quantities and lower-case symbols for those of the stars and the
  SMS. } $ m_0\simeq0.1\,\msun$ and radii of $r_0\simeq3\,\rsun$
\citep*[e.g.][]{1999MNRAS.310..360T,2016ARA&A..54..135H}.  Clusters
form with a range of masses and densities, but typical values we
consider are initial stellar densities within the half-mass radius
($\rhn$) of $\rhohn \simeq 10^3\,\msun/\pc^3$, such that a
proto-cluster with a stellar mass of $M_0\simeq10^5\,\msun$ has a
radius of $\rhn\simeq2.3\,\pc$. The total number of stars is then
$N=M_0/m_0 =10^6$.  Initially, the velocities of the stars are set by
the turbulent velocities in the cloud and the gas potential.
Subclusters of protostars quickly virialise in their own potentials,
because the local star formation efficiency is high
\citep{2012MNRAS.425..450M} and the subclusters contract as the result
of gas accretion and merge with each other to become a dense,
self-gravitating stellar system surrounded by lower density gas
\citep{2011MNRAS.410.2799M}.  The resulting dynamical time of the
virialised proto-cluster is $\tdynn \sim (GM_0/\rhn^3)^{-1/2}\simeq
0.16\,\myr$, with $G$ the gravitational constant. From here on we
consider the cluster to be a single entity, but we get back to the
possible importance of subclustering in the discussion
(Section~\ref{ssec:discreteness}).  In the next section we discuss the
evolution of the stellar cluster after gas accretion on its member
stars has commenced.

\subsection{Gas accretion and adiabatic contraction}
\label{ssec:accretion}
After the protostars formed, they grow in mass by accretion of gas
(e.g. \citealt{1998MNRAS.298...93B,2009Sci...323..754K,2016ARA&A..54..135H};
\citealt*{2017MNRAS.467.1313V}). We assume that this gas accretion is
fueled by gas flowing into the cluster with 100\% accretion efficiency
and that no new stars form (i.e. $N$ remains constant)\footnote{It
  does not matter for the response of the cluster whether the gas
  accretes on the stars, or on the molecular clumps (i.e. before the
  collapse).}.  We define the gas accretion timescale as
\begin{equation}
\tacc = \frac{M}{\dot{M}},
\label{eq:tacc}
\end{equation}
where $\dot{M}$ is the gas accretion rate onto the cluster and $M$ is
the instantaneous total mass in stars. We assume that $\dot{M}$ is
constant in time, such that $M$ and $\tacc$ increase linearly with
time during the accretion phase.  This is different from what is found
in models of the collapse of a self-gravitating cloud
\citep[$\dot{M}\propto t$,][]{2015ApJ...804...44M}, but a roughly
constant $\dot{M}$ in time (or slightly declining with time) was found
by \citet{2017ApJ...834...69L} in cosmological zoom-in simulations,
where the accretion rate is regulated by stellar feedback. In our model, $\dot{M}$ is a free parameter and
we assume that $\dot{M}$ is proportional to $N$,
i.e. $\dot{M}=\mdotaccav N$, where $\mdotaccav$ is the average
accretion rate on individual stars. Our assumption of $\dot{M}\propto
N$ means that $\mdotaccav$ does not depend on $N$.  To preserve the
shape of the stellar mass function and keep the total $\dot{M}$
constant we assume that $\mdotacc$ is proportional to the initial mass
of the star (i.e. $\mdotacc \propto m_0$)\footnote{This is not a
  critical assumption to approximately preserve the IMF slope, because
  for Bondi-Hoyle accretion (i.e. $\mdotacc\propto m^2$) the mass
  function evolves to a $-2$ power-law, i.e. close to Salpeter,
  independent of the initial functional form
  \citep{1982NYASA.395..226Z,2015MNRAS.452..566B}. }.  We assume a
typical value of $\mdotaccav\simeq0.1\,\msun/\myr$, which is similar
to what is found for young, low-mass pre-main sequence stars in the
Milky Way \citep[$\sim10^5\,$yr,
  $\sim0.7\,\msun$,][]{2016ARA&A..54..135H, 2017ApJ...846..110D} 
  and in magnetohydrodynamic simulations
  \citep[e.g.][]{2017ApJ...847..104O}.  For this accretion rate, the
initial accretion timescale is $\taccn\simeq1\,\myr$, which is
reasonable given our current understanding of GC formation timescales
\citep[i.e. few Myrs, see
  e.g.][]{2016ApJ...823...52K,2017ApJ...834...69L,2018MNRAS.474.4232K}.
For these values of the accretion rate, the mean mass of stars grows
to $0.6\,\msun$ in $5\,\myr$, roughly equal to the mean mass of a
\citet{2001MNRAS.322..231K} initial mass function (IMF, in the range
$0.1\,\msun$ to $100\,\msun$). We assume that the accretion is halted
at this time by stellar feedback \citep{2017ApJ...834...69L}.  We
  discuss the duration of the accretion phase and the start of the
  stellar evolution phase in more detail in
  Section~\ref{ssec:mod:cluster}.

For the adopted typical value of $\tacc$, we have $\tacc>>\tdyn$, such
that the angular momentum of stars plus gas is conserved during the
accretion and the stellar clusters responds adiabatically. We define
$\vrms$ as the one-dimensional velocity dispersion of the cluster, and
assume that the angular momentum of the inflowing gas is negligible,
such that the adiabatic invariant $M\vrms \rh$ of the star cluster is
preserved, while $M$ increases as the result of accretion
\citep{1986ApJ...301...27B}. Because $\tacc>>\tdyn$, virial
equilibrium is maintained and then $\vrms \simeq \sqrt{GM/(6\rh)}$ and
we find that the star cluster contracts as $\rh\propto M^{-3}$
\citep{1998MNRAS.298...93B}.  From this we see that the gas accretion
process leads to a rapid increase of the density ($\rhoh\propto
M^{10}$), and stellar collisions can become important already after a
modest amount of gas accretion \citep{2011MNRAS.410.2799M}.

The increasing density also leads to a decrease in the two-body
relaxation timescale.  When this timescale becomes shorter than
$\tacc$, two-body heating becomes important and the cluster starts
expanding \citep{2008MNRAS.388.1171C}, reducing the collision rate.
When these two timescales are of comparable magnitude, the cluster has
reached its maximum density and a requirement for SMS formation is
thus that sufficient stellar collisions need to happen before this
moment.  To quantify this condition, we need a measure of the
relaxation timescale, for which we use the definition of the half-mass
relaxation timescale from \citet{1971ApJ...164..399S}
\begin{align}
\trh \simeq 0.138\frac{N}{\psi\ln\Lambda}\left(\frac{\rh^3}{GM}\right)^{1/2}.
\label{eq:trh1}
\end{align}
Here $\ln\Lambda$ is the Coulomb logarithm and $\psi$ depends on the
stellar mass spectrum: $\psi=1$ for equal-mass clusters and
$\psi\simeq10-100$ for clusters with a full mass spectrum
\citep{2010MNRAS.408L..16G}. We adopt a constant $\ln\Lambda=10$ and
we assume $\psi=30$ to include the effect of a spectrum of masses for
the accreting protostars.

For a constant $N$ and the scaling $\rh\propto M^{-3}$, we find $\trh
\propto M^{-5}$ from equation~({\ref{eq:trh1}).  Combined with a
  constant $\dot{M}$, we have $\trh/\tacc \propto M^{-6}$.  The
  maximum cluster density is reached when $\tacc\simeq\nrel\trh$, with
  $\nrel\simeq10$ being the number of relaxation times that needs to
  elapse before collisional dynamics becomes important (see
  e.g. \citealt{1980ApJ...242..765C} for the idealised case of a
  single-mass cluster and \citealt{2014MNRAS.442.1265A} for multimass
  clusters). From the linear increase of $M(t)$ and the relations for
  $\trh$ and $\tacc$ we find that the end of the contraction phase is
  at time
 \begin{equation}
\tcontr = \taccn \left[\left( \frac{\nrel\trhn}{\taccn}\right)^{1/6} -1\right].
\label{eq:tcontr}
\end{equation}
We note that $\tcontr$ is also the time that $\nrel$ half-mass
relaxation timescales have elapsed (i.e. $\nrel = \int_0^{\tcontr} \dr
t^\prime/\trh(t^\prime)$), hence $\tcontr$ can be thought of as the
moment of core collapse. Because $\tcontr$ is more sensitive to
$\taccn$ than to $\trhn$, the start of the relaxation dominated phase
is triggered by gas accretion and $\tcontr$ is therefore much shorter
than the conventional core collapse of GCs, which happens on a
timescale of Gyrs\footnote{The reason that another core collapse
  occurs on this long timescale is because after $\sim3\,$Myr the
  cluster expands by a factor of a few as the result of stellar mass
  loss, thereby increasing $\trh$, and secondly because the massive
  stars die, such that $\psi$ reduces to $\psi\simeq 2$
  \citep*{1998ApJ...495..786K}, which also increases $\trh$ (see
  equation~\ref{eq:trh1}). For a more elaborate discussion on core
  collapse(s) in multimass systems we refer to
  \citet{2012MNRAS.425.2493B,2013MNRAS.432.2779B}.}.

At $\tcontr$, the total mass in stars has increased to
\citep{2010MNRAS.407..381D}
\begin{align}
M(\tcontr) = M_0\left(\frac{\nrel\trhn}{\taccn}\right)^{1/6}.
\end{align}
For the initial parameters of our fiducial cluster
(Section~\ref{ssec:typical}) we have $\trhn\simeq75\,\myr$ and
combined with $\taccn\simeq1\,\myr$ (Section~\ref{ssec:accretion}) we
find that the mass needs to increase by only a factor of $\sim3$
before two-body relaxation becomes important. The corresponding
$\tcontr\simeq2\,\myr$, which is before feedback from supernova
explosions becomes important, implying that the maximum density is
reached before gas accretion stops.  The weak dependence on $\trhn$
means that $\tcontr$ is relatively insensitive to $N$.  The next
question is whether stellar collisions can become important before
two-body relaxation stops the contraction of the cluster.  Before we
address this, we first introduce the properties of the SMS in the next
section.

\subsection{SMS properties}
\label{ssec:smsprop}
To be able to follow the growth of the SMS, we need to know how its
radius depends on its mass. The mass-radius relation for SMSs is very
uncertain, but we can be guided by our understanding of massive
stars. \citet{2010MNRAS.408..731C} present parameters for a sample of
stars with masses $\sim90-130\,\msun$, and find that they have radii
of $\sim30\,\rsun$.  We adopt a
mass-radius relation of the form
\begin{align}
\rsms = 30\,\rsun\,\left(\frac{\msms}{100\,\msun}\right)^\delta,
\label{eq:mr}
\end{align}
where $0<\delta\lesssim1$ for $\msms>100\,\msun$ and $\delta=0.5$ for
$\msms<100\,\msun$.  Smoothed Particle Hydrodynamics (SPH) models of
stellar collisions of solar type stars show that the collision product
is typically a factor 3-20 larger than a star with the same mass that
started unperturbed \citep{2003MNRAS.345..762L}. SPH models of
colliding massive stars ($\gtrsim100\,\msun$) show that right after
the collision the star can be 10-100 times larger than the equilibrium
$\rsms$, and that the star settles to a radius that is a few times the
equilibrium value after $\sim10^3\,\yr$
\citep{2007ApJ...668..435S}. The collision rate can be high enough to
have the next collision occurring before this settling
occurs. However, it is not clear how efficient a diffuse halo of a few
$10^3\,\rsun$ is in dragging other stars into the SMS.  We note that
collisions and the high radiation pressure of stars
$\gtrsim300\,\msun$ may lead to larger radii
(\citealt*{2012A&A...538A..40G}; \citealt{2015A&A...581A..15S}; Rob
Izzard, private communication) and we consider the uncertain radii by
varying $\delta$.

The luminosity of a SMS, $\lsms$, is close to its Eddington limit for
electron scattering: $\ledd/\lsun=3.7 \times 10^{4}\, \msms/ \msun$.
The stellar models of \citet{2005AstL...31..695N} show that $\Gamma =
\lsms/\ledd$ varies between 0.56 and 0.94 in the range of $3\times
10^2 < \msms / \msun < 10^4$ and we adopt $\Gamma =0.75$ such that

\begin{equation}
 \lsms\simeq2.8\times10^6\,\lsun\,\frac{\msms}{100\,\msun}.
 \label{eq:lsms}
\end{equation}
This agrees well with the observed values from
\citet{2010MNRAS.408..731C}, who find luminosities of
$(1.5-3)\times10^6\,\lsun$.  This is also in good agreement (by a
factor 2 to 3) with the luminosity of the zero age main sequence
models of SMS that we use for the nucleosynthesis discussion in Section~\ref{ssec:nucleo} and that were computed by Denissenkov (private
communication) with the evolution code {\sc mesa} \citep{2011ApJS..192....3P} with the same assumptions as in
\citet{2015MNRAS.448.3314D}.

With the Stefan-Boltzmann law (i.e. assuming black-body radiation),
the mass-radius relation (equation~\ref{eq:mr}) and the luminosity
(equation~\ref{eq:lsms}) we find that $\Teff \simeq 43\,{\rm
  kK}\,(\msms/100\,\msun)^{1/4-\delta/2}$. This $\Teff$ is in
  excellent agreement with what was found by
\citet{2010MNRAS.408..731C} for unperturbed massive stars
($\sim40\,$kK). The
$\delta$-dependence implies that the SMS has a mass-independent
$\Teff$ for $\delta=0.5$ and that SMSs are cooler than $100\,\msun$
stars for $\delta>0.5$.

Finally, we adopt an initial $\msms=5\,\msun$ before gas accretion,
which has a radius of $6.7\,\rsun$.  With a description for the
mass-radius relation in place, we can now proceed with the growth of
the SMS via collisions.

\subsection{Mass growth of a central SMS}
\label{ssec:massgrowth}
The collision rate, $\ncolldot$, experienced by a star with mass
$\msms$ and $\rsms$ in a system with stellar number density $n$ and
velocity dispersion $\vrms$, with other stars of mass $m$ and radius
$r$ is \citep[][Chapter 7]{1976ApL....17...87H, 2008gady.book.....B}
\begin{align}
\ncolldot = 2\sqrt{2\pi}\left(\frac{\msms+m}{\msms}\right)^{1/2}n\vrms d^2\left( 1 + \frac{G\msms}{d\vrms^2}\right),
\label{eq:tcoll}
\end{align}
where it is assumed that a collision occurs when the stars are at a
distance $d=r+\rsms$.  The final term within brackets has two
contributions: the first one is due to the geometrical cross section
of the star, and the second contribution is due to gravitational
focusing, which enhances the cross section of stars. For stellar
collisions, the gravitational focusing term dominates, because the
escape velocity from the stellar surface is much larger than the
typical velocities of the stars, and therefore $G\msms/(d\vrms^2)>>1$.
\citet{2010MNRAS.407..381D} assumed that all stars have the same mass
and radius to derive the total number of collisions in the contraction
phase: $\ncoll(\tcontr)\propto N^{5/3}\dot{M}^{2/3}$. The number of
collisions per stars is a factor $N$ smaller. Combined with our
assumption that $\dot{M}\propto N$, we find that the number of
collisions experienced by an individual star scales as $\propto
N^{4/3}$. This super-linear scaling of $\ncoll$ with $N$ shows that
the stellar collisions experienced by a star in the adiabatic
contraction phase are more important per unit of cluster mass in more
massive clusters. Combined with the absence of a dependence on any
other parameters, this is the first ingredient to explain the observed
increase of $\fenr$ and $\Delta Y$ with GC mass
(section~\ref{sec:model}).

After a collision between two stars, the collision product is more
massive and has a larger radius, which increases its cross section and
collision rate (see equation~\ref{eq:tcoll}). There is therefore a
high probability that the first collision product is involved in
subsequent collisions and typically one very massive star forms as the
result of stellar collisions ($\gtrsim 100\,\msun$,
e.g. \citealt{2002ApJ...576..899P, 2004Natur.428..724P};
\citealt*{2006MNRAS.368..141F,2016MNRAS.459.3432M}).  To understand
how the mass of a star grows as the result of continuous collisions,
we assume that $\msms>>m$ and $\rsms>>r$, which is true after several
collisions, and with $\mdotsmscoll = \langle m\rangle\ncolldot$ we
then see from equation~(\ref{eq:tcoll}) that the growth rate is
\begin{equation}
\mdotsmscoll \simeq 2\sqrt{2\pi} G\,\msms\rsms \frac{\rhoc}{\vrms},
\label{eq:mdot}
\end{equation}
where $\rhoc$ is the central mass density of the cluster, which we
relate to $\rhoh$ as $\rhoc=\fch\rhoh$, with\footnote{This density
  contrast depends on the density profile of the cluster, for which we
  have little guidance. For a \citet{1966AJ.....71...64K} model with
  dimensionless central potential of $W_0=9(10)$ it is roughly
  $\fch=440(2000)$.\label{fnote:rhoc}} $\fch=10^3$.  Note that the
dependence of $\mdotsmscoll$ on the central mass density means that a
star in a mass segregated cluster, with a similar $\rhoc$ and $\vrms$,
experiences the same $\mdotsmscoll$, as the result of fewer collisions
with more massive stars.  Defining the timescale for the growth of the
SMS as $\tsms \equiv \msms/\mdotsmscoll$, we have
\begin{equation}
\tsms = \frac{1}{2\sqrt{2\pi} G\,\rsms} \frac{\vrms}{\rhoc}.
\label{eq:tsms}
\end{equation}
In a contracting cluster, the ratio $\vrms/\rhoc\propto M^{-8}$,
strongly reducing $\tsms$ while the cluster grows in mass. For the
properties of our initial conditions (i.e. when the contraction
starts, Section~\ref{ssec:typical}) this timescale is $\tsmsn \simeq
1680\,\myr$  for a star of $5\,\msun$ and radius $6.7\,\rsun$,
i.e. collisions are irrelevant at the time of fragmentation and it
reduces to $\lesssim1\,\myr$ after the cluster mass has increased by a
factor of two.  In addition, $\tsms$ becomes shorter as collisions
proceed, because $\rsms$ increases (for $\delta>0$ in
equation~\ref{eq:mr}). In this section we ignore the dynamical
feedback of  binaries  on the cluster properties, which we discuss in
Section~\ref{ssec:cap}. Then, a runaway collision process can occur
and from integrating equation~(\ref{eq:mdot}) we find that the time when
$\msms\rightarrow\infty$ is
\begin{align}
\tinf \displaystyle&\simeq \taccn \left[ \left( \frac{9\tsmsn}{\delta\taccn} \right)^{1/9} -1\right].
\label{eq:tinf}
\end{align}
This relation holds for $\tsmsn/\taccn>>1$, which is satisfied because
for the typical parameters of Section~\ref{ssec:typical} we have
$\tsmsn/\taccn\simeq 700$.  For a cluster to be able to form a SMS
via stellar collisions, this runaway process needs to occur before the
end of the contraction phase derived in Section~\ref{ssec:accretion},
i.e.  $\tinf\lesssim\tcontr$. Using the expression for the end of the
contraction phase in equation~(\ref{eq:tcontr}) we find that this
criterion is met when
\begin{align}
N \gtrsim 1.4\times10^6\,\left(\frac{\dot{M}}{1.4\times10^5\,\msun/\myr}\right)^{-3/4}\finit,
\label{eq:ncrit}
\end{align}
where $\finit$ depends on the initial conditions of the cluster, the
protostars and the SMS (see Section~\ref{ssec:typical})
\begin{align}
\finit =
&\left(\frac{\psi}{30}\right)^{9/4}\,\left(\frac{\nrel}{10}\right)^{-9/4}\,\left(\frac{\delta}{0.5}\right)^{-3/2}\left(\frac{\fch}{10^3}\right)^{-3/2}
\left(\frac{\rhohn}{10^3\,\msun/\pc^3}\right)^{-1/8}\,\times\nonumber\\
&\left(\frac{m_0}{0.1\,\msun}\right)^{5/4}\,\left(\frac{\rsmsn}{6.7\,\rsun}\right)^{-3/2},
\label{eq:finit}
\end{align}
 where the adopted scaling $\rsmsn=6.7\,\rsun$ applies to
  $\msmsn=5\,\msun$ (see Section~\ref{ssec:smsprop}).  From
equation~(\ref{eq:ncrit}) we see that in the GC mass regime
($M\gtrsim{\rm few}\times10^5\,\msun$), a runaway stellar collision
process occurs before the contraction phase ends. The criterion is
surprisingly insensitive to $\rhohn$, implying that $N$ is the
dominant cluster property determining whether a runaway collision can
occur.  Apart from $\dot{M}$, all other parameters are similar in
different environments, making $\dot{M}$ the only environmental
parameter. From this simple argument we see that massive
(i.e. high $N$) clusters in environments with high gas accretion
rates (i.e. high $\dot{M}$) are the places in which runaway
collisions can take place to form a SMS. At high redshift, the high
gas densities and gas fractions enable the formation of massive GCs
\citep{1997ApJ...480..235E}. These conditions also lead to higher
accretion rates compared to the present day
\citep{2017ApJ...836...80E,2017ApJ...834...69L}. At this stage, we
have the first quantitative explanation for why SMS formation occurred
predominantly in massive, old GCs.

In Fig.~\ref{fig:msms_simple} we show the evolution of $\msms$,
obtained from solving coupled differential equations for $\dot{M}$,
$\rhdot$ and $\msmsdot$, where in the latter we use $\msmsn =
5\,\msun$, $\rsmsn = 6.7\,\rsun$ and $\delta=0.5$ (see
Section~\ref{ssec:smsprop}) and include both the gas accretion term
($\mdotacc=5\,\msun/\myr$) and the contribution from stellar
collision (equation~\ref{eq:mdot}). The models are solved until
$t=\min(\tcontr, \tinf)$. For the cluster with $N=10^5$ (blue, dotted
line) the runaway phase is interrupted because the relaxation driven
expansion starts before the collision runaway formation of the SMS
starts, i.e. $\tinf \gtrsim \tcontr$. Note that because gas accretion
is included in $\mdotsms$,  the critical $N$ is slightly lower ($N\simeq10^6$)
  than the estimate given  in equation~(\ref{eq:ncrit}) ($N\simeq1.4\times10^6$).  The two larger clusters experience
a runaway collision process before the contraction phase ends
(i.e. $\tinf < \tcontr$).

\begin{figure}
\includegraphics[width=8cm]{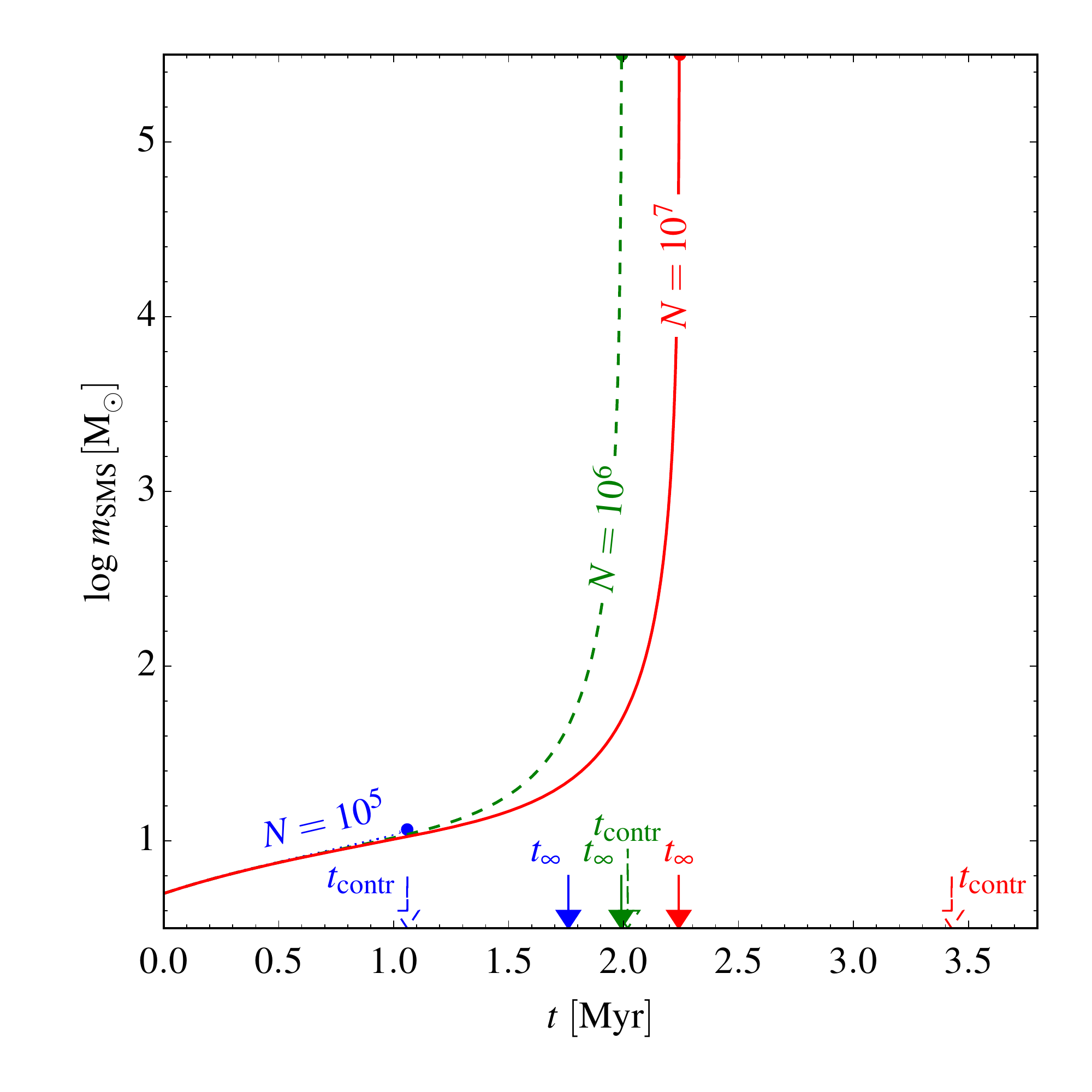}
\caption{The mass of a SMS that grows by gas accretion and stellar
  collisions in the centre of a gas accreting cluster, for clusters
  with different $N$. The values of $\msms(t)$ are found by
  numerically integrating $\mdotsms = \mdotacc + \mdotsmscoll$, with
  $\mdotacc = \msmsn/\myr$ and $\mdotsmscoll$ given by
  equation~(\ref{eq:mdot}). Lines are plotted until $t=\min(\tcontr,
  \tinf)$ and the effect of relaxation is not included. The end of the
  contraction phases ($\tcontr$) are also indicated and from this it
  can be seen that runaway collisions do not occur in clusters with
  $N\lesssim10^6$, because for those clusters $\tcontr<\tinf$, while
  $\tcontr>\tinf$ for the larger clusters and runaway collisions occur
  before the end of the contraction phase. }
\label{fig:msms_simple}
\end{figure}

\subsection{A capped collision rate}
\label{ssec:cap}
In the previous section we solved for $\msms(t)$ by ignoring the
feedback from binaries.  This allows $\msms$ to become as large as the
mass of the cluster.  As we will show here, a genuine runaway process
will not happen because of dynamical feedback from binaries involving
the SMS.  At high stellar densities, stars can become bound to the
SMS, either by triple interactions \citep{1975MNRAS.173..729H} or via
tidal capture \citep{1975MNRAS.172P..15F}. These binary systems are
efficient in heating the surrounding stars in interactions, directly
via accelerating stars and indirectly via ejecting stars \citep[see
  e.g.][]{1984ApJ...280..298G}. As a result of the interactions the
binary orbit shrinks, until the star collides with the SMS.  A similar
heating occurs when stars get captured by a massive central black hole
in the cluster core (\citealt*{1976ApJ...209..214B,
  2004ApJ...613.1133B}; \citealt{2007PASJ...59L..11H}).  This is why
the predicted exponential growth of the central star presented in the
previous section (Fig.~\ref{fig:msms_simple}) will not occur: at some
point the SMS-star binaries will generate enough energy to inflate the
core, thereby decreasing the collision rate until some equilibrium is
found \citep{2007PASJ...59L..11H}. We speculate that this is the
reason why numerical simulations of stellar collisions in dense
stellar cluster \citep[][]{2002ApJ...576..899P, 2004Natur.428..724P,
  2006MNRAS.368..141F,2016MNRAS.459.3432M} always find that the second
derivative of $\msms(t)$ is negative, which is not expected in a
runaway process.

The question is now: what sets the maximum collision rate?
\citet{1975MNRAS.173..729H} shows that the formation rate of hard
binaries in three-body interactions scales as $\msms^3n^2/\vrms^9$
(see also Chapter 22 \citealt{2003gmbp.book.....H} and \citealt{1986AcA....36...19S}),  while the
  collision rate goes as $\msms^{1+\delta}n/\vrms$
  (equation~\ref{eq:mdot}).  We estimate the three-body binary formation rate at the hard-soft boundary from equation~1 of \citet{1986AcA....36...19S} for our typical initial conditions in Section~\ref{ssec:typical} and for $\msms=100\,\msun$.  We can directly compare this binary formation rate to the collision rate following from equation~(\ref{eq:mdot}). We find that the binary formation rate is roughly a factor of two larger at the start.
  The stronger dependence on both $\msms$
  and the stellar number density in the binary formation rate means
  that this process becomes more important than direct collision
  as the SMS grows in mass and the cluster becomes denser.  We assume
that the SMS-star pairs form marginally bound and that the companion
star collides with the SMS when the pericentre distance of its orbit
equals $\rsms$.  The specific orbital energy depends on the semi-major
axis of the orbit, $a$, as $\Eorb \simeq -G\msms/(2a)$, where we
assumed that $\msms$ is much larger than the mass of the colliding
star. For the colliding orbit, $a = \rsms/(1-\varepsilon)$, where
$\varepsilon$ is the eccentricity of the orbit.  We then find $\Eorb
\simeq -G\msms(1-\varepsilon)/(2\rsms)$. A collision of a star with
mass $m$ is, therefore, accompanied by a dynamical energy supply of
$\Delta E = m|\Eorb|$.  Because the cluster dynamics becomes
collisional near the moment of maximum stellar density (i.e. two-body
effects become important), a natural assumption is that the rate at
which  dynamical energy is supplied to the cluster by binaries
($\Edotcoll$) can not exceed the flow of energy that can be
transported through $\rh$ by two-body relaxation, i.e.
\begin{equation}
\Edotcoll  \simeq \mdotsms|\Eorb|\lesssim \zeta \frac{|E|}{\trh},
\label{eq:edotcoll}
\end{equation}
where $E\simeq -GM^2/(4\rh)$ is the total energy of the cluster and
$\zeta\simeq\nrel^{-1}\simeq0.1$ \citep*{H61,
  H65,2011MNRAS.413.2509G,2012MNRAS.422.3415A}.  We then assume that
all stars that collide with the SMS undergo this binary hardening
phase in the collision phase, i.e. the collisions resulting from
coalescence following binary formation and hardening are more
efficient than direct stellar collisions, as the result of the
decreased core density. We  then find that the growth rate of the
SMS is

\begin{equation}
\mdotsmsrel =  \zeta\frac{\rsms }{\msms } \frac{M^2}{\rh\trh}.
\label{eq:mdotsmsrel}
\end{equation}
Here we assumed that the orbital eccentricity is $\varepsilon = 0.5$,
and $\mdotsmsrel$ would be higher for more eccentric orbits,  and
  if stars start off more bound to the SMS (i.e. $\Eorb <0 $ when the
  binary forms).  Comparing this with the expression for
$\mdotsmscoll$ (equation~\ref{eq:mdot}), we see that this collision
rate is less sensitive to the properties of the SMS, because it scales
with $\rsms/\msms$, as opposed to $\rsms\msms$. This is because the
collision rate is now set by how much energy can be transported by
two-body relaxation, and the liberated energy is proportional to
$\msms/\rsms$.

Assuming that the cluster spends most of its time in this regime
(i.e. $\tinf<<\tcontr$) and ignoring gas accretion on the SMS, we can
derive $\msms(\tcontr)$ by integrating equation~(\ref{eq:mdotsmsrel})
from $t=0$ to $t=\tcontr$ (i.e. similar to what
\citealt{2010MNRAS.407..381D} did to estimate the total number of
collisions in equal-mass clusters). Using the mass-radius relation for
the SMS (equation~\ref{eq:mr}), we then find
\begin{equation}
\msms(\tcontr) \propto \left( \frac{\rsmsn\msmsn^{5/6-\delta}}{\rhohn^{1/12}} N^{5/3}\dot{M}^{5/6}\right)^{1/(2-\delta)}.
\end{equation}
If we again assume $\dot{M} \propto N$ then for fixed $\msmsn$, $\rsmsn$ and $\rhohn$ we find
\begin{equation}
\msms(\tcontr) \propto N^{5/(4-2\delta)}=
\begin{cases}
N^{5/4}, &\delta=0,\\
N^{5/3}, &\delta=1/2,\\
N^{5/2}, &\delta=1.
\end{cases}
\end{equation} 
For all reasonable assumptions for the uncertain mass-radius evolution
of the SMS (i.e. the value of $\delta$), the scaling $\msms(N)$ is
super-linear. Although we have not folded in mass loss from the SMS,
it is encouraging that massive GCs can form more SMS mass per unit
 cluster mass, which is required to explain the observed
scalings of $\Delta Y$ and $\fenr$ with GC mass.  In
Fig.~\ref{fig:msms_simple_rel} we show the growth of $\msms$ for a
capped energy production rate, and otherwise the same conditions as in
Fig.~\ref{fig:msms_simple}.
\begin{figure}
\includegraphics[width=8cm]{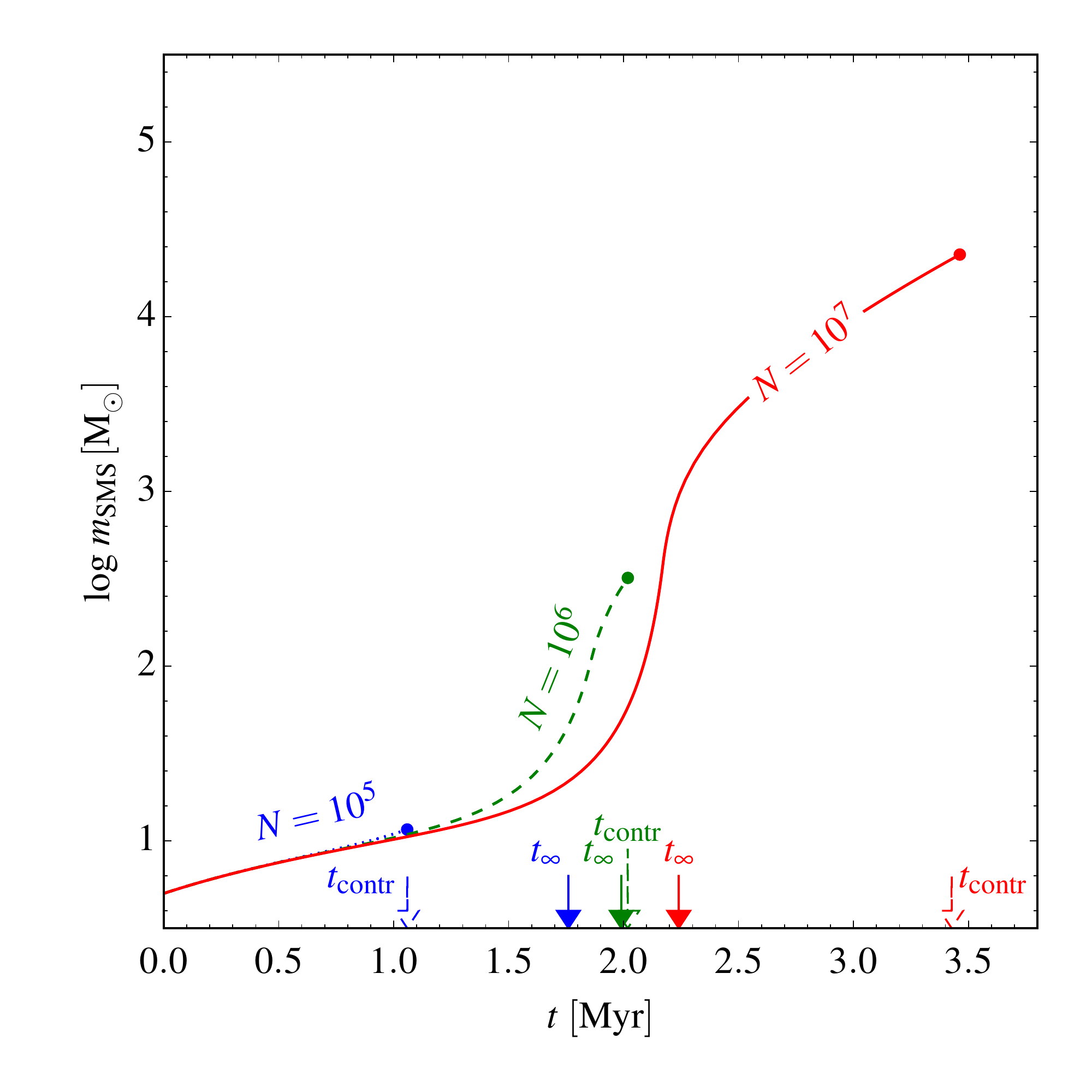}
\caption{As Fig.~\ref{fig:msms_simple}, but now $\mdotsmscoll$ is
  limited by the flow of energy through the cluster
  (equation~\ref{eq:mdotsmsrel}), which becomes relevant at
  $t\simeq\tinf$. As before, the $N=10^5$ cluster does not enter the
  runaway collision phase because $\tcontr<\tinf$. For the larger $N$
  clusters, the collision rate is capped by two-body relaxation
  (equation~\ref{eq:mdotsmsrel}) after $\sim1\,\myr$ ($N=10^6$) and
  $\sim1.5\,\myr$ ($N=10^7$).  }
\label{fig:msms_simple_rel}
\end{figure}

After collisions have become important, they are the dominant growth
term. In our model, gas accretion on the SMS is proportional to its
initial mass, but even if gas accretion is proportional to the
instantaneous mass it would contribute an order of magnitude less to
the growth of $\msms$ than binary coalescence.

We now have a simple model for the formation of SMS via stellar
collisions. We note that some aspects of this idea, such as the
efficiency of cluster contraction following mass accretion and the
collisional formation of SMSs, have already received some numerically
validation \citep{2010A&A...516A..73D, 2011MNRAS.410.2799M}.  There
are several uncertainties in the model presented in this section.
Firstly, the efficiency of the accretion of gas that flows into the
cluster onto the protostars. We assumed this to be 100\%, but this
needs to be verified with hydrodynamical simulations with high spatial
and temporal resolution
\citep[e.g.][]{2017MNRAS.467.1067D,2017MNRAS.464.3536R,
  2017MNRAS.467.1067D,2017MNRAS.472.4155G,
  2017MNRAS.472.4982S}. Secondly, $\rsms$ is critical in the success
of the enrichment model as we will discuss in the next
section. Finally, the growth rate of $\msms$ in the model relying on
the capped collision rate depends on $\Eorb$ when the stars are
captured, and the evolution of the eccentricity $\varepsilon$. Their
values affect the constants of proportionality, and not the scaling
with $M$ and $N$. Other effects, such as the interaction of multiple
companions of the SMS and the stellar mass function, which could
potentially affect the scaling relations presented here, need to be
understood as part of a detailed model for SMS growth. A numerical
validation, similar to the work done on stellar disruption rates by an
IMBH in GCs \citep[e.g.][]{2004ApJ...613.1133B, 2004ApJ...613.1143B},
would be an important future step. In the next section we include the
SMS formation model in a GC self-enrichment model.

\section{An early self-enrichment model}
\label{sec:model}
We use the SMS formation model of the previous section to develop a
model for GC self-enrichment in the first few Myr of its formation.
We start by putting together the different ingredients for GC and SMS
formation and evolution in Section~\ref{ssec:mod:cluster} and
\ref{ssec:mod:sms}. We add the effect of stellar winds in
Section~\ref{ssec:mod:winds}, which is required to pollute the
intra-cluster medium with hot-hydrogen burning products of the SMS
that end up in the chemical composition of the MSPs we observe today.
We present the predictions for the evolution of the global properties
of the SMS and their dependence with the total cluster mass in
Section~\ref{subsec:globalSMSproperties}. We discuss all the other
aspects of the early self-enrichment model in
Section~\ref{sec:discussion}.
\subsection{Cluster evolution}
\label{ssec:mod:cluster}
Cluster formation is complex because of the various physical
  processes at work on their respective timescales, which makes a
  definition of when the cluster forms (i.e. $t=0$) rather ambiguous.
  In our model, we define $t=0$ as the start of the gas accretion
  phase, i.e. when the cloud has just fragmented.
  Magnetohydrodynamical models of star formation in turbulent clouds
  by \citet[][their figure 13]{2014ApJ...797...32P} show that
  protostars of any mass may take $\sim1-2\,\myr$ to form, with large
  variations in the assembly timescale depending on the environment.
  Radiation hydrodynamics simulation of the collapse down to the
  protostar then show that this essentially proceeds on the free-fall
  time \citep{2016MNRAS.463.2553R}, which is significantly shorter
  than $1\,\myr$ for a clump massive enough to form a $100\,\msun$
  star. For simplicity, we assume that the gas accretion phase
  dominates in the first 2\,\myr, and that stars that are massive
  enough (typically $\geq 10\,\msun$) to go supernova reach zero-age
  main sequence at $t=2\,\myr$. This corresponds to the moment when
  the so-called stellar evolution phase starts (i.e. when H-burning
  ignites in the core of the stars more massive than $10\,\msun$; at
  that moment, low-mass protostars should still be on the Hayashi
  track, since typical PMS duration ranges between 10 and 100\,Myr,
  see
  e.g. \citealt{1994ApJS...90..467D,2002A&A...382..563B,2016A&A...587A.105A}). 
  We assume that gas accretion proceeds after
  this, although the efficiency may be reduced because of outflows
  \citep{2015MNRAS.450.4035F, 2017ApJ...847..104O}. This timeline is
  illustrated in Fig.~\ref{fig:model}.

We solve for the change in cluster stellar mass as the result of gas
accretion and mass loss from stellar evolution as:
\begin{align}
\dot{M} &= \mgcdotacc + \mdotsev,
\label{eq:mdotmod}
\end{align}
where $\mgcdotacc$ describes the gas accretion and is given by
\begin{align}
\mgcdotacc &=\begin{cases}
\mdotaccav\,N&t<5\,\myr\\
 0&t\ge5\,\myr,
\end{cases}
\end{align}
with $\mdotaccav=0.1\,\msun\,\myr^{-1}$. We assume that at
$t>5\,\myr$ the gas accretion is halted by stellar feedback. This time corresponds to the moment when massive stars have an age of $3\,\myr$, i.e.  close to the typical theoretical lifetime of a $120\,\msun$
star  ($\sim3.5\,\myr$, see e.g. \citealt{2013A&A...558A.103G}). After $t=5$\,Myr,
the stellar mass decreases as the result of stellar mass-loss.
  In the
first 10\,Myr of massive star evolution  (typical theoretical
  lifetime of a 15$\,\msun$ star, \citealt{2013A&A...558A.103G}), a
stellar population loses about 10\% of its mass, so we use
 \begin{align}
\mdotsev &=\begin{cases}
0\,&t<5\,\myr\\
 -0.1M/t&t\ge5\,\myr.
\end{cases}
\end{align}
 Note that in equation~(\ref{eq:mdotmod}) we do not subtract the
  mass of stars that end up on the SMS, because this (negative)
  contribution is negligible in most models. The half-mass radius of
the cluster evolves as
\begin{align}
\rhdot(N,M, \rh)=-3\frac{\dot{M}}{M}\rh + \zeta \frac{\rh}{\trh},
\label{eq:rmdot}
\end{align}
where the first contribution is due to the mass evolution and the
second contribution describes the expansion due to two-body
relaxation.  Before $t=5\,\myr$, the first term describes the adiabatic
contraction ($\dot{M}>0, \rhdot<0$) described in
Section~\ref{ssec:accretion}, and after $t=5\,\myr$ it describes the
expansion ($\dot{M}<0, \rhdot>0$) following stellar evolution
mass-loss.  For homologous and adiabatic mass-loss, the cluster
expands as $\rhdot/\rh = -\dot{M}/M$ \citep{1980ApJ...235..986H}, but
because most of the stellar mass-loss occurs in the centre where the
massive stars reside, the factor of 3  approximates the enhanced expansion
as the result of mass segregation.  We use equation~(\ref{eq:trh1})
for $\trh$, with the parameters described in
Section~\ref{ssec:accretion}. As long as $\tacc<<\trh$ in the first 5
Myr, the first term dominates and the cluster contracts, until
$\tacc\simeq\trh/\zeta$, after which relaxation starts to dominate the
evolution and the cluster expands as the result of two-body
heating  by SMS-star binaries. For clusters without gas accretion, this relaxation driven
expansion results in a radius evolution of the form $\rh\propto
t^{2/3}$ \citep{H61}, but in our case the radius expands more slowly
because of the gas accretion, which contributes negatively to
$\rhdot$.

With the evolution of $M$ and $\rh$ of the cluster, we can now
consider the evolution of the SMS in its centre.

\subsection{Hypothesis for the mass growth of the SMS (gas accretion and collisions)}
\label{ssec:mod:sms}
The SMS mass grows as the result of gas accretion and stellar
collisions. Its rate of growth as a function of the cluster properties
and its own properties is given by
\begin{align}
\mdotsms = \mdotacc + \min\left(\mdotsmscoll, \mdotsmsrel\right).
\label{eq:mdotsmsmod}
\end{align}
Here $\mdotacc$ is the rate of growth as the result of gas accretion,
and as discussed in Section~\ref{ssec:accretion}, we adopt an
accretion rate that is proportional to the stellar mass at the time of
fragmentation, i.e.  $\mdotacc = \msmsn\,\myr^{-1}$, and we adopt an
initial mass and radius of the SMS of  $\msmsn=5\,\msun$ and
$\rsmsn=6.7\,\rsun$ (see Section~\ref{ssec:smsprop}).  The exact value
of $\msmsn$ is not very important, because the onset of the runaway
collision process (i.e. $\tinf$) is very insensitive to $\msmsn$ and
$\rsmsn$ (see equation~\ref{eq:tinf}).  The second term contains the
two mass growth rates due to collisions discussed in
Section~\ref{sec:analytics}: $\mdotsmscoll$ is given by
equation~(\ref{eq:mdot}) and describes the growth of the SMS before it
is regulated (i.e. capped) by relaxation, which becomes important when
$\mdotsmsrel< \mdotsmscoll$ and then the SMS grows at a rate
$\mdotsmsrel$, given by equation~(\ref{eq:mdotsmsrel}).
 
The radius of the SMS is passively evolved via the mass-radius
relation of equation~(\ref{eq:mr}) and we adopt values of $\delta=0.5$
and $\delta=1$.

\subsection{Hypothesis for the mass loss of the SMS by stellar winds}
\label{ssec:mod:winds}
Hot massive stars experience a strong mass loss by stellar winds.  We
will estimate the radiation driven mass-loss rates for very massive
stars and SMSs based on the stellar models of SMSs with $\msms > 100\,
\msun$ from \citet{2005AstL...31..695N}. Very massive main sequence
stars are largely convective, due to their high luminosity. This
implies that their dimensionless structure is (almost) independent of
the energy source, which facilitates the calculation of the
mass-luminosity relation.  The main parameter that determines the
structure and luminosity of the star is $\mu^{2} \msms$, where $\mu$
is the mean particle mass ($\mu = 0.60$ for $X=0.75$, $Y=0.25$,
$Z=0.001$).

The total potential plus kinetic energy of a radiation driven wind can
not exceed the stellar luminosity.  This implies that $0.5 \mdotwind
(\vesc^2 + \vinf^2) < \lsms$ or $\mdotwind <
\lsms\rsms/(G\msms)$. Here $\vesc$ is the escape velocity from the
surface of the SMS and $\vinf$ is the terminal wind velocity.  This is
a severe upper-limit, because it implies that the total luminosity is
used to drive the wind and no light will leave the star. A more
realistic estimate is obtained if we assume that the total momentum of
all the photons is used to drive the wind. This results in
$\max(\mdotwind) = \lsms/(c \vinf)$.  Multiple scattering of photons,
as occurs e.g. in the winds of Wolf-Rayet stars, can increase this
upper-limit by at most a factor $f \simeq 3$
\citep{2011A&A...531A.132V}.  Terminal wind velocities of radiation
driven winds scale with $\vesc$, i.e. $\vinf = a \vesc$, with $a=2.6$
for stars with $\Teff > 20\,$kK and $a = 1.3$ for stars with $10 <
\Teff/{\rm kK} < 20$ \citep{1999isw..book.....L}.  This results in an
estimate of
\begin{align}
\max(\mdotwind)   &=   \frac{f}{a}  \frac{\lsms}{c} \sqrt { \frac{\rsms}{2G\msms(1-\Gamma)}},  \\
                 & \simeq 10^{-4}\,\msunyr \left(\frac{\msms}{100\,\msun}\right)^{1/2+\delta/2}.
                 \label{eq:mdotwind0}
\end{align}
The maximum mass-loss rates derived here agree with the mass-loss
rates calculated by \citet{2011A&A...531A.132V} for very massive stars
($\gtrsim100\,\msun$) with solar metallicity. However,
\citet{2001A&A...369..574V}  and \citet{2018arXiv180308042V}  have shown that the mass-loss rates depend
on metallicity as $\mdotwind \propto Z^{0.8}$. If this also holds for
supermassive stars, the mass-loss rates of low metallicity ZAMS stars
with $Z = 0.001$ may be 10 times smaller.

The estimates described above are for H-rich main sequence stars. As
the stars evolve almost homogeneously, because they are largely
convective, the H-abundance decreases and the He-abundance increases.
As the He-abundance increases, $\mu$ increases and the opacity
$\sigmae$ decreases due to the reduced abundance of electrons. This
results in an increase in luminosity and an increase in the Eddington
luminosity, because $\ledd \propto \msms/\sigmae$.
\citet{2011A&A...531A.132V} have shown that the mass-loss rate
increases strongly with the ratio $\Gamma=\lsms/\ledd$ for stars close
to their Eddington limit.  In fully convective stars, $\Gamma$
increases as a function of $\mu^2 \msms/\sigmae$
\citep{2005AstL...31..695N}.  Both factors $\msms$ and $\mu^2/\sigmae$
increase during the build-up of a SMS, so the luminosity approaches
the Eddington limit and the mass-loss rate will increase.

\begin{figure*}
    \centering
    \subfloat{%
        \includegraphics[width=0.47\linewidth]{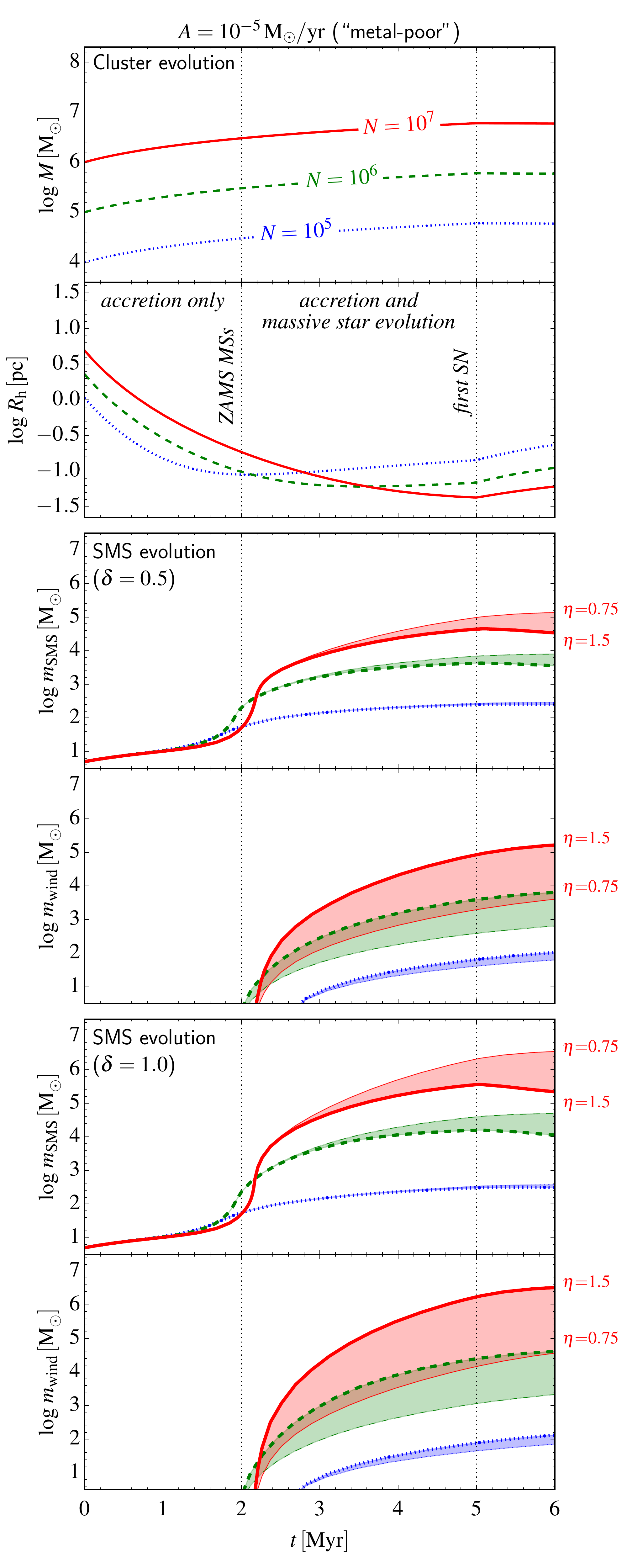}%
        \label{fig:a}%
        }%
    \hfill%
    \subfloat{%
        \includegraphics[width=0.47\linewidth]{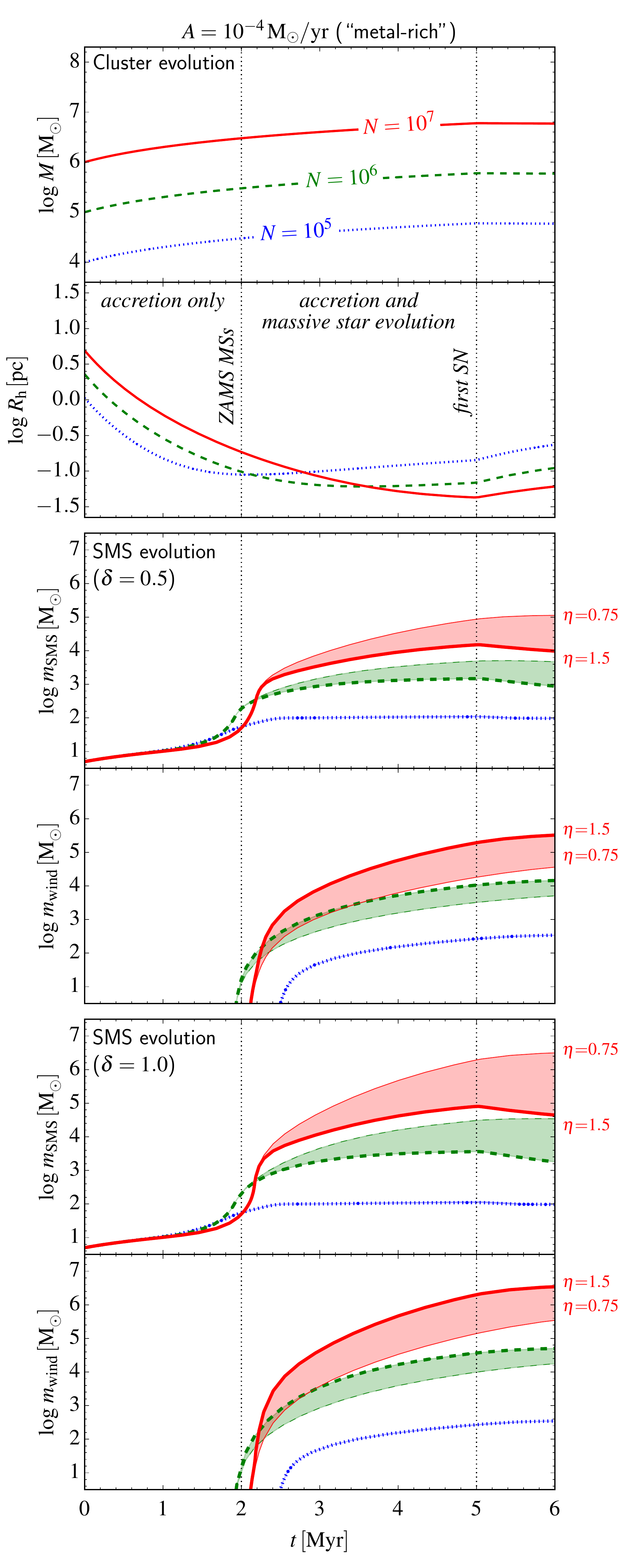}%
        \label{fig:b}%
        }%
    \caption{Model for SMS formation in a GC that accretes gas until
      $t = 5\,\myr$.  As stated in Section~\ref{ssec:mod:cluster},
      gas accretion and cluster contraction starts at $t=0$ and
          the first massive stars $> 10\,\msun$ reach the ZAMS at
          $t=2\,\myr$.  The possible occurrence of the first
          SNe (from stars of $\sim 120\,\msun$, main sequence lifetime
          $\sim 3$~Myr) is also indicated.  Left panels show the
      results for the metal-poor clusters/SMSs with the lower
      $\mdotwind$ ($A=10^{-5}\,\msunyr$ in
      equation~\ref{eq:mdotwind}), and the right panels show the
      result for the metal-rich case ($A=10^{-4}\,\msunyr$).  The
      cluster $M$ and $\rh$ evolve according to
      equations~(\ref{eq:mdotmod}) and (\ref{eq:rmdot}),
      respectively. The SMS mass and radius are evolved using
      equations~(\ref{eq:mdotsmsmod}) and (\ref{eq:mr}), respectively,
      for $\delta=0.5$ (middle panels) and $\delta=1$ (bottom
      panels). The thick lines show the result for $\eta=0.75$ in
      equation~(\ref{eq:mdotwind}) and the thin lines show the results
      for $\eta=1.5$.  The middle panels show the accumulated mass
      liberated by the SMS in winds (from equation~\ref{eq:mdotwind})
      for $\delta=0.5$, while the bottom two panels show the result
      for $\delta=1$. For stronger winds ($\eta=1.5$), $\msms$ grows
      more slowly, but more mass is liberated.}
\label{fig:model}
\end{figure*}

These estimates refer to stars in thermal and dynamical equilibrium.
However, as a SMS accretes a  massive star, it may be out of
equilibrium during approximately the Kelvin-Helmholtz timescale which
is about $10^4$ yrs for a $300\,\msun$ star at the main sequence and
$10^3$ yrs for a He star.  At that time the $\Teff$ may be
considerably lower than the value adopted above
\citep{2009A&A...497..255G}.  We can obtain an estimate of $\Teff$
from the study of Luminous Blue Variables (LBVs), which increase their
radius and decrease their $\Teff$ during outbursts to about 10\,kK
\citep{1994PASP..106.1025H}.  Substituting this low value of $\Teff$
in the formalism described above, and adopting the corresponding value
of $a=1.3$, we find a maximum radiation driven mass-loss rate that is
a factor five larger than predicted by equation~(\ref{eq:mdotwind0}).
This results in a mass-loss rate of $3 \times10^{-4}\, \msunyr$ for an
LBV of $60\,\msun$, which is close to the observed mass-loss rates of
LBVs during outburst.  If a SMS captures stars at a rate of order one
per $10^{4}$ yrs or faster, the star will be out of equilibrium for
most of its lifetime, so the time-averaged mass-loss rate will be
higher than for stars in equilibrium and may resemble that of LBVs.
Given these arguments, we adopt a simple relation
\begin{equation}
\mdotwind = A  \left(\frac{\msms}{100\,\msun}\right)^{\eta},
\label{eq:mdotwind}
\end{equation}
with $A = 10^{-4}\,\msunyr$ and $10^{-5}\,\msunyr$ to allow for the
possible mass loss reduction of SMS of low metallicity, and
$\eta=0.75$, corresponding to constant \Teff\ (i.e. $\delta=1/2$), and
$\eta= 1.5$ to account for the increasing mass-loss rate as the He
content increases.

\subsection{Predicted evolution of the SMS global properties, and dependence on the cluster mass}
\label{subsec:globalSMSproperties}
With all the ingredients of the previous sections in place, we are now
able to solve for the evolution of $M$ and $\rh$ and the SMS
properties $\msms$ and $\rsms$. We solve the coupled ordinary
differential equations of the previous sections with the `dopri5'
integrator \citep*{hairer1993solving}, which is a Runge-Kutta
integrator with adaptive step-size to calculate fourth and fifth order
accurate solutions. It is supplied by the \scipy\ sub-package
\integrate. The mass that is lost from the SMS by winds ($\mwind$) is
the amount of polluted material that is available for recycling into
MSPs.

The most uncertain parameters are $\delta$ in the SMS mass-radius
relation (equation~\ref{eq:mr}) and $A$ and $\eta$, which set the
strength of the stellar wind as a function of $\msms$
(equation~\ref{eq:mdotwind}). We therefore vary all three.
Fig.~\ref{fig:model} shows the result of the model for $\delta=0.5$
and $\delta=1$ for a relatively low $\mdotwind$ ($A=10^{-5}\,\msun/\yr$), corresponding to the
metal-poor conditions (left panels) and for a relatively high
$\mdotwind$  ($A=10^{-4}\,\msun/\yr$), corresponding to the metal-rich regions (right
panels). The two top panels show the evolution of cluster $M$ and
$\rh$ for clusters with different $N$. The middle panels show $\msms$
and $\mwind$ for $\delta=0.5$, where the thick lines show the result
of $\eta=0.75$ (i.e. $\mdotwind\propto \msms^{0.75}$) and the thin
lines are for $\eta=1.5$. For larger $\mdotwind$ (higher $\eta$), the
SMS reaches lower masses, but releases more mass in winds. The bottom
panels of Fig.~\ref{fig:model} shows similar results, but for
$\delta=1$.

The small difference in $\delta$ has a large effect on $\msms$ and
$\mwind$. Because for larger $\delta$ the SMS has a larger $\rsms$, it
has a larger cross section making collisions more frequent in the
early phases. At later stages, the effect of a larger $\rsms$ is that
the collisions result in a lower energy production, because the
colliding stars feel less of the gravitational potential of the SMS
when they collide. The most massive cluster ($N=10^7$) forms a SMS
reaching a mass of $\sim10\%$ of the cluster mass, whereas $\mwind$
can be an order of magnitude larger than $\max(\msms)$, because of
continuous rejuvenation of the SMS via collisions.  We notice that for
$\delta=1$, the total mass of processed material, $\mproc =
\msms+\mwind$, is insensitive to $A$ and $\eta$. For $N=[10^5, 10^6,
  10^7]$, we find $\mproc/\msun \simeq [870, 39\,000, 1\,629\,000]$,
or $\mproc/M \simeq [0.015, 0.057, 0.37]$. For $\delta=0.5$, models
with an order of magnitude larger $A$ result in 10\% larger $\mproc$
(for a given $N$), and varying $\eta$ from $0.75$ to $1.5$ results in
a maximum increase of $\mproc$ by a factor of two. In conclusion,
$\mproc$ is most sensitive to $N$ and $\delta$.
This `conveyor belt' production of material processed through the SMS
`nuclear reactor' allows this SMS scenario to overcome the mass budget
problem discussed in Section~\ref{sec:intro}.

Because the SMS wind is released while the cluster is still accreting
(cold) gas, the wind material can remain in
the cluster: the hot, processed material from the SMS wind mixes with
the cold pristine gas which subsequently accretes on the protostars
(see Section~\ref{ssec:mixing} for more details).  The bottom panels
of Fig.~\ref{fig:model} show that sufficient material can be produced
in a few Myrs to account for the observed proportions of chemically
anomalous low-mass stars in GCs (we discuss this further in
Sections~\ref{ssec:nucleo} and \ref{ssec:abundances}). A schematic representation of the GC
enrichment scenario is shown in Fig.~\ref{fig:schematic}. In the next
section we discuss several aspects of this GC self-enrichment
scenario in more detail.

\begin{figure*}
\begin{center}
\includegraphics[width=12cm]{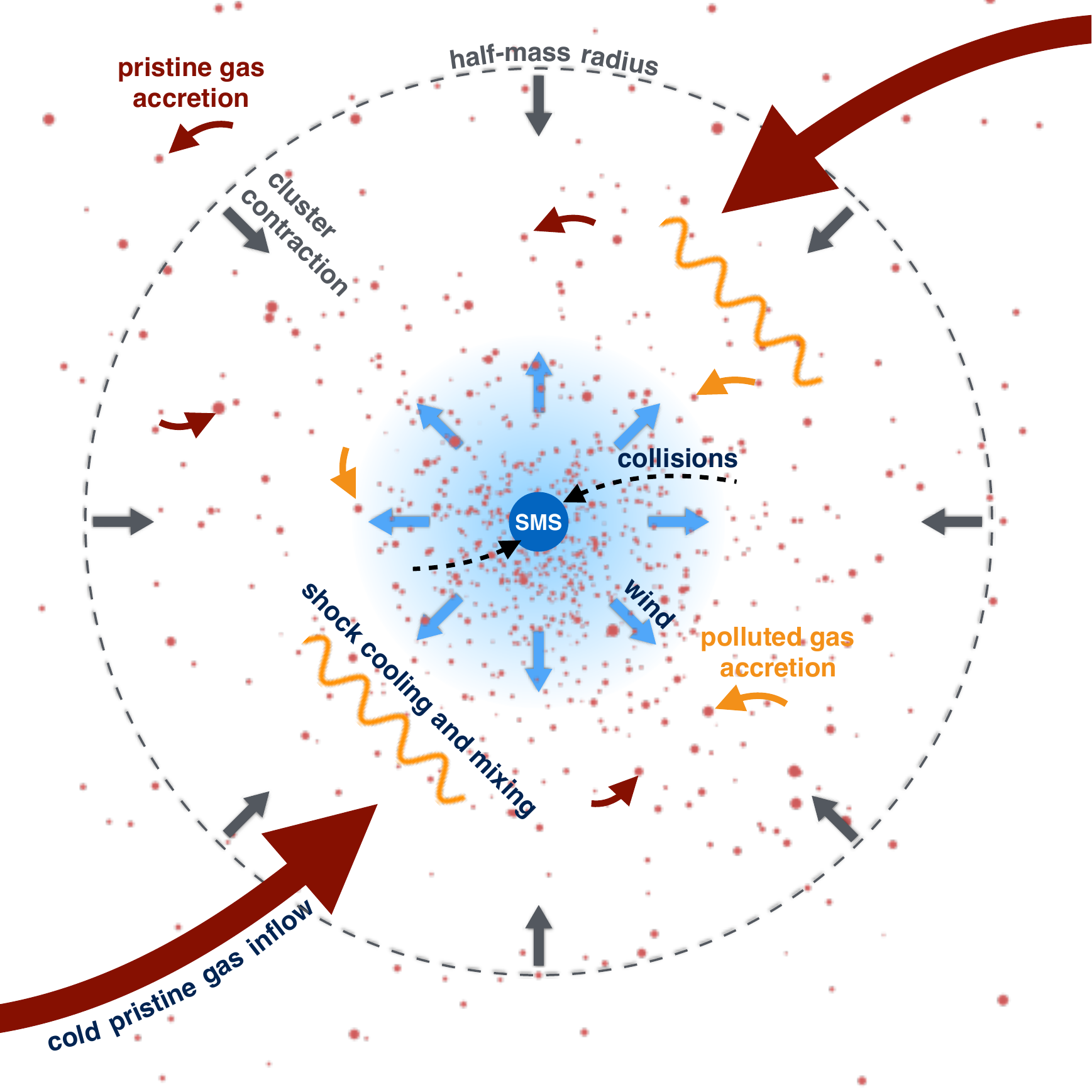}
\caption{Schematic picture of the enrichment scenario presented in
  Section~\ref{sec:model}. Cold, pristine gas accretes onto the stars
  in the cluster, causing the cluster to contract. The higher stellar
  density results in stellar collisions, forming a SMS in the cluster
  centre. The SMS blows a wind enriched in hot-hydrogen products,
  which interacts and mixes with the inflowing gas. The mixed material
  subsequently accretes onto the stars enriching them with SMS
  yields. }
\label{fig:schematic}
\end{center}
\end{figure*}

\section{Discussion}
\label{sec:discussion}

In this section we discuss several aspects and uncertainties of the
model presented in Section~\ref{sec:model} in more detail and we make
observational predictions.
\subsection{Mixing of SMS wind and pristine gas}
\label{ssec:mixing}
Ejecta in stellar winds in general and likely also in winds from SMSs
have high velocities, typically of the order of a $10^3\,\kms$
(e.g. \citealt{2012A&A...537A..37M}; also for our SMS model of
Section~\ref{ssec:smsprop} we find
$\vinf\simeq3000\,\kms(\msms/100\,\msun)^{1/2-\delta/2}$). This may
exceed the escape velocity of our modelled star clusters. For a
\citet{1966AJ.....71...64K} model with dimensionless central potential
$W_0=9(10)$ we find that the escape velocity from the centre of the
cluster is\footnote{We use the {\sc limepy} models
  \citep{2015MNRAS.454..576G} to compute $\vesccl(0)$.} $\vesccl(0)
\simeq490(545)\,\kms\sqrt{M/10^6\,\msun}\sqrt{0.1\,\pc/\rh}$.  The SMS
is assumed to be in the very centre, because this is where the
collision rate is highest and if it is displaced by collisions, it
rapidly sinks back via dynamical friction.  Interaction with ambient
gas slows down the ejecta so they can accrete on the low-mass
protostars.  Interestingly, we do not expect the stellar winds,
including the SMS wind to push out the dense gas from the cluster in
the case of massive, compact star clusters, as investigated in the
present study \citep{2016A&A...587A..53K}: for a Salpeter mass
function, stellar winds can not remove the gas if $\vrms$ exceeds a
critical value that is in the range\footnote{Note that
  \citet{2016A&A...587A..53K} use a compactness parameter
  $C_5=(M/10^5\msun)(\rh/\pc)^{-1}\propto \vrms^2$} $10-30\,\kms$,
depending on the star formation efficiency. This is still applicable
for our modelled clusters that contain SMS, because the mass-loss rate
of the SMS is comparable to or less than the combined mass-loss rate
of the other massive stars.

For our lowest mass cluster with $N=10^5$, i.e. initially $10^4\, \msun$
in stars, $\vrms$ stays between $8-15\,\kms$ (derived from
Fig.~\ref{fig:model}). So as star formation proceeds, we expect some
gas to be removed. The higher mass clusters have $\vrms>30\,\kms$
throughout and will therefore not lose gas due to the stellar winds.

Can the ejecta mix with pristine gas? This should not simply be
assumed, as for example in the Milky Way, observations of radioactive
massive star ejecta show that there is no mixing for at least
$\sim1\,\myr$ after ejection
\citep{2013A&A...559A..99K,2015A&A...578A.113K}.  The SMS wind is,
however, efficiently caught in such a star cluster.  To see this, we
first calculate the stall radius of the SMS wind where the bulk
velocity of the wind $\vinf=v_3\,10^3\,\kms$ is thermalised at the
inner shock. The stall radius, $\rstall$, is defined by the ram
pressure, which declines with distance $r$ from the SMS as $r^{-2}$,
matching the ambient pressure $P$. The general pressure level is set
by the gravitational potential, $P=fGM^2/\rh^4$, where $M=10^6 M_6
\msun$ is the total mass of the system and $f\lesssim1$ is a function
of the detailed mass distribution. Ram pressure balance, $\rho \vinf^2
= P$, then implies
\begin{align} 
\rstall = 10^{-5}\,\mathrm{pc}\, |\dot{M}_2|^{1/2} v_3^{1/2} R_\mathrm{h,-1}^2 f^{-1/2} M_6^{-1},
\end{align}
where $\rh = 0.1 R_\mathrm{h,-1}\,\pc$ and $\dot{M} =
10^2\,\dot{M}_2\,\msun/\myr$.  This is very small compared to the size
of our cluster for all reasonable choices of parameters. The SMS wind
density at the stall radius is given by
\begin{align}
\rho = 10^{-15} \mathrm{g\, cm}^{-3} \, v_3^{-2}M_6^2 R_\mathrm{h,-1}^{-4}.
\end{align}
This is so high that the ejecta, which will first be heated to keV
temperatures at the wind shock, will cool down instantly via emission
of bremsstrahlung and collisionally excited line emission, and clump
via the thermal instability. The resulting cold gas will integrate
naturally into the ongoing accretion flows onto the low-mass
stars/clumps.  The recent proposal by \citet*{2017arXiv171104007S}
that polluted stars might be formed in the photoionised shells of very
massive supergiants is conceptually similar, also relying on high gas
densities.  Low-mass stars observed with extreme (low) oxygen
abundances formed out of essentially pure hot hydrogen burning ejecta,
i.e. with a dilution factor with pristine material less than $10-30\%$ \citep[respective values for NGC~2808 and NGC~6752,
  see][]{2017A&A...608A..28}.  Complete mixing before accretion from
the intracluster medium can therefore not take place (details in
Section~\ref{ssec:nucleo}). Instead we have to postulate that extreme stars
form close to the stall radius where the intracluster medium should
predominantly consist of SMS ejecta.  Details are complex and are
beyond the scope of the present work.  Importantly, the inability of
the system to lose dense gas via stellar feedback (compare above) will
ensure that all the cold gas, including the cooled-down ejecta will
end up in low-mass stars.

\subsection{Nucleosynthesis in SMS,  composition of their ejecta, and dilution with pristine gas}
\label{ssec:nucleo}
The key points of the `conveyor belt' mechanism for the production of
the observed abundance anti-correlations and the moderate enrichment
of He by a SMS are the following: (1) a SMS is a fully convective
object, with a convection flow timescale ($\tau_{\rm conv}\leq \yr$)
many orders of magnitudes shorter than the timescale for nuclear
fusion; (2) the nuclear reactions that produce the anti-correlations
reach an equilibrium on a timescale much shorter ($\tau_{\rm NaO}\sim
\yr$) than the main sequence lifetime for the full conversion of H
into He ($\thelium\sim 0.2 - 2\, \myr$, depending on the SMS mass);
(3) the products of nuclear fusion are spread throughout the star by
convection so that a large fraction of the SMS is quickly enriched and
partly ejected in the wind; (4) collisions constantly refurnish the
stellar interior with low-He content material on an intermediate
time-scale ($\tsms\sim0.1\,\myr$). The relative time-scales,
i.e. $\tau_{\rm conv}\lesssim\tau_{\rm NaO} \ll \tsms<\tau_{\rm He}$,
ensures the efficient production and release of CNONaMgAl processed
material, whilst keeping low the He abundance inside the SMS and its
wind.

Consequently, we can discuss nucleosynthesis in SMSs by focusing on
their central temperature and assume that the material they eject at a
given evolution time has the same chemical composition as the central
regions where hot H-burning occurs, with the difference that He
remains relatively constant. Here we base our discussion on SMS models
presented in \citet{2015MNRAS.448.3314D} and in
\citet{2017A&A...608A..28} as well as additional models for the
metallicity range relevant for GCs computed by P.~Denissenkov with the
same stellar evolution code {\sc mesa}
\citep{2011ApJS..192....3P,2013ApJS..208....4P} and the same input
physics (Denissenkov, private communication). These models cover the
mass range between 10$^3$ to $7 \times 10^4\,\msun$, and
$-2.2<\feh<-0.5$. Accretion and rotation are not accounted for in the
computations; we assume that this does not affect the nucleosynthesis
picture, nor the timescales.

Over the whole mass and metallicity range considered, SMS models reach
high enough central temperature ($\geq 60\,$MK) to run CNO and NeNa at
equilibrium already at the very beginning of the main sequence (see
figure~4 of \citealt{2017A&A...608A..28}, for models with
$\feh=-1.5$). Therefore, in all SMS models the C--N and O--Na
anti-correlations build up immediately for very low He enrichment
($\Delta Y<0.01$, in agreement with the observationally inferred
variation in the He content of most GCs).  However, a minimum SMS mass
of $\sim 5 \times 10^3\,\msun$ is required to reach Mg-burning
temperature ($\sim 75 - 80\,$MK) early on the main sequence with
similarly low He enrichment (figure~4 of
\citealt{2017A&A...608A..28}).  Because of the dependence between the
mass and the central temperature of the stars, we thus expect the
Mg--Al anti-correlation to be present only in the most massive
clusters where the more massive SMS can form (see
Section~\ref{subsec:globalSMSproperties}).

A metallicity dependence is also expected, as stars of a given mass
reach hotter temperature on the early main sequence when metallicity
decreases. Additionally, $\mdotwind$ is lower at low metallicity,
leading to larger $\msms$. Thus for a given cluster mass, more extreme
Mg-depletion is foreseen at lower metallicity. This nicely explains
the finding of a bivariate relation of the Mg spread as a function of
$M$ and $\feh$ in Galactic GCs  (\citealt{2009A&A...505..139C};\citealt{2017A&A...601A.112P}).

In the current SMS models, which are computed without accretion nor
rejuvenation by stellar collisions, the central temperature strongly
increases at the end of the main sequence, up to values where Na and
Al efficiently burn through p-captures. This is predicted to occur for
$\Delta Y\gtrsim0.55$ (see figure~4 of \citealt{2017A&A...608A..28}),
i.e. for He enrichment more extreme than allowed by the current
photometric constraints ($\Delta Y$ between 0.01 and 0.18, see
references in \citealt{2017A&A...608A..28}).  Therefore, SMSs are in
principle successful to get the right chemistry to explain the whole
observed abundance patterns, but only in their early main sequence
nuclear burning configuration. In the `conveyor belt' scheme we
actually expect SMS to reach an equilibrium state that allows them to
stay in these conditions as collisions constantly bring fresh material
with low He content that is instantaneously mixed within the
convective interior of the SMS.  Varying dilution factors between SMS
ejecta with low He content and material with pristine composition
nicely accounts for the whole range of abundances observed along the
anti-correlations (e.g. figure~1 in \citealt{2015MNRAS.448.3314D} and
figure~7 in \citealt{2015MNRAS.449.3333B}).  These variations should
be directly related to the varying amounts of SMS ejecta that are
accreted by individual low-mass protostars. In particular, one can
speculate that the clumps located in the more central regions closer
to the SMS form the stars with the most extreme abundance variations,
while those in the more external regions of the cluster form stars
with chemical composition closer to the original one. This accounts
for the fact that polluted stars are generally more concentrated in
the innermost region than pristine stars
\citep[e.g.][]{2011A&A...525A.114L,2016MNRAS.463..449S}, although
there are clusters, such as M15, for which the relative radial
distribution may be reversed in the very inner parts
(\citealt{2015ApJ...804...71L}, but see
\citealt{2018MNRAS.tmp..706N}). Because of the short relaxation
time-scale in the centre, the radial distribution of stars in the core
is actually unlikely to be the result of the initial conditions.
Similarly, the relative amounts of pristine and processed gas present
in the cluster should be a function of time, with more pristine gas
being available in the beginning. This should also contribute to
the spread in the abundance pattern \citep{2007A&A...475..859D}.
Estimates of the total dilution factor using different constraints
(e.g. the extent and shape of the O--Na and Li--Na anti-correlations)
indicate that the overall process operated approximately in the $1:1$
dilution regime \citep{2017A&A...608A..28}. In the next section we
discuss several elemental abundances in more detail.

\subsection{Abundances of various elements}
\label{ssec:abundances}

\subsubsection{Sodium}
\label{ssec:sodium}
Figure~6 of \citet{2009A&A...505..117C} shows the Na--O
anti-correlation for about 2000 stars in 19 different GCs. From this
we see that a typical GC star has a Na abundance that is about 2.5
times higher than that of pristine stars in the Milky Way
halo. Considering that the material in the stars is the result of
material with pristine abundance $\xprist$ and processed material with
abundance $\xenr$, then the resulting abundance of the mixed material
$\xmixed$ can be written as

\begin{equation}
\xmixed= \frac{\xenr + f\xprist}{1+f},
\end{equation}
where $f$ is the dilution factor, i.e. one part of processed material
is mixed with $f$ parts of pristine material
\citep[e.g.][]{2017A&A...608A..28}, which in our model can be written
as $f = M/\mwind-1$. For $\xmixed/\xprist=2.5$, as estimated from the
Carretta et al. results, we find $f=(\xenr/\xprist -2.5)/1.5$, or
$f\simeq 5$ for $\xenr/\xprist\simeq10$ (see figure~1 in
\citealt{2015MNRAS.448.3314D}). Using the bottom panels of
Fig.~\ref{fig:model}, we see the required $M/\mwind=6$ (i.e. $f = 5$)
is feasible. For $\eta=1.5$, we find $M/\mwind \simeq 10(2)$ for
$N=10^6(10^7)$ (almost independently of $A$).  From this we see that
it is in principle possible to produce sufficient amounts of Na in
this SMS scenario.  
We note that the dilution tracks of SMS yields reproduce the shape of the O--Na anti-correlation \citep{2014MNRAS.437L..21D}. As explained in \citet{2007A&A...470..179P,2017A&A...608A..28}, O is at equilibrium at the temperature where Na is produced, and the minimum dilution factor along the O--Na anti-correlation is actually set by the observed extreme O abundances. Therefore, reproducing the Na enrichment implies that we also get the correct amount of O depletion.

\subsubsection{Helium}
\label{ssec:helium}

\citet{2007ApJ...661L..53P} estimate the amount of additional He that
is needed to explain the two main sequences blue-wards of the pristine
main sequence in NGC\,2808, one of the clusters displaying some of the
most extreme multiple population features \citep[see
  also][]{2015ApJ...808...51M}. One of the populations needs
$2.2\times10^4\,\msun$ additional He and the second one needs
$9.1\times10^3\,\msun$ additional He. For a mass fraction of $Y\simeq
0.4$ this implies a total mass of SMS processed material of
$\mwind\simeq(2.2+0.9)/0.6\simeq5\times10^4\,\msun$. This assumes that
the SMS ejecta have $Y=0.4$ independent of time, which is of course
oversimplified. However, in Section~\ref{ssec:mod:winds} we discussed
that $\mdotwind$ goes up quickly if the He abundance increases and
$\msms$ increases, hence most material will be ejected in the later
stages of the SMS evolution.  From Fig.~\ref{fig:model} we see that
this is achievable even with less favourable choice of $\eta=0.75$ and
$\delta=0.5$ for $N=10^7$.  \citet{2014ApJ...785...21M} shows $\Delta
Y$ as a function of $M_V$ and finds that a typical cluster with
$M_V\simeq-7.5$ has $\Delta Y \simeq 0.03$. If we again assume that
the processed material has $Y=0.4$, then for a present day GC mass of
$2\times10^5\,\msun$ we (only) need $1500\,\msun$ in He, or
$\mwind\simeq2500\,\msun$. From Fig.~\ref{fig:model} we see that
this is also achievable for lower-mass GCs with $N=10^6$. Because the
exact He abundance of the wind is very uncertain, because of the
sensitive dependence on various time-scales, it is difficult to make
strong predictions at this point. However, from the rough estimates
here we conclude that it is possible for a single SMS to produce
sufficient amount of He, even for the most extreme (i.e. the most
massive) clusters.

\subsubsection{Lithium}
\label{Li}
In our model, H-processed material from the SMS wind mixes with
inflowing pristine material. Because the CNONaMgAl abundance patterns
are produced at high temperatures, all the fragile Li (burning
temperature of $\sim 2.5\,$MK) is destroyed inside the SMS and the
wind material is Li free. However, thanks to dilution of SMS ejecta
with pristine material (Section~\ref{ssec:nucleo}), it is expected
that right after cluster formation there is a Li--Na anti-correlation
in the material that can be accreted by the newly forming protostars
\citep[e.g.][]{2011A&A...527A.148L}.  In this framework, the maximum
initial Li abundance in a given star at birth obviously depends on the
Li content of the pristine gas, on the amount of dilution with SMS
ejecta, and on the moment when polluted material is accreted by the
newly forming star (very fast accretion is actually needed to maintain
a non-negligible lithium content in the accreting star;
\citealt{2014MNRAS.443.3302D}, \citealt{2014A&A...566A.109S}).

Importantly, the photospheric Li abundance of GC low-mass stars is
expected to decrease as a result of the combination of several
processes at act in their interiors along their evolution. This
includes nuclear burning in the fully convective phase and
thermohaline mixing induced by accretion once a radiative core has
developed on the pre-main sequence, atomic diffusion in the presence
of weak turbulence, mass loss, rotation-induced mixing, and internal
gravity waves along the main sequence, and dredge-up and thermohaline
mixing along the red giant branch (for references see
e.g. \citealt{2000IAUS..198...74P} and
\citealt{2016EAS....80..177C}). These processes are known to modify
the surface Li abundance of Population I main sequence and giant stars
(including the Sun and field and open cluster stars). They could thus
potentially explain why the primordial Li abundance, as derived from
Planck \citep{2016RvMP...88a5004C} is about three times larger than
what is found in Pop II stars in the Milky Way
\citep[e.g.][]{1982A&A...115..357S,2005A&A...442..961C,2010A&A...522A..26S,2010A&A...515L...3M}
and its GCs
\citep{2006Natur.442..657K,2007ApJ...671..402K,2006Natur.442..636C,2010A&A...519L...3M,2011MNRAS.412...81M,2016A&A...589A..61G}.
These processes could also blur the initial Li--Na anti-correlation,
through differential effects all along the evolution of GC stars born
with different initial masses, chemical compositions, and rotation
rates.

As a matter of fact, observational hints for a Li--Na anti-correlation
were found in old GCs, but only in the case of studies focusing on
main sequence turnoff stars
\citep{2005A&A...441..549P,2007A&A...470..153B,2009A&A...503..545L}. We
argue that the absence of a Li--Na anti-correlation in studies
focusing on red giant GC stars \citep[e.g.][]{2015MNRAS.449.4038D}
where the effects of dilution and thermohaline mixing add to the
complexity, can therefore not be used as evidence for an absence of an
anti-correlation of the initial Li--Na abundances. Stellar evolution
models including all the relevant internal processes are thus urgently
needed to explain the Li data in GC stars all along their evolution
and to definitively assess the validity of an initial Li--Na
anti-correlation right after cluster formation.

\subsection{The final fate of the SMS}
\label{ssec:fate}
In practice, our model requires the conveyor belt mechanism to cease,
before the SMS releases He-burning products or supernova ejecta. At
this point we can only speculate why this should be the case.  One
possibility is that gas accretion into the cluster stops during the
H-burning phase. The gas that remains in the cluster accretes onto the
stars and once the cluster is gas-free, stellar winds will no longer
be bound by the cluster, but spread their material on a larger scale.

Strongly depending on metallicity and mass loss, black holes of up to
280~$\msun$ have been proposed in the literature for stars up to
350~$\msun$ \citep{2017MNRAS.470.4739S}.  It is hence not
inconceivable that an intermediate mass black hole (IMBH) of the order
of $10^3\,\msun$ might result as the remnant of SMS.  It is equally
well possible that the SMS is completely disrupted by a
pair-instability supernova after sufficient mass loss
\citep[e.g.][]{2008A&A...477..223Y}.

Although there is a lot of debate on whether IMBHs are present in GCs
(see e.g. \citealt*{2008ApJ...676.1008N};
\citealt{2011A&A...533A..36L} vs.  \citealt{2010ApJ...710.1032A,
  2013ApJ...769..107L}; \citealt*{2017MNRAS.468.4429Z};
\citealt{2018MNRAS.473.4832G}), an upper limit of $10^3\,\msun$ is
currently allowed by the $N$-body models of Milky Way GCs of
\citet{2017MNRAS.464.2174B}. However, we note that clusters form with
high densities in our model, such that a small black hole can grow to
IMBH masses via stellar collisions on a Gyr timescale
\citep{2015MNRAS.454.3150G}.  A better understanding of the remnant
masses in combination with improved constraints on IMBHs in GCs hence
have the potential to rule out the presence of SMSs in proto-GCs.

\subsection{Discreteness}
\label{ssec:discreteness}
In the majority of clusters the main sequence is unimodal, and
broadened, but the most massive GCs display two or more distinct main
sequences, which implies that there the He abundances are discrete
\citep[e.g.][and references therein]{2015AJ....149...91P}. In this
section we suggest two ideas that could lead to discrete He abundances
in the SMS-enrichment model.  An increase in the He abundance in more
massive GCs may result from the (relative) longevity of the
contraction phase in massive GCs, resulting in more massive SMSs. More
massive SMSs have shorter He production timescales
($\thelium\lesssim\,\myr$) and will therefore produce more He in a
fixed time. This, however, only explains why $\Delta Y$ increases with
GC mass, but not why it is bimodal. This may be understood by
considering the He production in the SMS as a function of time. As the
SMS grows, $Y$ goes up and $\mdotwind$ increases, because
$\mdotwind\propto \mu^2$ and because $\mdotwind\propto
\msms^{0.75-1.5}$ (see Section~\ref{ssec:mod:winds}). These two
effects lead to an increase of the He production rate and the He
release rate with time. It remains to be demonstrated whether this
material is also efficiently accreted on the low-mass stars, but it
provides a first step to understanding a bimodality.

The arguments above can not be used to understand the presence of more
than two main sequences. For this, we need to relax our assumption of
monolithic GC formation with a single SMS.  In reality, GC formation
is hierarchical (see
e.g. \citealt*{2007arXiv0712.0828K,2008ApJ...686.1174O};
\citealt{2012ApJ...754L..37S, 2017MNRAS.472.4982S}) and massive GCs
form from the merger of several smaller stellar clumps.  If the total
GC mass is large enough, it is possible that more than one of the
sub-clumps are massive enough to form their own SMS, each providing
material with different levels of enrichment (depending on the mass of
the clump), which then form a cluster with distinct levels of chemical
enrichment via dry mergers. Also, numerical models of collisional
runaway formation of SMSs have shown that a high primordial binary
fraction can result in more than one SMS \citep{2006ApJ...640L..39G}.

An alternative route to discreteness -- which may also operate in
conjunction with the subclustering -- is the slow-Jeans instability
that SMSs may undergo as the result of their high radiation pressure
\citep{2008ApJ...684..212T}.  This instability and subsequent
fragmentation develops on a Kelvin-Helmholtz timescale, $\tkh$.  To
see whether this instability may be important for the timescales we
consider here, we consider again our model from
Fig.~\ref{fig:msms_simple_rel} and now assume that this instability is
important for SMSs with masses in excess of $10^3\,\msun$.  We assume
that after $10\,\tkh$ the SMS fragments and leaves a core of
$100\,\msun$.  The factor of 10 is based on the fact that $\tkh$ is
about 10 times longer in the core compared to the surface (see
figure~3 in \citealt{2014MNRAS.437L..21D}).  After the fragmentation
the SMS can grow again via stellar collisions. In
Fig.~\ref{fig:msms_simple_tkh} we show the result of this model. The
SMS in the most massive cluster undergoes several `restarts', while in
the smaller cluster there is not enough time for the instability to
develop because the SMS does not exceed $10^3\,\msun$.  The pollution
of the intra-cluster medium halts until the SMS becomes massive enough
again to blow a strong wind, with higher helium abundance than before
the instability, thereby creating a distinct population. As long as
the cluster is dense enough to re-grow the SMS via collisions, this
process can repeat several times (see
Fig.~\ref{fig:msms_simple_tkh}). Because of our limited understanding
of this instability, this idea for realising discreteness is
speculative.

\begin{figure}
\includegraphics[width=8cm]{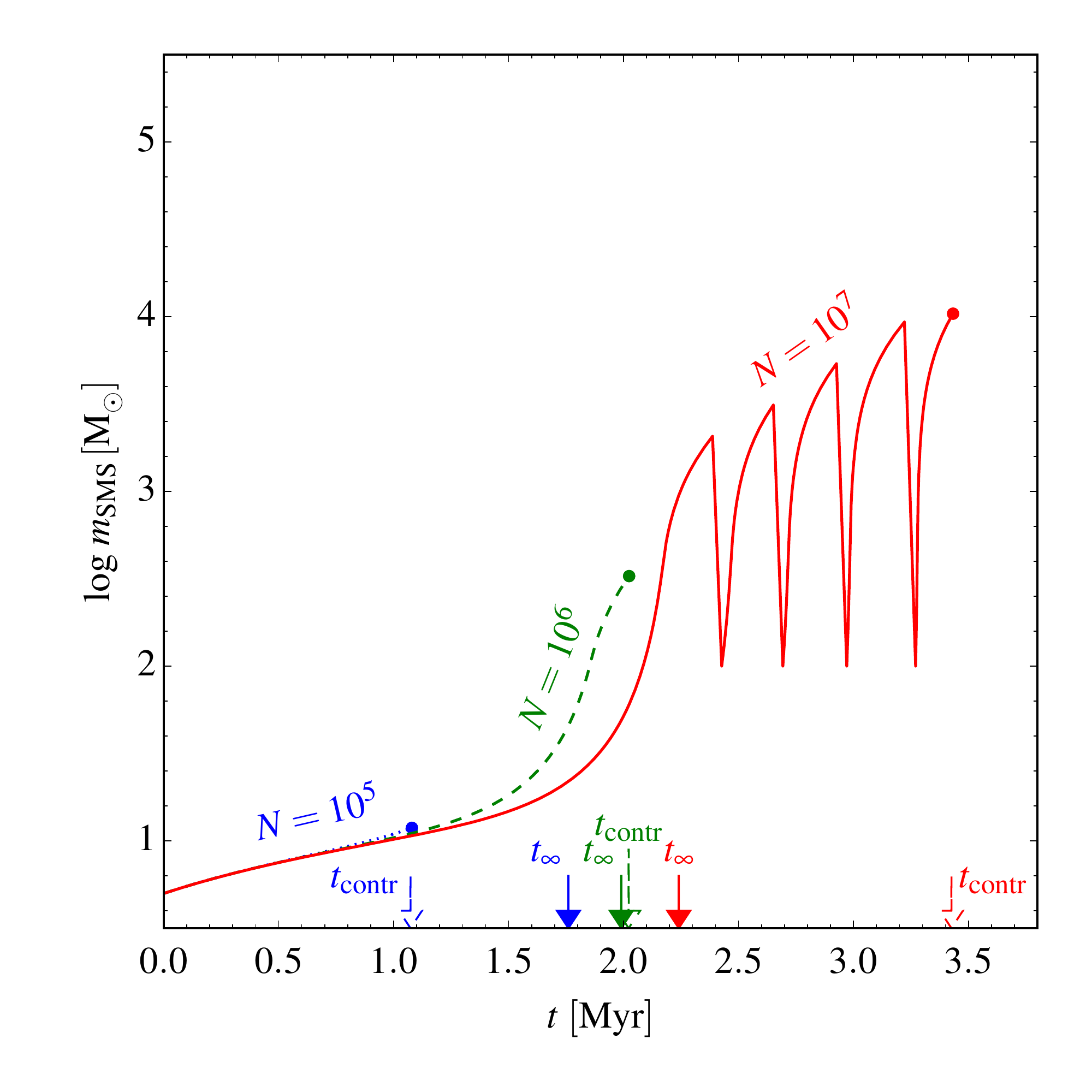}
\caption{As Fig.~\ref{fig:msms_simple_rel}, but now including the
  effect of the gravitational instability due to the high radiation
  pressure \citep{2008ApJ...684..212T}. For $\msms>10^3\,\msun$, we
  assume that the SMS fragments and leaves behind a core of
  $100\,\msun$ after $10\,\tkh$ have elapsed. The SMS in the most
  massive cluster ($N=10^7$) lives long enough to be affected by this,
  while the SMS in the smaller clusters never reaches the critical
  mass and evolutionary phase. }
\label{fig:msms_simple_tkh}
\end{figure}

We therefore prefer the stellar wind argument above that can explain
bi-modal He abundances and the stellar subclustering to create more
than two populations.  Combined with the fact that we expect the SMS
to be able to form sufficient amounts of He (see
Section~\ref{ssec:helium}), we conclude that the observed discreteness
of MSPs is not an obstacle to further progress and that the SMS
formation provides promising possibilities.

\subsection{Age dependence}
\label{ssec:age}
As mentioned in Section~\ref{sec:intro}, there are clusters with ages
as young as 2\,\gyr\ in the Magellanic Clouds displaying MSPs in the form of N spreads. There are no conditions
in our model that are unique to the early Universe, because the
threshold for multiple population formation is set by $N$ and
$\dot{M}$. The latter is likely to be higher in gas-rich environments,
making the early Universe conditions favourable for the formation of
SMSs in GCs, but our model does not restrict MSPs to
GCs that form in this epoch. There are no Galactic open clusters with
similar ages and masses to see whether the MSPs
feature is unique to the Magellanic Clouds, or in fact is a common
feature among star clusters with similar ages and masses. One possible
explanation for the presence of MSPs in relatively
young clusters in the Magellanic Clouds is that dwarf irregular
galaxies have long gas-consumption timescales, resulting in gas-rich
galactic environments at relatively low redshift.

 It is also important to keep in mind that a SMS is not required to
 produce N (nor Na) enrichment \citep[e.g.][and reference
   therein]{2017A&A...608A..28}. It may be possible that material from
 fast-rotating massive stars \citep{2007A&A...464.1029D,2013A&A...552A.121K} or very massive stars \citep[$10^2-10^3\,\msun$,][]{2018arXiv180308042V}, enriched in N and Na, cools when it interacts
 with cold, pristine gas and accretes on the low-mass stars, in a
 low-density cluster. More details on the mass budget of the N
 enrichment are required to assess this idea.

\subsection{Cluster structure}
One of the implications of this GC formation model is that GCs with
MSPs form with very high densities and with a nearly constant $\rh$,
or massive clusters being even slightly smaller than low-mass clusters
after a few Myr (see Fig.~\ref{fig:model}). An inverted $M-\rh$
relation, or nearly constant $\rh$ is indeed found for young massive
clusters \citep[e.g.][]{2006A&A...448.1031K,2010ARA&A..48..431P}.  Old
GCs have larger $\rh$ than predicted by our model, but a direct
comparison can not be made because the present-day masses and radii
are strongly affected by dynamical evolution. In fact, the properties
of nearly all Milky Way GCs are affected by relaxation driven
evolution \citep{H61, 2011MNRAS.413.2509G}, which means that their
present-day $\rh$ are almost independent of their initial $\rh$. We
can therefore only conclude that the present-day $\rh$ of GCs does not
exclude high initial densities.  From models of the radius
distribution of GCs, \citet{2013MNRAS.432L...1A} conclude that high
initial densities are preferred. Also, support for very high initial
densities of GCs comes from dynamical Monte Carlo models of individual
GCs \citep[e.g.][]{2009MNRAS.395.1173G, 2011MNRAS.410.2698G} and the
Milky Way GC population \citep{2017MNRAS.464L..36A}.  Clusters with
$\trh$ much longer than their ages have most likely not formed very
dense. The low-density, intermediate age SMC cluster Lindsay~1 has
MSPs. \citet{2011AJ....142...36G} show that $\trh\simeq8\,\gyr$,
i.e. very similar to its age.  If this cluster has not undergone any
processes that have inflated its radius in addition to two-body
relaxation (i.e. cluster mergers, tidal interaction, etc.), it is
difficult to understand the MSPs of this cluster with our model.
However, we recall that a SMS is not required to produce N, and it may
be that winds of O-stars or very massive stars were trapped in
low-density clusters (see Section~\ref{ssec:age} and
\citealt{2018arXiv180308042V}).
 
\subsection{SMS radius}
\label{ssec:smsradius}
In Section~\ref{subsec:globalSMSproperties} we show that the amount of
polluted material that is released is sensitive to the uncertain
mass-radius relation of SMSs.  The radius of the SMS, $\rsms$, is
controlled by the index $\delta$ in the mass-radius relation
(equation~\ref{eq:mr}). For $\msms=10^4\,\msun$ we obtain a radius of
$\rsms\simeq300\,\rsun(3000\,\rsun)$ for $\delta=0.5(1)$, which is
larger than the radii of the SMS in the models of
\citet{2014MNRAS.437L..21D} (few$\times10^2\,\rsun$, Pavel
Denissenkov, private communication), but the exact structure of the
SMS is sensitive to many of the assumptions that have to be made and
models with larger $\rsms$ exist. For example, the models of SMSs with
masses up to $10^3\,\msun$ and solar abundance of
\citet{2008A&A...477..223Y} show a nearly linear mass-radius relation
(i.e. $\delta=1$), which would have $\rsms\simeq10^3\,\rsun$ when
extrapolating their mass-radius relation to
$\msms=10^4\,\msun$. \citet{1999PASJ...51..417I} find even larger
radii for stars of solar abundance and $10^3\,\msun$. These large
radii may be the result of the high metallicity, and more metal-poor
stars are likely smaller.  
 We note that the radii we explored here are all below the maximum radius that is set by the Hayashi limit, which is defined by $\Teff \simeq 3000\,$K and is due to the steep drop in H$^{-}$ opacity with temperature. For the luminosity of equation~(\ref{eq:lsms}), we find that this maximum radius is $\rsms \simeq6.2\times10^3\,\rsun\,(\msms/100\,\msun)^{1/2}$. 
The amount of polluted material produced is sensitive to $\rsms$, therefore it will be critical to improve in the near future our
understanding of the structure of SMSs up to $\sim10^5\,\msun$,
including rotation, general relativistic effects and disequilibrium
evolution as the result of collisions.

\subsection{Disruptive collisions}
\label{ssec:collisions}
In our model we have not included the effect of mass loss following a
collision. In SPH models of massive star collisions,
\citet{2008MNRAS.383L...5G} find that less than 10\% of the mass is
ejected from the collision product, and even lower for higher mass
ratios. However, head-on collisions, with high relative velocity,
could be disruptive for both stars \citep{2005MNRAS.358.1133F}.
Because the SMS has a large radius, even grazing collisions could be
disruptive and shed parts of the envelope. This may reduce the growth
of $\msms$, and liberate additional material from the SMS to pollute
the protostars.

Collisions are also more disruptive if gravitational focusing is
unimportant.  However, for our adopted mass-radius relation, the
escape velocity from the surface of the SMS is
$\vesc\simeq1100\,\kms(\msms/100\,\msun)^{(1-\delta)/2}$, such that
for $\delta\le1$, the escape velocity is always larger than
$1100\,\kms$ and we are always in the regime that gravitational
focusing is important and little mass is lost in collisions
\citep{2005MNRAS.358.1133F}.

\subsection{Observational predictions}
\label{ssec:obspred}
Here we provide several predictions that can be used to test the SMS
formation model and the associated GC self-enrichment scenario
presented in this paper.
\begin{itemize}
\item {\bf Star formation at high redshift:} Depending on the
  mass-radius relation (i.e. the value of $\delta$), the outer parts
  of the SMS could be relatively cold.  For $\delta=0.5$ we estimate
  $\Teff=43\,$kK, and for $\delta=1$ we estimate
  $7.6\lesssim\Teff/{\rm kK}\lesssim13.6$, for
  $10^5\gtrsim\msms/\msun\gtrsim10^4$ (note that more massive $\msms$
  implies cooler $\Teff$ for $\delta>0.5$). Therefore, the SMS could
  be almost as cool as red supergiants, which normally appear in
  stellar populations of 10-15\,Myr. Combined with the strong wind
  mass-loss, we expect to see P Cygni profiles or emission lines in a
  spectrum with relatively low temperatures (e.g. H$\alpha$ or CaII in
  emission depending on the effective temperature). The CaII features
  could be comparable to those exhibited by the coolest LBVs or by the
  peculiar SN~2002bu and SN~2010dn in their late and coolest phases
  \citep[e.g.][]{2011MNRAS.415..773S}.  Because of the high luminosity
  of the SMS \citep[$10^{9-10}\,\lsun$,][]{2014MNRAS.437L..21D}, these
  spectroscopic features may be observable in star forming regions at
  a redshift of $z\gtrsim2$ ($V\lesssim26\,$mag for $10^{10}\,\lsun$)
  with future 30-metre class telescopes and the {\it James Webb Space
    Telescope (JWST)}.  With existing facilities these features could
  be looked for in gravitationally lensed star forming galaxies
  \citep[e.g.][]{2015ApJ...808..139S}, where individual clumps can be
  resolved \citep[e.g.][]{2017MNRAS.467.4304V}. These spectroscopic
  features in the optical should be visible in combination with signs
  of inflow of cold molecular gas, potentially observable with {\it
    ALMA}.  The spectroscopic feature of the cool SMS will be super
  imposed on stellar populations features from very young stellar
  populations ($\lesssim3\,\myr$), such as hot O-stars in excess of
  $100\,\msun$ that may display He II $\lambda 1640$ in emission.
  Strong He II $\lambda 1640$ emission lines were found in
  gravitationally lensed star forming regions at redshifts $z\simeq
  6-7$ \citep[e.g.][]{2015ApJ...808..139S}, which have also attributed
  to Pop III stars \citep{2003A&A...397..527S}. In the resolved
  starburst cluster R136 in 30 Doradus, this emission line is entirely
  produced by the seven stars with masses $\gtrsim100\,\msun$
  \citep{2016MNRAS.458..624C}, which have strong, but slow winds
  because of their near Eddington luminosity, resulting in narrow
  emission lines \citep{2015A&A...578L...2G}. A discussion on the
  different contributions to He II, and other lines, is presented in
  \citet{2017MNRAS.472.2608S}. Unfortunately, the high gas densities
  may obscure the SMS and stellar populations for a significant
  fraction of their life, making it impossible to make solid
  predictions for the number of SMSs that should be observable.
\item {\bf GC kinematics:} Because the SMS grows via stellar
  collisions, the angular momentum of the star builds up in a random
  walk process, and the SMS will therefore have a random spin
  direction with respect to the angular momentum of the pristine
  population, and a rotation velocity of order $\sim100\,\kms$.
  Because of angular momentum conservation, this rotational velocity
  becomes negligible once the wind reaches $\sim1\,$pc. Assuming that
  the stars that accrete more processed material than their seed mass
  inherit the angular momentum of the SMS wind, we expect the stars
  with most extreme abundances to have low streaming motions in the
  cluster, and could be counter-rotating or co-rotating with the
  pristine population. Also, the spin axes of the pristine population
  and (extreme) polluted populations do not need to be aligned.  The
  prediction for the low orbital spin for the polluted stars is
  opposite to that of the MGMs, because there the polluted population
  forms out of material from a first population, and when the material
  cools to form new stars, the rotational velocity goes up because of
  angular momentum conservation. As a result, on the MGMs the angular
  momentum vectors are aligned.  In the fast rotating massive star
  scenario of \citet{2013A&A...552A.121K}, the second generation of
  stars is formed as companions in the decretion discs of massive
  first generation stars and hence share their kinematics.  In the
  early disc accretion model of \citet{2013MNRAS.436.2398B} the
  polluted population is expected to rotate slower
  \citep{2015MNRAS.450.1164H}, but have aligned spin axes.  In M13,
  \citet{2017MNRAS.465.3515C} find that the polluted population
  rotates faster, and the relative angle between the spin axes is
  between 0-45 degrees.  Contrary to M13, in $\omega$ Cen the polluted
  population rotates slower than the pristine population
  \citep{2018ApJ...853...86B}. This cluster-to-cluster variation of
  the relative rotation speeds of the two populations is expected in
  our model, but further studies of differential rotation of multiple
  population in GCs are needed to shed light on the magnitude and
  orientation of the orbital angular momentum vectors of the different
  populations.  Several studies have found that the polluted stars
  have more radially anisotropic orbits (see
  \citealt{2013ApJ...771L..15R} for 47 Tuc,
  \citealt{2015ApJ...810L..13B} for NGC\,2808 and
  \citealt{2018ApJ...853...86B} for $\omega$ Cen). In our model we
  expect this to be the case, because the polluted stars are initially
  more centrally concentrated, and during their evolution they are
  expected to be scattered to wider, radial orbits
  \citep{2015MNRAS.450.1164H}. However, this is not a unique kinematic
  signature, because in every other scenario the polluted stars are
  also initially more centrally concentrated.
\item {\bf [Fe/H] dependence:} At lower $\feh$ the SMS is hotter (for
  a given $\msms$). This predicts that the extent of the Mg--Al
  anti-correlation is larger at lower $\feh$.  Because we also find a
  strong (super-linear) correlation between $\msms$ and cluster mass,
  we expect preferentially the massive and metal-poor GCs to contain
  more pronounced Mg--Al anti-correlations.  Indications for such a
  dependence of the Mg--Al anti-correlation on both $M$ and $\feh$
  were found in the Galactic GCs by {\bf \citet{2009A&A...505..139C}
    and} \citet{2017A&A...601A.112P}.  A $M$ and [Fe/H] dependence was
  also found for the slope of the O--Na anti-correlation
  \citep{2009A&A...505..117C}, in the sense that a shallower slope
  (i.e. a lower minimum O abundance and maximum Na abundance) was
  found for more massive, metal-poor GCs. A shallower slope of the
  O--Na anti-correlation is what is expected from the yields of more
  massive SMSs \citep[see figure~1 of][]{2015MNRAS.448.3314D} and the
  results of \citet{2009A&A...505..117C} therefore supports the fact
  that massive, metal-poor GCs had more massive SMSs.  However, we
  caution that there are other metallicity effects: the wind mass-loss
  rates increase with $\feh$, which may imply higher $\msms$ at lower
  $\feh$ and work in the same direction as the temperature
  dependence. But this also predicts a higher fraction of polluted
  material in more metal-rich GCs, while \citet{2017MNRAS.464.3636M}
  find no such correlation.  However, we note that metal-rich stars
  are larger, and a larger cross section leads to a higher collision
  rate, working in the opposite direction (more massive SMS at higher
  $\feh$ and therefore higher $\mdotwind$). The sensitivity to our
  poorly-constrained model parameters $\delta$ and $\eta$ (see
  Fig.~\ref{fig:model}), which may both depend on $\feh$, complicates
  the discussion on $\feh$ dependence.
\item {\bf GCs without MSPs:} All clusters with MSPs should have $\trh\lesssim\,{\rm Age}$. It
  would therefore be interesting to look for signatures of MSPs, in
  the extended clusters in M31 \citep[e.g.][]{2005MNRAS.360.1007H} or
  some of the low-density outer-halo clusters in the Milky Way, such
  as Crater or the Palomar clusters with $\trh>\,{\rm Age}$. It would
  in particular be interesting to look for the Mg--Al anti-correlation
  in these clusters, because this can not be produced with ordinary
  O-stars and therefore a SMS is required and it is not expected to
  form in these clusters in our model.
\item {\bf Young massive clusters:} Finally, it may be worthwhile to look for low-redshift analogues
  of this GC formation model. Signatures of high gas inflow rates in
  star forming regions in nearby star burst galaxies have been
  reported \citep{2015Natur.519..331T, 2017ApJ...849L...1O}. If the
  mass accretion rates are high enough, these regions may harbour an
  obscured SMS.
\end{itemize}

\section{Conclusions}
\label{sec:conclusions}
We present a model for the concurrent formation of SMSs and GCs, and
use this to explain the abundance anomalies that are observed in old,
Fe-normal GCs, and intermediate aged massive star clusters
\citep[$\gtrsim2\,\gyr$,][]{2018MNRAS.473.2688M}. In our model, the
SMS forms via stellar collisions, which are triggered by a contraction
of the proto-GC following gas accretion onto its member protostars.  A
formation mechanism in which dense, massive clusters result from gas
inflow and hierarchical cluster assembly is supported by both
hydrodynamical simulations of globally collapsing molecular clouds
\citep{2017MNRAS.467.1313V} and observations
\citep[e.g.][]{2014prpl.conf..291L, 2016MNRAS.457.4536W}\footnote{We
  note that these authors refer to this formation mechanism as the
  `conveyor-belt mode', which in their work refers to continuous gas
  inflow during cluster formation. In our model, this gas inflow is a
  requirement to activate the conveyor-belt production of hot-hydrogen
  burning products from the SMS.}.  The formation of SMSs via stellar
collisions has been addressed with numerical simulations
(\citealt{2002ApJ...576..899P, 2004Natur.428..724P,
  2006MNRAS.368..141F}; \citealt*{2015MNRAS.451.2352K};
\citealt{2016MNRAS.459.3432M, 2017MNRAS.472.1677S}), which all
conclude that the rate of growth of the SMS is sensitive to the
adopted initial density of the cluster. In our model, $\msms$ is
insensitive to the initial  cluster density, because the cluster
density increases as the result of gas accretion, until two-body
relaxation becomes important and reduces the density.  The resulting
$\msms$ is therefore only a function of the total number of stars $N$
and the gas accretion rate $\dot{M}$, in the sense that massive
clusters ($N\gtrsim10^6$) experiencing a high accretion rate
($\dot{M}\gtrsim 10^5\,\msun/\myr$) are able form a SMS before
relaxation becomes important.

Because SMSs are convective objects, we argue that the fuelling of
pristine material via collisions allows the stars to remain close to
their early-main sequence configuration (i.e. the central temperature
remains in the 70 - 80\,MK range and the central He content does not
increase, as it would in the absence of collisions). Therefore, we assume that
all along the accretion phase the He content of the ejecta remains at
low values and that the CNONaMgAl patterns are preserved and similar
to the values given by the current models at that phase until the
stars succumb to their winds or to the Jeans instability.  The
corresponding yields of SMSs show excellent agreement with the
abundances of anomalous stars in GCs once dilution with pristine gas
is accounted for \citep{2014MNRAS.437L..21D}.  The relation we obtain
between the mass of the cluster and the maximum mass of the SMS,
together with the dependence between the central temperature of the
SMS and its mass and metallicity is well supported by the indications
of a bivariate relation of the Mg depletion with the Galactic GCs
masses and [Fe/H].

Our model provides a scenario for the formation of a SMS, but also for
the pollution of the low-mass protostars in the cluster: the SMS wind
interacts with the inflowing cold gas, subsequently cools and accretes
onto the protostars in the cluster. In our model we are able to
overcome the so-called mass budget problem, since the accumulated mass
in SMS winds can supersede the maximum mass of the SMS itself by more
than an order of magnitude, because it is continuously rejuvenated
with fresh hydrogen by stellar collisions.  This avoids the need for
cluster birth masses that are more than an order of magnitude larger
than the present day GC masses (\citealt{2008MNRAS.391..825D};
\citealt{2011MNRAS.413.2297S}; \citealt{2012ApJ...758...21C}), which
is at tension with UV-luminosity functions of high redshift star
forming regions \citep{mbk17} and the (low) number of field stars in
dwarf galaxies with GCs
\citep{2012A&A...544L..14L,2014A&A...565A..98L}.  More importantly,
our model predicts a super-linear relation between the amount of
processed material and GC mass, providing an explanation for the
observed increase of the fraction of polluted stars and helium with GC
mass \citep{2014ApJ...785...21M,2017MNRAS.464.3636M}.

In this study we focussed on Fe-simple GCs, and provide a model that
gives rise to CNONaMgAl and He variations in a single cluster
formation event. It has been shown that each Fe sub-populations in
Fe-complex GCs displays CNONaMgAl variations (see
\citealt{2010A&A...520A..95C} for M54 and
\citealt{2011ApJ...731...64M} for $\omega$ Cen). A straight-forward
explanation for this is that each (unrelated) star formation event in
the nucleus of a galaxy results in the formation of a SMS, producing
the light-element variations for that sub-population. This idea needs
to be scrutinized in future models.

Apart from the observational tests we propose in
Section~\ref{ssec:obspred}, another important next step is to validate
the scaling relations proposed in this work with numerical
simulations. \citet{2017MNRAS.467.3775P} present results of
collisional $N$-body simulations in which very massive stars form and
rejuvenate via stellar collisions. To test our model, the
hydrodynamical effect of gas accretion and stellar wind interaction
needs to be combined with such collisional $N$-body simulations. This
will allow to increase our understanding of the formation and
evolution of the SMS and its host GC and the pollution scenario with
the various scaling relations proposed here.

\section*{Acknowledgements}
We thank Alvio Renzini and an anonymous referee for reports
that helped to improve the paper.  We are grateful to Douglas Heggie,
Abbas Askar and Mirek Giersz for insightful feedback on an earlier
version of the paper and discussions on binary heating and IMBH
formation.  We thank Rob Izzard and Pavel Denissenkov for discussions
on the structure and temperature of SMSs, Daniel Schaerer for an
insight on the spectroscopic features of SMSs, and Nikos Prantzos for
discussions on nucleosynthesis.   We thank Jorick Vink for discussions on stellar winds. MG acknowledges financial support
from the Royal Society (University Research Fellowship) and the
European Research Council (ERC StG-335936, CLUSTERS). VHB acknowledges
support from the NRC-Canada Plaskett Fellowship and from the Radboud
Excellence Initiative.  CC acknowledges support from the Swiss
National Science Foundation (SNF) for the Projects 200020-159543
`Multiple stellar populations in massive star clusters - Formation,
evolution, dynamics, impact on galactic evolution" and 200020-169125
`Globular cluster archeology'. OA acknowledges support from the
Swedish Research Council (grant 2014-5791) and the Knut and Alice
Wallenberg Foundation. NB thanks the ERC for support (CoG-646928).  We
thank the International Space Science Institute (ISSI, Bern, CH) for
welcoming the activities of the Team 271 "Massive Star Clusters Across
the Hubble Time" (2013 - 2016; team leader CC). Most of the processing
of the results has been done using the {\sc python} programming
language and the following open source modules: {\sc
  numpy}\footnote{http://www.numpy.org}, {\sc
  scipy}\footnote{http://www.scipy.org}, {\sc
  matplotlib}\footnote{http://matplotlib.sourceforge.net}.


\bibliographystyle{mn2e}

\begin{thebibliography}{221}
\expandafter\ifx\csname natexlab\endcsname\relax\def\natexlab#1{#1}\fi

\bibitem[{{Abel} {et~al}\mbox{.}(2002){Abel}, {Bryan}, \&
  {Norman}}]{2002Sci...295...93A}
{Abel} T., {Bryan} G.~L., {Norman} M.~L., 2002, Science, 295, 93

\bibitem[{{Alexander} \& {Gieles}(2012)}]{2012MNRAS.422.3415A}
{Alexander} P.~E.~R., {Gieles} M., 2012, \mnras, 422, 3415

\bibitem[{{Alexander} \& {Gieles}(2013)}]{2013MNRAS.432L...1A}
{Alexander} P.~E.~R., {Gieles} M., 2013, \mnras, 432, L1

\bibitem[{{Alexander} {et~al}\mbox{.}(2014){Alexander}, {Gieles}, {Lamers}, \&
  {Baumgardt}}]{2014MNRAS.442.1265A}
{Alexander} P.~E.~R., {Gieles} M., {Lamers} H.~J.~G.~L.~M., {Baumgardt} H.,
  2014, \mnras, 442, 1265

\bibitem[{{Amard} {et~al}\mbox{.}(2016){Amard}, {Palacios}, {Charbonnel},
  {Gallet}, \& {Bouvier}}]{2016A&A...587A.105A}
{Amard} L., {Palacios} A., {Charbonnel} C., {Gallet} F., {Bouvier} J., 2016,
  \aap, 587, A105

\bibitem[{{Anderson}(2002)}]{2002ASPC..265...87A}
{Anderson} J., 2002, in Astronomical Society of the Pacific Conference Series,
  Vol. 265, Omega Centauri, A Unique Window into Astrophysics, {van Leeuwen}
  F., {Hughes} J.~D., {Piotto} G., eds., p.~87

\bibitem[{{Anderson} \& {van der Marel}(2010)}]{2010ApJ...710.1032A}
{Anderson} J., {van der Marel} R.~P., 2010, \apj, 710, 1032

\bibitem[{{Askar} {et~al}\mbox{.}(2017){Askar}, {Szkudlarek},
  {Gondek-Rosi{\'n}ska}, {Giersz}, \& {Bulik}}]{2017MNRAS.464L..36A}
{Askar} A., {Szkudlarek} M., {Gondek-Rosi{\'n}ska} D., {Giersz} M., {Bulik} T.,
  2017, \mnras, 464, L36

\bibitem[{{Bahcall} \& {Wolf}(1976)}]{1976ApJ...209..214B}
{Bahcall} J.~N., {Wolf} R.~A., 1976, \apj, 209, 214

\bibitem[{{Ballesteros-Paredes} {et~al}\mbox{.}(2015){Ballesteros-Paredes},
  {Hartmann}, {P{\'e}rez-Goytia}, \& {Kuznetsova}}]{2015MNRAS.452..566B}
{Ballesteros-Paredes} J., {Hartmann} L.~W., {P{\'e}rez-Goytia} N., {Kuznetsova}
  A., 2015, \mnras, 452, 566

\bibitem[{{Baraffe} {et~al}\mbox{.}(2002){Baraffe}, {Chabrier}, {Allard}, \&
  {Hauschildt}}]{2002A&A...382..563B}
{Baraffe} I., {Chabrier} G., {Allard} F., {Hauschildt} P.~H., 2002, \aap, 382,
  563

\bibitem[{{Bastian}(2017)}]{2017IAUS..316..302B}
{Bastian} N., 2017, in IAU Symposium, Vol. 316, Formation, Evolution, and
  Survival of Massive Star Clusters, {Charbonnel} C., {Nota} A., eds., pp.
  302--309

\bibitem[{{Bastian} {et~al}\mbox{.}(2015){Bastian}, {Cabrera-Ziri}, \&
  {Salaris}}]{2015MNRAS.449.3333B}
{Bastian} N., {Cabrera-Ziri} I., {Salaris} M., 2015, \mnras, 449, 3333

\bibitem[{{Bastian} {et~al}\mbox{.}(2013){Bastian}, {Lamers}, {de Mink},
  {Longmore}, {Goodwin}, \& {Gieles}}]{2013MNRAS.436.2398B}
{Bastian} N., {Lamers} H.~J.~G.~L.~M., {de Mink} S.~E., {Longmore} S.~N.,
  {Goodwin} S.~P., {Gieles} M., 2013, \mnras, 436, 2398

\bibitem[{{Bastian} \& {Lardo}(2015)}]{2015MNRAS.453..357B}
{Bastian} N., {Lardo} C., 2015, \mnras, 453, 357

\bibitem[{{Bastian} \& {Lardo}(2018)}]{BL18}
{Bastian} N., {Lardo} C., 2018, ARAA, arXiv:1712.01286

\bibitem[{{Baumgardt}(2017)}]{2017MNRAS.464.2174B}
{Baumgardt} H., 2017, \mnras, 464, 2174

\bibitem[{{Baumgardt} {et~al}\mbox{.}(2004{\natexlab{a}}){Baumgardt}, {Makino},
  \& {Ebisuzaki}}]{2004ApJ...613.1133B}
{Baumgardt} H., {Makino} J., {Ebisuzaki} T., 2004{\natexlab{a}}, \apj, 613,
  1133

\bibitem[{{Baumgardt} {et~al}\mbox{.}(2004{\natexlab{b}}){Baumgardt}, {Makino},
  \& {Ebisuzaki}}]{2004ApJ...613.1143B}
{Baumgardt} H., {Makino} J., {Ebisuzaki} T., 2004{\natexlab{b}}, \apj, 613,
  1143

\bibitem[{{Bedin} {et~al}\mbox{.}(2004){Bedin}, {Piotto}, {Anderson},
  {Cassisi}, {King}, {Momany}, \& {Carraro}}]{2004ApJ...605L.125B}
{Bedin} L.~R., {Piotto} G., {Anderson} J., {Cassisi} S., {King} I.~R., {Momany}
  Y., {Carraro} G., 2004, \apjl, 605, L125

\bibitem[{{Bellini} {et~al}\mbox{.}(2018){Bellini}, {Libralato}, {Bedin},
  {Milone}, {van der Marel}, {Anderson}, {Apai}, {Burgasser}, {Marino}, \&
  {Rees}}]{2018ApJ...853...86B}
{Bellini} A. {et~al.}, 2018, \apj, 853, 86

\bibitem[{{Bellini} {et~al}\mbox{.}(2015){Bellini}, {Vesperini}, {Piotto},
  {Milone}, {Hong}, {Anderson}, {van der Marel}, {Bedin}, {Cassisi},
  {D'Antona}, {Marino}, \& {Renzini}}]{2015ApJ...810L..13B}
{Bellini} A. {et~al.}, 2015, \apjl, 810, L13

\bibitem[{{Binney} \& {Tremaine}(2008)}]{2008gady.book.....B}
{Binney} J., {Tremaine} S., 2008, {Galactic Dynamics: Second Edition}.
  Princeton University Press

\bibitem[{{Blumenthal} {et~al}\mbox{.}(1986){Blumenthal}, {Faber}, {Flores}, \&
  {Primack}}]{1986ApJ...301...27B}
{Blumenthal} G.~R., {Faber} S.~M., {Flores} R., {Primack} J.~R., 1986, \apj,
  301, 27

\bibitem[{{Bonifacio} {et~al}\mbox{.}(2007){Bonifacio}, {Pasquini}, {Molaro},
  {Carretta}, {Fran{\c c}ois}, {Gratton}, {James}, {Sbordone}, {Spite}, \&
  {Zoccali}}]{2007A&A...470..153B}
{Bonifacio} P. {et~al.}, 2007, \aap, 470, 153

\bibitem[{{Bonnell} {et~al}\mbox{.}(1998){Bonnell}, {Bate}, \&
  {Zinnecker}}]{1998MNRAS.298...93B}
{Bonnell} I.~A., {Bate} M.~R., {Zinnecker} H., 1998, \mnras, 298, 93

\bibitem[{{Boylan-Kolchin}(2017)}]{mbk17}
{Boylan-Kolchin} M., 2017, arXiv:1711.00009

\bibitem[{{Breen} \& {Heggie}(2012)}]{2012MNRAS.425.2493B}
{Breen} P.~G., {Heggie} D.~C., 2012, \mnras, 425, 2493

\bibitem[{{Breen} \& {Heggie}(2013)}]{2013MNRAS.432.2779B}
{Breen} P.~G., {Heggie} D.~C., 2013, \mnras, 432, 2779

\bibitem[{{Cabrera-Ziri} {et~al}\mbox{.}(2015){Cabrera-Ziri}, {Bastian},
  {Longmore}, {Brogan}, {Hollyhead}, {Larsen}, {Whitmore}, {Johnson},
  {Chandar}, {Henshaw}, {Davies}, \& {Hibbard}}]{2015MNRAS.448.2224C}
{Cabrera-Ziri} I. {et~al.}, 2015, \mnras, 448, 2224

\bibitem[{{Cabrera-Ziri} {et~al}\mbox{.}(2016){Cabrera-Ziri}, {Lardo},
  {Davies}, {Bastian}, {Beccari}, {Larsen}, \&
  {Hernandez}}]{2016MNRAS.460.1869C}
{Cabrera-Ziri} I., {Lardo} C., {Davies} B., {Bastian} N., {Beccari} G.,
  {Larsen} S.~S., {Hernandez} S., 2016, \mnras, 460, 1869

\bibitem[{{Carretta} {et~al}\mbox{.}(2009{\natexlab{a}}){Carretta},
  {Bragaglia}, {Gratton}, \& {Lucatello}}]{2009A&A...505..139C}
{Carretta} E., {Bragaglia} A., {Gratton} R., {Lucatello} S.,
  2009{\natexlab{a}}, \aap, 505, 139

\bibitem[{{Carretta} {et~al}\mbox{.}(2010{\natexlab{a}}){Carretta},
  {Bragaglia}, {Gratton}, {Lucatello}, {Bellazzini}, {Catanzaro}, {Leone},
  {Momany}, {Piotto}, \& {D'Orazi}}]{2010A&A...520A..95C}
{Carretta} E. {et~al.}, 2010{\natexlab{a}}, \aap, 520, A95

\bibitem[{{Carretta} {et~al}\mbox{.}(2009{\natexlab{b}}){Carretta},
  {Bragaglia}, {Gratton}, {Lucatello}, {Catanzaro}, {Leone}, {Bellazzini},
  {Claudi}, {D'Orazi}, {Momany}, {Ortolani}, {Pancino}, {Piotto},
  {Recio-Blanco}, \& {Sabbi}}]{2009A&A...505..117C}
{Carretta} E. {et~al.}, 2009{\natexlab{b}}, \aap, 505, 117

\bibitem[{{Carretta} {et~al}\mbox{.}(2010{\natexlab{b}}){Carretta},
  {Bragaglia}, {Gratton}, {Recio-Blanco}, {Lucatello}, {D'Orazi}, \&
  {Cassisi}}]{2010A&A...516A..55C}
{Carretta} E., {Bragaglia} A., {Gratton} R.~G., {Recio-Blanco} A., {Lucatello}
  S., {D'Orazi} V., {Cassisi} S., 2010{\natexlab{b}}, \aap, 516, A55

\bibitem[{{Charbonnel}(2006)}]{2006Natur.442..636C}
{Charbonnel} C., 2006, \nat, 442, 636

\bibitem[{{Charbonnel}(2016)}]{2016EAS....80..177C}
{Charbonnel} C., 2016, in EAS Publications Series, Vol.~80, EAS Publications
  Series, {Moraux} E., {Lebreton} Y., {Charbonnel} C., eds., pp. 177--226

\bibitem[{{Charbonnel} {et~al}\mbox{.}(2014){Charbonnel}, {Chantereau},
  {Krause}, {Primas}, \& {Wang}}]{2014A&A...569L...6C}
{Charbonnel} C., {Chantereau} W., {Krause} M., {Primas} F., {Wang} Y., 2014,
  \aap, 569, L6

\bibitem[{{Charbonnel} \& {Primas}(2005)}]{2005A&A...442..961C}
{Charbonnel} C., {Primas} F., 2005, \aap, 442, 961

\bibitem[{{Clarke} \& {Bonnell}(2008)}]{2008MNRAS.388.1171C}
{Clarke} C.~J., {Bonnell} I.~A., 2008, \mnras, 388, 1171

\bibitem[{{Cohn}(1980)}]{1980ApJ...242..765C}
{Cohn} H., 1980, \apj, 242, 765

\bibitem[{{Conroy}(2012)}]{2012ApJ...758...21C}
{Conroy} C., 2012, \apj, 758, 21

\bibitem[{{Cordero} {et~al}\mbox{.}(2017){Cordero}, {H{\'e}nault-Brunet},
  {Pilachowski}, {Balbinot}, {Johnson}, \& {Varri}}]{2017MNRAS.465.3515C}
{Cordero} M.~J., {H{\'e}nault-Brunet} V., {Pilachowski} C.~A., {Balbinot} E.,
  {Johnson} C.~I., {Varri} A.~L., 2017, \mnras, 465, 3515

\bibitem[{{Crowther} {et~al}\mbox{.}(2016){Crowther}, {Caballero-Nieves},
  {Bostroem}, {Ma{\'{\i}}z Apell{\'a}niz}, {Schneider}, {Walborn}, {Angus},
  {Brott}, {Bonanos}, {de Koter}, {de Mink}, {Evans}, {Gr{\"a}fener},
  {Herrero}, {Howarth}, {Langer}, {Lennon}, {Puls}, {Sana}, \&
  {Vink}}]{2016MNRAS.458..624C}
{Crowther} P.~A. {et~al.}, 2016, \mnras, 458, 624

\bibitem[{{Crowther} {et~al}\mbox{.}(2010){Crowther}, {Schnurr}, {Hirschi},
  {Yusof}, {Parker}, {Goodwin}, \& {Kassim}}]{2010MNRAS.408..731C}
{Crowther} P.~A., {Schnurr} O., {Hirschi} R., {Yusof} N., {Parker} R.~J.,
  {Goodwin} S.~P., {Kassim} H.~A., 2010, \mnras, 408, 731

\bibitem[{{Cyburt} {et~al}\mbox{.}(2016){Cyburt}, {Fields}, {Olive}, \&
  {Yeh}}]{2016RvMP...88a5004C}
{Cyburt} R.~H., {Fields} B.~D., {Olive} K.~A., {Yeh} T.-H., 2016, Reviews of
  Modern Physics, 88, 015004

\bibitem[{{Dale}(2017)}]{2017MNRAS.467.1067D}
{Dale} J.~E., 2017, \mnras, 467, 1067

\bibitem[{{D'Antona} {et~al}\mbox{.}(2005){D'Antona}, {Bellazzini}, {Caloi},
  {Pecci}, {Galleti}, \& {Rood}}]{2005ApJ...631..868D}
{D'Antona} F., {Bellazzini} M., {Caloi} V., {Pecci} F.~F., {Galleti} S., {Rood}
  R.~T., 2005, \apj, 631, 868

\bibitem[{{D'Antona} \& {Mazzitelli}(1994)}]{1994ApJS...90..467D}
{D'Antona} F., {Mazzitelli} I., 1994, \apjs, 90, 467

\bibitem[{{D'Antona} {et~al}\mbox{.}(2014){D'Antona}, {Ventura}, {Decressin},
  {Vesperini}, \& {D'Ercole}}]{2014MNRAS.443.3302D}
{D'Antona} F., {Ventura} P., {Decressin} T., {Vesperini} E., {D'Ercole} A.,
  2014, \mnras, 443, 3302

\bibitem[{{Davis} {et~al}\mbox{.}(2010){Davis}, {Clarke}, \&
  {Freitag}}]{2010MNRAS.407..381D}
{Davis} O., {Clarke} C.~J., {Freitag} M., 2010, \mnras, 407, 381

\bibitem[{{De Marchi} {et~al}\mbox{.}(2017){De Marchi}, {Panagia}, \&
  {Beccari}}]{2017ApJ...846..110D}
{De Marchi} G., {Panagia} N., {Beccari} G., 2017, \apj, 846, 110

\bibitem[{{de Mink} {et~al}\mbox{.}(2009){de Mink}, {Pols}, {Langer}, \&
  {Izzard}}]{2009A&A...507L...1D}
{de Mink} S.~E., {Pols} O.~R., {Langer} N., {Izzard} R.~G., 2009, \aap, 507, L1

\bibitem[{{Decressin} {et~al}\mbox{.}(2010){Decressin}, {Baumgardt},
  {Charbonnel}, \& {Kroupa}}]{2010A&A...516A..73D}
{Decressin} T., {Baumgardt} H., {Charbonnel} C., {Kroupa} P., 2010, \aap, 516,
  A73

\bibitem[{{Decressin} {et~al}\mbox{.}(2007{\natexlab{a}}){Decressin},
  {Charbonnel}, \& {Meynet}}]{2007A&A...475..859D}
{Decressin} T., {Charbonnel} C., {Meynet} G., 2007{\natexlab{a}}, \aap, 475,
  859

\bibitem[{{Decressin} {et~al}\mbox{.}(2009){Decressin}, {Charbonnel}, {Siess},
  {Palacios}, {Meynet}, \& {Georgy}}]{2009A&A...505..727D}
{Decressin} T., {Charbonnel} C., {Siess} L., {Palacios} A., {Meynet} G.,
  {Georgy} C., 2009, \aap, 505, 727

\bibitem[{{Decressin} {et~al}\mbox{.}(2007{\natexlab{b}}){Decressin}, {Meynet},
  {Charbonnel}, {Prantzos}, \& {Ekstr{\"o}m}}]{2007A&A...464.1029D}
{Decressin} T., {Meynet} G., {Charbonnel} C., {Prantzos} N., {Ekstr{\"o}m} S.,
  2007{\natexlab{b}}, \aap, 464, 1029

\bibitem[{{Denisenkov} \& {Denisenkova}(1990)}]{1990SvAL...16..275D}
{Denisenkov} P.~A., {Denisenkova} S.~N., 1990, Soviet Astronomy Letters, 16,
  275

\bibitem[{{Denissenkov} \& {Hartwick}(2014)}]{2014MNRAS.437L..21D}
{Denissenkov} P.~A., {Hartwick} F.~D.~A., 2014, \mnras, 437, L21

\bibitem[{{Denissenkov} \& {Herwig}(2003)}]{2003ApJ...590L..99D}
{Denissenkov} P.~A., {Herwig} F., 2003, \apjl, 590, L99

\bibitem[{{Denissenkov} {et~al}\mbox{.}(2015){Denissenkov}, {VandenBerg},
  {Hartwick}, {Herwig}, {Weiss}, \& {Paxton}}]{2015MNRAS.448.3314D}
{Denissenkov} P.~A., {VandenBerg} D.~A., {Hartwick} F.~D.~A., {Herwig} F.,
  {Weiss} A., {Paxton} B., 2015, \mnras, 448, 3314

\bibitem[{{D'Ercole} {et~al}\mbox{.}(2008){D'Ercole}, {Vesperini}, {D'Antona},
  {McMillan}, \& {Recchi}}]{2008MNRAS.391..825D}
{D'Ercole} A., {Vesperini} E., {D'Antona} F., {McMillan} S.~L.~W., {Recchi} S.,
  2008, \mnras, 391, 825

\bibitem[{{Doherty} {et~al}\mbox{.}(2014){Doherty}, {Gil-Pons}, {Lau},
  {Lattanzio}, \& {Siess}}]{2014MNRAS.437..195D}
{Doherty} C.~L., {Gil-Pons} P., {Lau} H.~H.~B., {Lattanzio} J.~C., {Siess} L.,
  2014, \mnras, 437, 195

\bibitem[{{D'Orazi} {et~al}\mbox{.}(2015){D'Orazi}, {Gratton}, {Angelou},
  {Bragaglia}, {Carretta}, {Lattanzio}, {Lucatello}, {Momany}, {Sollima}, \&
  {Beccari}}]{2015MNRAS.449.4038D}
{D'Orazi} V. {et~al.}, 2015, \mnras, 449, 4038

\bibitem[{{Elmegreen}(2017)}]{2017ApJ...836...80E}
{Elmegreen} B.~G., 2017, \apj, 836, 80

\bibitem[{{Elmegreen} \& {Efremov}(1997)}]{1997ApJ...480..235E}
{Elmegreen} B.~G., {Efremov} Y.~N., 1997, \apj, 480, 235

\bibitem[{{Fabian} {et~al}\mbox{.}(1975){Fabian}, {Pringle}, \&
  {Rees}}]{1975MNRAS.172P..15F}
{Fabian} A.~C., {Pringle} J.~E., {Rees} M.~J., 1975, \mnras, 172, 15p

\bibitem[{{Federrath}(2015)}]{2015MNRAS.450.4035F}
{Federrath} C., 2015, \mnras, 450, 4035

\bibitem[{{Forestini} \& {Charbonnel}(1997)}]{1997A&AS..123..241F}
{Forestini} M., {Charbonnel} C., 1997, \aaps, 123

\bibitem[{{Freitag} \& {Benz}(2005)}]{2005MNRAS.358.1133F}
{Freitag} M., {Benz} W., 2005, \mnras, 358, 1133

\bibitem[{{Freitag} {et~al}\mbox{.}(2006){Freitag}, {G{\"u}rkan}, \&
  {Rasio}}]{2006MNRAS.368..141F}
{Freitag} M., {G{\"u}rkan} M.~A., {Rasio} F.~A., 2006, \mnras, 368, 141

\bibitem[{{Gaburov} {et~al}\mbox{.}(2008){Gaburov}, {Lombardi}, \& {Portegies
  Zwart}}]{2008MNRAS.383L...5G}
{Gaburov} E., {Lombardi} J.~C., {Portegies Zwart} S., 2008, \mnras, 383, L5

\bibitem[{{Gavagnin} {et~al}\mbox{.}(2017){Gavagnin}, {Bleuler}, {Rosdahl}, \&
  {Teyssier}}]{2017MNRAS.472.4155G}
{Gavagnin} E., {Bleuler} A., {Rosdahl} J., {Teyssier} R., 2017, \mnras, 472,
  4155

\bibitem[{{Georgy} {et~al}\mbox{.}(2013){Georgy}, {Ekstr{\"o}m}, {Eggenberger},
  {Meynet}, {Haemmerl{\'e}}, {Maeder}, {Granada}, {Groh}, {Hirschi}, {Mowlavi},
  {Yusof}, {Charbonnel}, {Decressin}, \& {Barblan}}]{2013A&A...558A.103G}
{Georgy} C. {et~al.}, 2013, \aap, 558, A103

\bibitem[{{Gieles} {et~al}\mbox{.}(2018){Gieles}, {Balbinot}, {Yaaqib},
  {H{\'e}nault-Brunet}, {Zocchi}, {Peuten}, \& {Jonker}}]{2018MNRAS.473.4832G}
{Gieles} M., {Balbinot} E., {Yaaqib} R.~I.~S.~M., {H{\'e}nault-Brunet} V.,
  {Zocchi} A., {Peuten} M., {Jonker} P.~G., 2018, \mnras, 473, 4832

\bibitem[{{Gieles} {et~al}\mbox{.}(2010){Gieles}, {Baumgardt}, {Heggie}, \&
  {Lamers}}]{2010MNRAS.408L..16G}
{Gieles} M., {Baumgardt} H., {Heggie} D.~C., {Lamers} H.~J.~G.~L.~M., 2010,
  \mnras, 408, L16

\bibitem[{{Gieles} {et~al}\mbox{.}(2011){Gieles}, {Heggie}, \&
  {Zhao}}]{2011MNRAS.413.2509G}
{Gieles} M., {Heggie} D.~C., {Zhao} H., 2011, \mnras, 413, 2509

\bibitem[{{Gieles} \& {Zocchi}(2015)}]{2015MNRAS.454..576G}
{Gieles} M., {Zocchi} A., 2015, \mnras, 454, 576

\bibitem[{{Giersz} \& {Heggie}(2009)}]{2009MNRAS.395.1173G}
{Giersz} M., {Heggie} D.~C., 2009, \mnras, 395, 1173

\bibitem[{{Giersz} \& {Heggie}(2011)}]{2011MNRAS.410.2698G}
{Giersz} M., {Heggie} D.~C., 2011, \mnras, 410, 2698

\bibitem[{{Giersz} {et~al}\mbox{.}(2015){Giersz}, {Leigh}, {Hypki},
  {L{\"u}tzgendorf}, \& {Askar}}]{2015MNRAS.454.3150G}
{Giersz} M., {Leigh} N., {Hypki} A., {L{\"u}tzgendorf} N., {Askar} A., 2015,
  \mnras, 454, 3150

\bibitem[{{Glatt} {et~al}\mbox{.}(2011){Glatt}, {Grebel}, {Jordi}, {Gallagher},
  {Da Costa}, {Clementini}, {Tosi}, {Harbeck}, {Nota}, {Sabbi}, \&
  {Sirianni}}]{2011AJ....142...36G}
{Glatt} K. {et~al.}, 2011, \aj, 142, 36

\bibitem[{{Glebbeek} {et~al}\mbox{.}(2009){Glebbeek}, {Gaburov}, {de Mink},
  {Pols}, \& {Portegies Zwart}}]{2009A&A...497..255G}
{Glebbeek} E., {Gaburov} E., {de Mink} S.~E., {Pols} O.~R., {Portegies Zwart}
  S.~F., 2009, \aap, 497, 255

\bibitem[{{Goodman}(1984)}]{1984ApJ...280..298G}
{Goodman} J., 1984, \apj, 280, 298

\bibitem[{{Gr{\"a}fener} {et~al}\mbox{.}(2012){Gr{\"a}fener}, {Owocki}, \&
  {Vink}}]{2012A&A...538A..40G}
{Gr{\"a}fener} G., {Owocki} S.~P., {Vink} J.~S., 2012, \aap, 538, A40

\bibitem[{{Gr{\"a}fener} \& {Vink}(2015)}]{2015A&A...578L...2G}
{Gr{\"a}fener} G., {Vink} J.~S., 2015, \aap, 578, L2

\bibitem[{{Gruyters} {et~al}\mbox{.}(2016){Gruyters}, {Lind}, {Richard},
  {Grundahl}, {Asplund}, {Casagrande}, {Charbonnel}, {Milone}, {Primas}, \&
  {Korn}}]{2016A&A...589A..61G}
{Gruyters} P. {et~al.}, 2016, \aap, 589, A61

\bibitem[{{G{\"u}rkan} {et~al}\mbox{.}(2006){G{\"u}rkan}, {Fregeau}, \&
  {Rasio}}]{2006ApJ...640L..39G}
{G{\"u}rkan} M.~A., {Fregeau} J.~M., {Rasio} F.~A., 2006, \apjl, 640, L39

\bibitem[{{Guszejnov} \& {Hopkins}(2015)}]{2015MNRAS.450.4137G}
{Guszejnov} D., {Hopkins} P.~F., 2015, \mnras, 450, 4137

\bibitem[{Hairer {et~al}\mbox{.}(1993)Hairer, N{\o}rsett, \&
  Wanner}]{hairer1993solving}
Hairer E., N{\o}rsett S., Wanner G., 1993, Solving Ordinary Differential
  Equations I: Nonstiff Problems, Solving Ordinary Differential Equations.
  Springer

\bibitem[{{Hartmann} {et~al}\mbox{.}(2016){Hartmann}, {Herczeg}, \&
  {Calvet}}]{2016ARA&A..54..135H}
{Hartmann} L., {Herczeg} G., {Calvet} N., 2016, \araa, 54, 135

\bibitem[{{Heggie} \& {Hut}(2003)}]{2003gmbp.book.....H}
{Heggie} D., {Hut} P., 2003, {The Gravitational Million-Body Problem: A
  Multidisciplinary Approach to Star Cluster Dynamics.} Cambridge Univ. Press,
  Cambridge

\bibitem[{{Heggie}(1975)}]{1975MNRAS.173..729H}
{Heggie} D.~C., 1975, \mnras, 173, 729

\bibitem[{{Heggie} {et~al}\mbox{.}(2007){Heggie}, {Hut}, {Mineshige}, {Makino},
  \& {Baumgardt}}]{2007PASJ...59L..11H}
{Heggie} D.~C., {Hut} P., {Mineshige} S., {Makino} J., {Baumgardt} H., 2007,
  \pasj, 59, L11

\bibitem[{{H{\'e}nault-Brunet} {et~al}\mbox{.}(2015){H{\'e}nault-Brunet},
  {Gieles}, {Agertz}, \& {Read}}]{2015MNRAS.450.1164H}
{H{\'e}nault-Brunet} V., {Gieles} M., {Agertz} O., {Read} J.~I., 2015, \mnras,
  450, 1164

\bibitem[{{Hennebelle} \& {Chabrier}(2008)}]{2008ApJ...684..395H}
{Hennebelle} P., {Chabrier} G., 2008, \apj, 684, 395

\bibitem[{{H{\'e}non}(1961)}]{H61}
{H{\'e}non} M., 1961, Annales d'Astrophysique, 24, 369; English translation:
  arXiv:1103.3499

\bibitem[{{H{\'e}non}(1965)}]{H65}
{H{\'e}non} M., 1965, Ann. Astrophys, 28, 62; English translation:
  ArXiv:1103.3498

\bibitem[{{Hills}(1980)}]{1980ApJ...235..986H}
{Hills} J.~G., 1980, \apj, 235, 986

\bibitem[{{Hills} \& {Day}(1976)}]{1976ApL....17...87H}
{Hills} J.~G., {Day} C.~A., 1976, \aplett, 17, 87

\bibitem[{{Hollyhead} {et~al}\mbox{.}(2017){Hollyhead}, {Kacharov}, {Lardo},
  {Bastian}, {Hilker}, {Rejkuba}, {Koch}, {Grebel}, \&
  {Georgiev}}]{2017MNRAS.465L..39H}
{Hollyhead} K. {et~al.}, 2017, \mnras, 465, L39

\bibitem[{{Humphreys} \& {Davidson}(1994)}]{1994PASP..106.1025H}
{Humphreys} R.~M., {Davidson} K., 1994, \pasp, 106, 1025

\bibitem[{{Huxor} {et~al}\mbox{.}(2005){Huxor}, {Tanvir}, {Irwin}, {Ibata},
  {Collett}, {Ferguson}, {Bridges}, \& {Lewis}}]{2005MNRAS.360.1007H}
{Huxor} A.~P., {Tanvir} N.~R., {Irwin} M.~J., {Ibata} R., {Collett} J.~L.,
  {Ferguson} A.~M.~N., {Bridges} T., {Lewis} G.~F., 2005, \mnras, 360, 1007

\bibitem[{{Ishii} {et~al}\mbox{.}(1999){Ishii}, {Ueno}, \&
  {Kato}}]{1999PASJ...51..417I}
{Ishii} M., {Ueno} M., {Kato} M., 1999, \pasj, 51, 417

\bibitem[{{Karakas} \& {Lattanzio}(2007)}]{2007PASA...24..103K}
{Karakas} A., {Lattanzio} J.~C., 2007, \pasa, 24, 103

\bibitem[{{Karakas} {et~al}\mbox{.}(2006){Karakas}, {Fenner}, {Sills},
  {Campbell}, \& {Lattanzio}}]{2006ApJ...652.1240K}
{Karakas} A.~I., {Fenner} Y., {Sills} A., {Campbell} S.~W., {Lattanzio} J.~C.,
  2006, \apj, 652, 1240

\bibitem[{{Katz} {et~al}\mbox{.}(2015){Katz}, {Sijacki}, \&
  {Haehnelt}}]{2015MNRAS.451.2352K}
{Katz} H., {Sijacki} D., {Haehnelt} M.~G., 2015, \mnras, 451, 2352

\bibitem[{{Khalaj} \& {Baumgardt}(2015)}]{2015MNRAS.452..924K}
{Khalaj} P., {Baumgardt} H., 2015, \mnras, 452, 924

\bibitem[{{Kim} {et~al}\mbox{.}(2018){Kim}, {Ma}, {Grudi{\'c}}, {Hopkins},
  {Hayward}, {Wetzel}, {Faucher-Gigu{\`e}re}, {Kere{\v s}}, {Garrison-Kimmel},
  \& {Murray}}]{2018MNRAS.474.4232K}
{Kim} J.-h. {et~al.}, 2018, \mnras, 474, 4232

\bibitem[{{Kim} {et~al}\mbox{.}(1998){Kim}, {Lee}, \&
  {Goodman}}]{1998ApJ...495..786K}
{Kim} S.~S., {Lee} H.~M., {Goodman} J., 1998, \apj, 495, 786

\bibitem[{{Kimm} {et~al}\mbox{.}(2016){Kimm}, {Cen}, {Rosdahl}, \&
  {Yi}}]{2016ApJ...823...52K}
{Kimm} T., {Cen} R., {Rosdahl} J., {Yi} S.~K., 2016, \apj, 823, 52

\bibitem[{{King}(1966)}]{1966AJ.....71...64K}
{King} I.~R., 1966, \aj, 71, 64

\bibitem[{{Kissler-Patig} {et~al}\mbox{.}(2006){Kissler-Patig}, {Jord{\'a}n},
  \& {Bastian}}]{2006A&A...448.1031K}
{Kissler-Patig} M., {Jord{\'a}n} A., {Bastian} N., 2006, \aap, 448, 1031

\bibitem[{{Korn} {et~al}\mbox{.}(2006){Korn}, {Grundahl}, {Richard}, {Barklem},
  {Mashonkina}, {Collet}, {Piskunov}, \& {Gustafsson}}]{2006Natur.442..657K}
{Korn} A.~J., {Grundahl} F., {Richard} O., {Barklem} P.~S., {Mashonkina} L.,
  {Collet} R., {Piskunov} N., {Gustafsson} B., 2006, \nat, 442, 657

\bibitem[{{Korn} {et~al}\mbox{.}(2007){Korn}, {Grundahl}, {Richard},
  {Mashonkina}, {Barklem}, {Collet}, {Gustafsson}, \&
  {Piskunov}}]{2007ApJ...671..402K}
{Korn} A.~J., {Grundahl} F., {Richard} O., {Mashonkina} L., {Barklem} P.~S.,
  {Collet} R., {Gustafsson} B., {Piskunov} N., 2007, \apj, 671, 402

\bibitem[{{Krause} {et~al}\mbox{.}(2013){Krause}, {Charbonnel}, {Decressin},
  {Meynet}, \& {Prantzos}}]{2013A&A...552A.121K}
{Krause} M., {Charbonnel} C., {Decressin} T., {Meynet} G., {Prantzos} N., 2013,
  \aap, 552, A121

\bibitem[{{Krause} {et~al}\mbox{.}(2012){Krause}, {Charbonnel}, {Decressin},
  {Meynet}, {Prantzos}, \& {Diehl}}]{2012A&A...546L...5K}
{Krause} M., {Charbonnel} C., {Decressin} T., {Meynet} G., {Prantzos} N.,
  {Diehl} R., 2012, \aap, 546, L5

\bibitem[{{Krause} {et~al}\mbox{.}(2016){Krause}, {Charbonnel}, {Bastian}, \&
  {Diehl}}]{2016A&A...587A..53K}
{Krause} M.~G.~H., {Charbonnel} C., {Bastian} N., {Diehl} R., 2016, \aap, 587,
  A53

\bibitem[{{Krause} {et~al}\mbox{.}(2015){Krause}, {Diehl}, {Bagetakos},
  {Brinks}, {Burkert}, {Gerhard}, {Greiner}, {Kretschmer}, \&
  {Siegert}}]{2015A&A...578A.113K}
{Krause} M.~G.~H. {et~al.}, 2015, \aap, 578, A113

\bibitem[{{Kretschmer} {et~al}\mbox{.}(2013){Kretschmer}, {Diehl}, {Krause},
  {Burkert}, {Fierlinger}, {Gerhard}, {Greiner}, \&
  {Wang}}]{2013A&A...559A..99K}
{Kretschmer} K., {Diehl} R., {Krause} M., {Burkert} A., {Fierlinger} K.,
  {Gerhard} O., {Greiner} J., {Wang} W., 2013, \aap, 559, A99

\bibitem[{{Kroupa}(2001)}]{2001MNRAS.322..231K}
{Kroupa} P., 2001, \mnras, 322, 231

\bibitem[{{Krumholz} \& {Bonnell}(2007)}]{2007arXiv0712.0828K}
{Krumholz} M.~R., {Bonnell} I.~A., 2007, ArXiv:0712.0828

\bibitem[{{Krumholz} {et~al}\mbox{.}(2009){Krumholz}, {Klein}, {McKee},
  {Offner}, \& {Cunningham}}]{2009Sci...323..754K}
{Krumholz} M.~R., {Klein} R.~I., {McKee} C.~F., {Offner} S.~S.~R., {Cunningham}
  A.~J., 2009, Science, 323, 754

\bibitem[{{Lamers} \& {Cassinelli}(1999)}]{1999isw..book.....L}
{Lamers} H.~J.~G.~L.~M., {Cassinelli} J.~P., 1999, {Introduction to Stellar
  Winds}. Cambridge, UK: Cambridge University Press, p. 452

\bibitem[{{Langer} {et~al}\mbox{.}(1993){Langer}, {Hoffman}, \&
  {Sneden}}]{1993PASP..105..301L}
{Langer} G.~E., {Hoffman} R., {Sneden} C., 1993, \pasp, 105, 301

\bibitem[{{Lanzoni} {et~al}\mbox{.}(2013){Lanzoni}, {Mucciarelli}, {Origlia},
  {Bellazzini}, {Ferraro}, {Valenti}, {Miocchi}, {Dalessandro}, {Pallanca}, \&
  {Massari}}]{2013ApJ...769..107L}
{Lanzoni} B. {et~al.}, 2013, \apj, 769, 107

\bibitem[{{Lardo} {et~al}\mbox{.}(2011){Lardo}, {Bellazzini}, {Pancino},
  {Carretta}, {Bragaglia}, \& {Dalessandro}}]{2011A&A...525A.114L}
{Lardo} C., {Bellazzini} M., {Pancino} E., {Carretta} E., {Bragaglia} A.,
  {Dalessandro} E., 2011, \aap, 525, A114

\bibitem[{{Larsen} {et~al}\mbox{.}(2015){Larsen}, {Baumgardt}, {Bastian},
  {Brodie}, {Grundahl}, \& {Strader}}]{2015ApJ...804...71L}
{Larsen} S.~S., {Baumgardt} H., {Bastian} N., {Brodie} J.~P., {Grundahl} F.,
  {Strader} J., 2015, \apj, 804, 71

\bibitem[{{Larsen} {et~al}\mbox{.}(2014){Larsen}, {Brodie}, {Forbes}, \&
  {Strader}}]{2014A&A...565A..98L}
{Larsen} S.~S., {Brodie} J.~P., {Forbes} D.~A., {Strader} J., 2014, \aap, 565,
  A98

\bibitem[{{Larsen} {et~al}\mbox{.}(2012){Larsen}, {Strader}, \&
  {Brodie}}]{2012A&A...544L..14L}
{Larsen} S.~S., {Strader} J., {Brodie} J.~P., 2012, \aap, 544, L14

\bibitem[{{Li} {et~al}\mbox{.}(2017){Li}, {Gnedin}, {Gnedin}, {Meng},
  {Semenov}, \& {Kravtsov}}]{2017ApJ...834...69L}
{Li} H., {Gnedin} O.~Y., {Gnedin} N.~Y., {Meng} X., {Semenov} V.~A., {Kravtsov}
  A.~V., 2017, \apj, 834, 69

\bibitem[{{Lind} {et~al}\mbox{.}(2011){Lind}, {Charbonnel}, {Decressin},
  {Primas}, {Grundahl}, \& {Asplund}}]{2011A&A...527A.148L}
{Lind} K., {Charbonnel} C., {Decressin} T., {Primas} F., {Grundahl} F.,
  {Asplund} M., 2011, \aap, 527, A148

\bibitem[{{Lind} {et~al}\mbox{.}(2009){Lind}, {Primas}, {Charbonnel},
  {Grundahl}, \& {Asplund}}]{2009A&A...503..545L}
{Lind} K., {Primas} F., {Charbonnel} C., {Grundahl} F., {Asplund} M., 2009,
  \aap, 503, 545

\bibitem[{{Lombardi} {et~al}\mbox{.}(2003){Lombardi}, {Thrall}, {Deneva},
  {Fleming}, \& {Grabowski}}]{2003MNRAS.345..762L}
{Lombardi} J.~C., {Thrall} A.~P., {Deneva} J.~S., {Fleming} S.~W., {Grabowski}
  P.~E., 2003, \mnras, 345, 762

\bibitem[{{Longmore} {et~al}\mbox{.}(2014){Longmore}, {Kruijssen}, {Bastian},
  {Bally}, {Rathborne}, {Testi}, {Stolte}, {Dale}, {Bressert}, \&
  {Alves}}]{2014prpl.conf..291L}
{Longmore} S.~N. {et~al.}, 2014, Protostars and Planets VI, 291

\bibitem[{{L{\"u}tzgendorf} {et~al}\mbox{.}(2011){L{\"u}tzgendorf},
  {Kissler-Patig}, {Noyola}, {Jalali}, {de Zeeuw}, {Gebhardt}, \&
  {Baumgardt}}]{2011A&A...533A..36L}
{L{\"u}tzgendorf} N., {Kissler-Patig} M., {Noyola} E., {Jalali} B., {de Zeeuw}
  P.~T., {Gebhardt} K., {Baumgardt} H., 2011, \aap, 533, A36

\bibitem[{{Maeder} \& {Meynet}(2006)}]{2006A&A...448L..37M}
{Maeder} A., {Meynet} G., 2006, \aap, 448, L37

\bibitem[{{Mapelli}(2016)}]{2016MNRAS.459.3432M}
{Mapelli} M., 2016, \mnras, 459, 3432

\bibitem[{{Marino} {et~al}\mbox{.}(2011){Marino}, {Milone}, {Piotto},
  {Villanova}, {Gratton}, {D'Antona}, {Anderson}, {Bedin}, {Bellini},
  {Cassisi}, {Geisler}, {Renzini}, \& {Zoccali}}]{2011ApJ...731...64M}
{Marino} A.~F. {et~al.}, 2011, \apj, 731, 64

\bibitem[{{Martocchia} {et~al}\mbox{.}(2018){Martocchia}, {Cabrera-Ziri},
  {Lardo}, {Dalessandro}, {Bastian}, {Kozhurina-Platais}, {Usher},
  {Niederhofer}, {Cordero}, {Geisler}, {Hollyhead}, {Kacharov}, {Larsen}, {Li},
  {Mackey}, {Hilker}, {Mucciarelli}, {Platais}, \&
  {Salaris}}]{2018MNRAS.473.2688M}
{Martocchia} S. {et~al.}, 2018, \mnras, 473, 2688

\bibitem[{{Mel{\'e}ndez} {et~al}\mbox{.}(2010){Mel{\'e}ndez}, {Casagrande},
  {Ram{\'{\i}}rez}, {Asplund}, \& {Schuster}}]{2010A&A...515L...3M}
{Mel{\'e}ndez} J., {Casagrande} L., {Ram{\'{\i}}rez} I., {Asplund} M.,
  {Schuster} W.~J., 2010, \aap, 515, L3

\bibitem[{{M{\'e}sz{\'a}ros} {et~al}\mbox{.}(2015){M{\'e}sz{\'a}ros},
  {Martell}, {Shetrone}, {Lucatello}, {Troup}, {Bovy}, {Cunha},
  {Garc{\'{\i}}a-Hern{\'a}ndez}, {Overbeek}, {Allende Prieto}, {Beers},
  {Frinchaboy}, {Garc{\'{\i}}a P{\'e}rez}, {Hearty}, {Holtzman}, {Majewski},
  {Nidever}, {Schiavon}, {Schneider}, {Sobeck}, {Smith}, {Zamora}, \&
  {Zasowski}}]{2015AJ....149..153M}
{M{\'e}sz{\'a}ros} S. {et~al.}, 2015, \aj, 149, 153

\bibitem[{{Milone} {et~al}\mbox{.}(2014){Milone}, {Marino}, {Dotter}, {Norris},
  {Jerjen}, {Piotto}, {Cassisi}, {Bedin}, {Recio Blanco}, {Sarajedini},
  {Asplund}, {Monelli}, \& {Aparicio}}]{2014ApJ...785...21M}
{Milone} A.~P. {et~al.}, 2014, \apj, 785, 21

\bibitem[{{Milone} {et~al}\mbox{.}(2013){Milone}, {Marino}, {Piotto}, {Bedin},
  {Anderson}, {Aparicio}, {Bellini}, {Cassisi}, {D'Antona}, {Grundahl},
  {Monelli}, \& {Yong}}]{2013ApJ...767..120M}
{Milone} A.~P. {et~al.}, 2013, \apj, 767, 120

\bibitem[{{Milone} {et~al}\mbox{.}(2012){Milone}, {Marino}, {Piotto}, {Bedin},
  {Anderson}, {Aparicio}, {Cassisi}, \& {Rich}}]{2012ApJ...745...27M}
{Milone} A.~P., {Marino} A.~F., {Piotto} G., {Bedin} L.~R., {Anderson} J.,
  {Aparicio} A., {Cassisi} S., {Rich} R.~M., 2012, \apj, 745, 27

\bibitem[{{Milone} {et~al}\mbox{.}(2015){Milone}, {Marino}, {Piotto},
  {Renzini}, {Bedin}, {Anderson}, {Cassisi}, {D'Antona}, {Bellini}, {Jerjen},
  {Pietrinferni}, \& {Ventura}}]{2015ApJ...808...51M}
{Milone} A.~P. {et~al.}, 2015, \apj, 808, 51

\bibitem[{{Milone} {et~al}\mbox{.}(2017){Milone}, {Piotto}, {Renzini},
  {Marino}, {Bedin}, {Vesperini}, {D'Antona}, {Nardiello}, {Anderson}, {King},
  {Yong}, {Bellini}, {Aparicio}, {Barbuy}, {Brown}, {Cassisi}, {Ortolani},
  {Salaris}, {Sarajedini}, \& {van der Marel}}]{2017MNRAS.464.3636M}
{Milone} A.~P. {et~al.}, 2017, \mnras, 464, 3636

\bibitem[{{Moeckel} \& {Clarke}(2011)}]{2011MNRAS.410.2799M}
{Moeckel} N., {Clarke} C.~J., 2011, \mnras, 410, 2799

\bibitem[{{Moeckel} {et~al}\mbox{.}(2012){Moeckel}, {Holland}, {Clarke}, \&
  {Bonnell}}]{2012MNRAS.425..450M}
{Moeckel} N., {Holland} C., {Clarke} C.~J., {Bonnell} I.~A., 2012, \mnras, 425,
  450

\bibitem[{{Monaco} {et~al}\mbox{.}(2010){Monaco}, {Bonifacio}, {Sbordone},
  {Villanova}, \& {Pancino}}]{2010A&A...519L...3M}
{Monaco} L., {Bonifacio} P., {Sbordone} L., {Villanova} S., {Pancino} E., 2010,
  \aap, 519, L3

\bibitem[{{Mucciarelli} {et~al}\mbox{.}(2008){Mucciarelli}, {Carretta},
  {Origlia}, \& {Ferraro}}]{2008AJ....136..375M}
{Mucciarelli} A., {Carretta} E., {Origlia} L., {Ferraro} F.~R., 2008, \aj, 136,
  375

\bibitem[{{Mucciarelli} {et~al}\mbox{.}(2011){Mucciarelli}, {Salaris},
  {Lovisi}, {Ferraro}, {Lanzoni}, {Lucatello}, \&
  {Gratton}}]{2011MNRAS.412...81M}
{Mucciarelli} A., {Salaris} M., {Lovisi} L., {Ferraro} F.~R., {Lanzoni} B.,
  {Lucatello} S., {Gratton} R.~G., 2011, \mnras, 412, 81

\bibitem[{{Muijres} {et~al}\mbox{.}(2012){Muijres}, {Vink}, {de Koter},
  {M{\"u}ller}, \& {Langer}}]{2012A&A...537A..37M}
{Muijres} L.~E., {Vink} J.~S., {de Koter} A., {M{\"u}ller} P.~E., {Langer} N.,
  2012, \aap, 537, A37

\bibitem[{{Murray} \& {Chang}(2015)}]{2015ApJ...804...44M}
{Murray} N., {Chang} P., 2015, \apj, 804, 44

\bibitem[{{Nadyozhin} \& {Razinkova}(2005)}]{2005AstL...31..695N}
{Nadyozhin} D.~K., {Razinkova} T.~L., 2005, Astronomy Letters, 31, 695

\bibitem[{{Nardiello} {et~al}\mbox{.}(2018){Nardiello}, {Milone}, {Piotto},
  {Anderson}, {Bedin}, {Bellini}, {Cassisi}, {Libralato}, \&
  {Marino}}]{2018MNRAS.tmp..706N}
{Nardiello} D. {et~al.}, 2018, \mnras, ArXiv:1803.05979

\bibitem[{{Nardiello} {et~al}\mbox{.}(2015){Nardiello}, {Piotto}, {Milone},
  {Marino}, {Bedin}, {Anderson}, {Aparicio}, {Bellini}, {Cassisi}, {D'Antona},
  {Hidalgo}, {Ortolani}, {Pietrinferni}, {Renzini}, {Salaris}, {Marel}, \&
  {Vesperini}}]{2015MNRAS.451..312N}
{Nardiello} D. {et~al.}, 2015, \mnras, 451, 312

\bibitem[{{Niederhofer} {et~al}\mbox{.}(2017){Niederhofer}, {Bastian},
  {Kozhurina-Platais}, {Larsen}, {Hollyhead}, {Lardo}, {Cabrera-Ziri},
  {Kacharov}, {Platais}, {Salaris}, {Cordero}, {Dalessandro}, {Geisler},
  {Hilker}, {Li}, {Mackey}, \& {Mucciarelli}}]{2017MNRAS.465.4159N}
{Niederhofer} F. {et~al.}, 2017, \mnras, 465, 4159

\bibitem[{{Norris}(2004)}]{2004ApJ...612L..25N}
{Norris} J.~E., 2004, \apjl, 612, L25

\bibitem[{{Noyola} {et~al}\mbox{.}(2008){Noyola}, {Gebhardt}, \&
  {Bergmann}}]{2008ApJ...676.1008N}
{Noyola} E., {Gebhardt} K., {Bergmann} M., 2008, \apj, 676, 1008

\bibitem[{{Oey} {et~al}\mbox{.}(2017){Oey}, {Herrera}, {Silich}, {Reiter},
  {James}, {Jaskot}, \& {Micheva}}]{2017ApJ...849L...1O}
{Oey} M.~S., {Herrera} C.~N., {Silich} S., {Reiter} M., {James} B.~L., {Jaskot}
  A.~E., {Micheva} G., 2017, \apjl, 849, L1

\bibitem[{{Offner} \& {Chaban}(2017)}]{2017ApJ...847..104O}
{Offner} S.~S.~R., {Chaban} J., 2017, \apj, 847, 104

\bibitem[{{Offner} {et~al}\mbox{.}(2008){Offner}, {Klein}, \&
  {McKee}}]{2008ApJ...686.1174O}
{Offner} S.~S.~R., {Klein} R.~I., {McKee} C.~F., 2008, \apj, 686, 1174

\bibitem[{{Padoan} {et~al}\mbox{.}(2014){Padoan}, {Haugb{\o}lle}, \&
  {Nordlund}}]{2014ApJ...797...32P}
{Padoan} P., {Haugb{\o}lle} T., {Nordlund} {\AA}., 2014, \apj, 797, 32

\bibitem[{{Padoan} \& {Nordlund}(2002)}]{2002ApJ...576..870P}
{Padoan} P., {Nordlund} {\AA}., 2002, \apj, 576, 870

\bibitem[{{Pancino} {et~al}\mbox{.}(2017){Pancino}, {Romano}, {Tang},
  {Tautvai{\v s}ien{\.e}}, {Casey}, {Gruyters}, {Geisler}, {San Roman},
  {Randich}, {Alfaro}, {Bragaglia}, {Flaccomio}, {Korn}, {Recio-Blanco},
  {Smiljanic}, {Carraro}, {Bayo}, {Costado}, {Damiani}, {Jofr{\'e}}, {Lardo},
  {de Laverny}, {Monaco}, {Morbidelli}, {Sbordone}, {Sousa}, \&
  {Villanova}}]{2017A&A...601A.112P}
{Pancino} E. {et~al.}, 2017, \aap, 601, A112

\bibitem[{{Pasquini} {et~al}\mbox{.}(2005){Pasquini}, {Bonifacio}, {Molaro},
  {Francois}, {Spite}, {Gratton}, {Carretta}, \& {Wolff}}]{2005A&A...441..549P}
{Pasquini} L., {Bonifacio} P., {Molaro} P., {Francois} P., {Spite} F.,
  {Gratton} R.~G., {Carretta} E., {Wolff} B., 2005, \aap, 441, 549

\bibitem[{{Paxton} {et~al}\mbox{.}(2011){Paxton}, {Bildsten}, {Dotter},
  {Herwig}, {Lesaffre}, \& {Timmes}}]{2011ApJS..192....3P}
{Paxton} B., {Bildsten} L., {Dotter} A., {Herwig} F., {Lesaffre} P., {Timmes}
  F., 2011, \apjs, 192, 3

\bibitem[{{Paxton} {et~al}\mbox{.}(2013){Paxton}, {Cantiello}, {Arras},
  {Bildsten}, {Brown}, {Dotter}, {Mankovich}, {Montgomery}, {Stello}, {Timmes},
  \& {Townsend}}]{2013ApJS..208....4P}
{Paxton} B. {et~al.}, 2013, \apjs, 208, 4

\bibitem[{{Petts} \& {Gualandris}(2017)}]{2017MNRAS.467.3775P}
{Petts} J.~A., {Gualandris} A., 2017, \mnras, 467, 3775

\bibitem[{{Pinsonneault} {et~al}\mbox{.}(2000){Pinsonneault}, {Charbonnel}, \&
  {Deliyannis}}]{2000IAUS..198...74P}
{Pinsonneault} M.~H., {Charbonnel} C., {Deliyannis} C.~P., 2000, in IAU
  Symposium, Vol. 198, The Light Elements and their Evolution, {da Silva} L.,
  {de Medeiros} R., {Spite} M., eds., p.~74

\bibitem[{{Piotto} {et~al}\mbox{.}(2007){Piotto}, {Bedin}, {Anderson}, {King},
  {Cassisi}, {Milone}, {Villanova}, {Pietrinferni}, \&
  {Renzini}}]{2007ApJ...661L..53P}
{Piotto} G. {et~al.}, 2007, \apjl, 661, L53

\bibitem[{{Piotto} {et~al}\mbox{.}(2015){Piotto}, {Milone}, {Bedin},
  {Anderson}, {King}, {Marino}, {Nardiello}, {Aparicio}, {Barbuy}, {Bellini},
  {Brown}, {Cassisi}, {Cool}, {Cunial}, {Dalessandro}, {D'Antona}, {Ferraro},
  {Hidalgo}, {Lanzoni}, {Monelli}, {Ortolani}, {Renzini}, {Salaris},
  {Sarajedini}, {van der Marel}, {Vesperini}, \&
  {Zoccali}}]{2015AJ....149...91P}
{Piotto} G. {et~al.}, 2015, \aj, 149, 91

\bibitem[{{Portegies Zwart} {et~al}\mbox{.}(2004){Portegies Zwart},
  {Baumgardt}, {Hut}, {Makino}, \& {McMillan}}]{2004Natur.428..724P}
{Portegies Zwart} S.~F., {Baumgardt} H., {Hut} P., {Makino} J., {McMillan}
  S.~L.~W., 2004, \nat, 428, 724

\bibitem[{{Portegies Zwart} \& {McMillan}(2002)}]{2002ApJ...576..899P}
{Portegies Zwart} S.~F., {McMillan} S.~L.~W., 2002, \apj, 576, 899

\bibitem[{{Portegies Zwart} {et~al}\mbox{.}(2010){Portegies Zwart}, {McMillan},
  \& {Gieles}}]{2010ARA&A..48..431P}
{Portegies Zwart} S.~F., {McMillan} S.~L.~W., {Gieles} M., 2010, \araa, 48, 431

\bibitem[{{Prantzos} \& {Charbonnel}(2006)}]{2006A&A...458..135P}
{Prantzos} N., {Charbonnel} C., 2006, \aap, 458, 135

\bibitem[{{Prantzos} {et~al}\mbox{.}(2007){Prantzos}, {Charbonnel}, \&
  {Iliadis}}]{2007A&A...470..179P}
{Prantzos} N., {Charbonnel} C., {Iliadis} C., 2007, \aap, 470, 179

\bibitem[{{Prantzos} {et~al}\mbox{.}(2017){Prantzos}, {Charbonnel}, \&
  {Iliadis}}]{2017A&A...608A..28}
{Prantzos} N., {Charbonnel} C., {Iliadis} C., 2017, \aap, 608, A28

\bibitem[{{Regan} \& {Haehnelt}(2009)}]{2009MNRAS.396..343R}
{Regan} J.~A., {Haehnelt} M.~G., 2009, \mnras, 396, 343

\bibitem[{{Renzini} {et~al}\mbox{.}(2015){Renzini}, {D'Antona}, {Cassisi},
  {King}, {Milone}, {Ventura}, {Anderson}, {Bedin}, {Bellini}, {Brown},
  {Piotto}, {van der Marel}, {Barbuy}, {Dalessandro}, {Hidalgo}, {Marino},
  {Ortolani}, {Salaris}, \& {Sarajedini}}]{2015MNRAS.454.4197R}
{Renzini} A. {et~al.}, 2015, \mnras, 454, 4197

\bibitem[{{Rey-Raposo} {et~al}\mbox{.}(2017){Rey-Raposo}, {Dobbs}, {Agertz}, \&
  {Alig}}]{2017MNRAS.464.3536R}
{Rey-Raposo} R., {Dobbs} C., {Agertz} O., {Alig} C., 2017, \mnras, 464, 3536

\bibitem[{{Richer} {et~al}\mbox{.}(2013){Richer}, {Heyl}, {Anderson},
  {Kalirai}, {Shara}, {Dotter}, {Fahlman}, \& {Rich}}]{2013ApJ...771L..15R}
{Richer} H.~B., {Heyl} J., {Anderson} J., {Kalirai} J.~S., {Shara} M.~M.,
  {Dotter} A., {Fahlman} G.~G., {Rich} R.~M., 2013, \apjl, 771, L15

\bibitem[{{Rosen} {et~al}\mbox{.}(2016){Rosen}, {Krumholz}, {McKee}, \&
  {Klein}}]{2016MNRAS.463.2553R}
{Rosen} A.~L., {Krumholz} M.~R., {McKee} C.~F., {Klein} R.~I., 2016, \mnras,
  463, 2553

\bibitem[{{Sabbi} {et~al}\mbox{.}(2012){Sabbi}, {Lennon}, {Gieles}, {de Mink},
  {Walborn}, {Anderson}, {Bellini}, {Panagia}, {van der Marel}, \& {Ma{\'{\i}}z
  Apell{\'a}niz}}]{2012ApJ...754L..37S}
{Sabbi} E. {et~al.}, 2012, \apjl, 754, L37

\bibitem[{{Sakurai} {et~al}\mbox{.}(2017){Sakurai}, {Yoshida}, {Fujii}, \&
  {Hirano}}]{2017MNRAS.472.1677S}
{Sakurai} Y., {Yoshida} N., {Fujii} M.~S., {Hirano} S., 2017, \mnras, 472, 1677

\bibitem[{{Salaris} \& {Cassisi}(2014)}]{2014A&A...566A.109S}
{Salaris} M., {Cassisi} S., 2014, \aap, 566, A109

\bibitem[{{Salpeter}(1955)}]{1955ApJ...121..161S}
{Salpeter} E.~E., 1955, \apj, 121, 161

\bibitem[{{Sandage} \& {Wildey}(1967)}]{1967ApJ...150..469S}
{Sandage} A., {Wildey} R., 1967, \apj, 150, 469

\bibitem[{{Sbordone} {et~al}\mbox{.}(2010){Sbordone}, {Bonifacio}, {Caffau},
  {Ludwig}, {Behara}, {Gonz{\'a}lez Hern{\'a}ndez}, {Steffen}, {Cayrel},
  {Freytag}, {van't Veer}, {Molaro}, {Plez}, {Sivarani}, {Spite}, {Spite},
  {Beers}, {Christlieb}, {Fran{\c c}ois}, \& {Hill}}]{2010A&A...522A..26S}
{Sbordone} L. {et~al.}, 2010, \aap, 522, A26

\bibitem[{{Schaerer}(2003)}]{2003A&A...397..527S}
{Schaerer} D., 2003, \aap, 397, 527

\bibitem[{{Schaerer} \& {Charbonnel}(2011)}]{2011MNRAS.413.2297S}
{Schaerer} D., {Charbonnel} C., 2011, \mnras, 413, 2297

\bibitem[{{Senchyna} {et~al}\mbox{.}(2017){Senchyna}, {Stark},
  {Vidal-Garc{\'{\i}}a}, {Chevallard}, {Charlot}, {Mainali}, {Jones},
  {Wofford}, {Feltre}, \& {Gutkin}}]{2017MNRAS.472.2608S}
{Senchyna} P. {et~al.}, 2017, \mnras, 472, 2608

\bibitem[{{Siess}(2010)}]{2010A&A...512A..10S}
{Siess} L., 2010, \aap, 512, A10

\bibitem[{{Simioni} {et~al}\mbox{.}(2016){Simioni}, {Milone}, {Bedin},
  {Aparicio}, {Piotto}, {Vesperini}, \& {Hong}}]{2016MNRAS.463..449S}
{Simioni} M., {Milone} A.~P., {Bedin} L.~R., {Aparicio} A., {Piotto} G.,
  {Vesperini} E., {Hong} J., 2016, \mnras, 463, 449

\bibitem[{{Smilgys} \& {Bonnell}(2017)}]{2017MNRAS.472.4982S}
{Smilgys} R., {Bonnell} I.~A., 2017, \mnras, 472, 4982

\bibitem[{{Smith} {et~al}\mbox{.}(2011){Smith}, {Li}, {Silverman},
  {Ganeshalingam}, \& {Filippenko}}]{2011MNRAS.415..773S}
{Smith} N., {Li} W., {Silverman} J.~M., {Ganeshalingam} M., {Filippenko} A.~V.,
  2011, \mnras, 415, 773

\bibitem[{{Sobral} {et~al}\mbox{.}(2015){Sobral}, {Matthee}, {Darvish},
  {Schaerer}, {Mobasher}, {R{\"o}ttgering}, {Santos}, \&
  {Hemmati}}]{2015ApJ...808..139S}
{Sobral} D., {Matthee} J., {Darvish} B., {Schaerer} D., {Mobasher} B.,
  {R{\"o}ttgering} H.~J.~A., {Santos} S., {Hemmati} S., 2015, \apj, 808, 139

\bibitem[{{Spera} \& {Mapelli}(2017)}]{2017MNRAS.470.4739S}
{Spera} M., {Mapelli} M., 2017, \mnras, 470, 4739

\bibitem[{{Spite} \& {Spite}(1982)}]{1982A&A...115..357S}
{Spite} F., {Spite} M., 1982, \aap, 115, 357

\bibitem[{{Spitzer} \& {Hart}(1971)}]{1971ApJ...164..399S}
{Spitzer} L.~J., {Hart} M.~H., 1971, \apj, 164, 399

\bibitem[{{Stodolkiewicz}(1986)}]{1986AcA....36...19S}
{Stodolkiewicz} J.~S., 1986, \actaa, 36, 19

\bibitem[{{Suzuki} {et~al}\mbox{.}(2007){Suzuki}, {Nakasato}, {Baumgardt},
  {Ibukiyama}, {Makino}, \& {Ebisuzaki}}]{2007ApJ...668..435S}
{Suzuki} T.~K., {Nakasato} N., {Baumgardt} H., {Ibukiyama} A., {Makino} J.,
  {Ebisuzaki} T., 2007, \apj, 668, 435

\bibitem[{{Sz{\'e}csi} {et~al}\mbox{.}(2015){Sz{\'e}csi}, {Langer}, {Yoon},
  {Sanyal}, {de Mink}, {Evans}, \& {Dermine}}]{2015A&A...581A..15S}
{Sz{\'e}csi} D., {Langer} N., {Yoon} S.-C., {Sanyal} D., {de Mink} S., {Evans}
  C.~J., {Dermine} T., 2015, \aap, 581, A15

\bibitem[{{Sz{\'e}csi} {et~al}\mbox{.}(2017){Sz{\'e}csi}, {Mackey}, \&
  {Langer}}]{2017arXiv171104007S}
{Sz{\'e}csi} D., {Mackey} J., {Langer} N., 2017, ArXiv:1711.04007

\bibitem[{{Thompson}(2008)}]{2008ApJ...684..212T}
{Thompson} T.~A., 2008, \apj, 684, 212

\bibitem[{{Tout} {et~al}\mbox{.}(1999){Tout}, {Livio}, \&
  {Bonnell}}]{1999MNRAS.310..360T}
{Tout} C.~A., {Livio} M., {Bonnell} I.~A., 1999, \mnras, 310, 360

\bibitem[{{Turner} {et~al}\mbox{.}(2015){Turner}, {Beck}, {Benford},
  {Consiglio}, {Ho}, {Kov{\'a}cs}, {Meier}, \& {Zhao}}]{2015Natur.519..331T}
{Turner} J.~L., {Beck} S.~C., {Benford} D.~J., {Consiglio} S.~M., {Ho}
  P.~T.~P., {Kov{\'a}cs} A., {Meier} D.~S., {Zhao} J.-H., 2015, \nat, 519, 331

\bibitem[{{Vanzella} {et~al}\mbox{.}(2017){Vanzella}, {Calura}, {Meneghetti},
  {Mercurio}, {Castellano}, {Caminha}, {Balestra}, {Rosati}, {Tozzi}, {De
  Barros}, {Grazian}, {D'Ercole}, {Ciotti}, {Caputi}, {Grillo}, {Merlin},
  {Pentericci}, {Fontana}, {Cristiani}, \& {Coe}}]{2017MNRAS.467.4304V}
{Vanzella} E. {et~al.}, 2017, \mnras, 467, 4304

\bibitem[{{V{\'a}zquez-Semadeni} {et~al}\mbox{.}(2017){V{\'a}zquez-Semadeni},
  {Gonz{\'a}lez-Samaniego}, \& {Col{\'{\i}}n}}]{2017MNRAS.467.1313V}
{V{\'a}zquez-Semadeni} E., {Gonz{\'a}lez-Samaniego} A., {Col{\'{\i}}n} P.,
  2017, \mnras, 467, 1313

\bibitem[{{Ventura} {et~al}\mbox{.}(2001){Ventura}, {D'Antona}, {Mazzitelli},
  \& {Gratton}}]{2001ApJ...550L..65V}
{Ventura} P., {D'Antona} F., {Mazzitelli} I., {Gratton} R., 2001, \apjl, 550,
  L65

\bibitem[{{Ventura} {et~al}\mbox{.}(2013){Ventura}, {Di Criscienzo}, {Carini},
  \& {D'Antona}}]{2013MNRAS.431.3642V}
{Ventura} P., {Di Criscienzo} M., {Carini} R., {D'Antona} F., 2013, \mnras,
  431, 3642

\bibitem[{{Vink}(2018)}]{2018arXiv180308042V}
{Vink} J.~S., 2018, ArXiv:1803.08042

\bibitem[{{Vink} {et~al}\mbox{.}(2001){Vink}, {de Koter}, \&
  {Lamers}}]{2001A&A...369..574V}
{Vink} J.~S., {de Koter} A., {Lamers} H.~J.~G.~L.~M., 2001, \aap, 369, 574

\bibitem[{{Vink} {et~al}\mbox{.}(2011){Vink}, {Muijres}, {Anthonisse}, {de
  Koter}, {Gr{\"a}fener}, \& {Langer}}]{2011A&A...531A.132V}
{Vink} J.~S., {Muijres} L.~E., {Anthonisse} B., {de Koter} A., {Gr{\"a}fener}
  G., {Langer} N., 2011, \aap, 531, A132

\bibitem[{{Walker} {et~al}\mbox{.}(2016){Walker}, {Longmore}, {Bastian},
  {Kruijssen}, {Rathborne}, {Galv{\'a}n-Madrid}, \&
  {Liu}}]{2016MNRAS.457.4536W}
{Walker} D.~L., {Longmore} S.~N., {Bastian} N., {Kruijssen} J.~M.~D.,
  {Rathborne} J.~M., {Galv{\'a}n-Madrid} R., {Liu} H.~B., 2016, \mnras, 457,
  4536

\bibitem[{{Wijnen} {et~al}\mbox{.}(2016){Wijnen}, {Pols}, {Pelupessy}, \&
  {Portegies Zwart}}]{2016A&A...594A..30W}
{Wijnen} T.~P.~G., {Pols} O.~R., {Pelupessy} F.~I., {Portegies Zwart} S., 2016,
  \aap, 594, A30

\bibitem[{{Yong} {et~al}\mbox{.}(2015){Yong}, {Grundahl}, \&
  {Norris}}]{2015MNRAS.446.3319Y}
{Yong} D., {Grundahl} F., {Norris} J.~E., 2015, \mnras, 446, 3319

\bibitem[{{Yungelson} {et~al}\mbox{.}(2008){Yungelson}, {van den Heuvel},
  {Vink}, {Portegies Zwart}, \& {de Koter}}]{2008A&A...477..223Y}
{Yungelson} L.~R., {van den Heuvel} E.~P.~J., {Vink} J.~S., {Portegies Zwart}
  S.~F., {de Koter} A., 2008, \aap, 477, 223

\bibitem[{{Zinnecker}(1982)}]{1982NYASA.395..226Z}
{Zinnecker} H., 1982, Annals of the New York Academy of Sciences, 395, 226

\bibitem[{{Zocchi} {et~al}\mbox{.}(2017){Zocchi}, {Gieles}, \&
  {H{\'e}nault-Brunet}}]{2017MNRAS.468.4429Z}
{Zocchi} A., {Gieles} M., {H{\'e}nault-Brunet} V., 2017, \mnras, 468, 4429

\end{thebibliography}

\bsp	
\label{lastpage}
\end{document}